\documentclass{aa}  

\usepackage{graphicx}
\usepackage{longtable}
\usepackage{txfonts}
\begin{document}

   \title{Chemical analysis of early-type stars with planets
   }

   
   \titlerunning{Early-type stars with planets}
   \authorrunning{Saffe et al.}

   \author{C. Saffe\inst{1,2,6}, P. Miquelarena\inst{1,2,6}, J. Alacoria\inst{1,6},
           M. Flores\inst{1,2,6}, M. Jaque Arancibia\inst{4,5},
           D. Calvo\inst{2}, G. Mart\'in Girardi\inst{2},
           M. Grosso\inst{1,2,6} \and A. Collado\inst{1,2,6} 
           }

\institute{Instituto de Ciencias Astron\'omicas, de la Tierra y del Espacio (ICATE-CONICET), C.C 467, 5400, San Juan, Argentina.
               \email{csaffe@conicet.gov.ar}
         \and Universidad Nacional de San Juan (UNSJ), Facultad de Ciencias Exactas, F\'isicas y Naturales (FCEFN), San Juan, Argentina.
         \and Observatorio Astron\'omico de C\'ordoba (OAC), Laprida 854, X5000BGR, C\'ordoba, Argentina.
         \and{Instituto de Investigaci\'on Multidisciplinar en Ciencia y Tecnolog\'ia, Universidad de La Serena, Ra\'ul Bitr\'an 1305, La Serena, Chile}
         \and Departamento de F\'isica y Astronom\'ia, Universidad de La Serena, Av. Cisternas 1200 N, La Serena, Chile.
        \and Consejo Nacional de Investigaciones Cient\'ificas y T\'ecnicas (CONICET), Argentina
         }

   


   \date{Received xxx, xxx ; accepted xxxx, xxxx}

 
  \abstract
   {
    }
   {To explore the chemical pattern of early-type stars with planets, searching for a possible signature of planet formation.
   In particular, we study a likely relation between the $\lambda$ Bo\"otis chemical pattern and the presence of giant planets.
   }
   {We performed a detailed abundance determination in a sample of early-type stars with and without planets via spectral synthesis.
   Fundamental parameters were initially estimated using Str\"omgren photometry or literature values and then
   refined by requiring excitation and ionization balances of Fe lines.
   We derived chemical abundances for 23 different species by fitting observed spectra with an iterative process.
   Synthetic spectra were calculated using the program SYNTHE together with local thermodynamic equilibrium (LTE)
   ATLAS12 model atmospheres.
   We used specific opacities calculated for each star, depending on the individual composition and microturbulence
   velocity v$_\mathrm{micro}$ through the opacity sampling (OS) method.
   The complete chemical pattern of the stars were then compared to those of $\lambda$ Bo\"otis stars and other
   chemically peculiar stars.
   }
   {
   We compared the chemical pattern of the stars in our sample (13 stars with planets and 24 stars without detected planets)
   with those of $\lambda$ Bo\"otis and other chemically peculiar stars.
   We have found four $\lambda$ Bo\"otis stars in our sample, two of which
   present planets and circumstellar disks (HR 8799 and HD 169142) and one without planets detected (HD 110058).
   We have also identified the first $\lambda$ Bo\"otis star orbited by a brown dwarf ($\zeta$ Del).
   This interesting pair $\lambda$ Bo\"otis star + brown dwarf could help to test stellar formation scenarios. 
   We found no unique chemical pattern for the group of early-type stars bearing giant planets.
   However, our results support, in principle, a suggested scenario in which giant planets orbiting pre-main-sequence stars
   possibly block the dust of the disk and result in a $\lambda$ Bo\"otis-like pattern.
   On the other hand, we do not find a $\lambda$ Bo\"otis pattern in different hot-Jupiter planet host stars,
   which do not support the idea of possible accretion from the winds of hot-Jupiters,
   recently proposed in the literature.
   Then, other mechanisms should account for the presence of the $\lambda$ Bo\"otis pattern between main-sequence stars.
   Finally, we suggest that the formation of planets around $\lambda$ Bo\"otis stars such as HR 8799 and HD 169142
   is also possible through the core accretion process and not only gravitational instability.
   }
   {}
   

   \maketitle
%

\section{Introduction}

The $\lambda$ Bo\"otis stars are a group of chemically peculiar objects on the upper 
main-sequence, showing underabundances ($\sim$1-2 dex) of iron-peak elements and near-solar abundances
of \ion{C}{}, \ion{N}{}, \ion{O}{} and \ion{S}{} \citep[e.g. ][]{kamp01,heiter02,andrievsky02}.
The class was discovered by \citet{morgan43} and named following the bright prototype $\lambda$ Bo\"otis,
which is one extreme member of the class.
This group of refractory-poor objects comprises about 2$\%$ of early B through early F stars
\citep{gray-corbally98,paunzen01}.
However, among the pre-main-sequence Herbig Ae/Be stars, the progenitors of A-type stars,
the $\lambda$ Bo\"otis-like fraction is about 33\% \citep{folsom12}.
The origin of the peculiarity still remains a puzzle, as we can read in the recent discussion of
\citet{murphy-paunzen17}.
Unlike common chemical peculiarities seen in Am and Ap stars, $\lambda$ Bo\"otis stars are not
constrained to slow rotation \citep[][]{abt-morrell95,murphy15}.
\citet{cowley82} first suggested that $\lambda$ Bo\"otis could possibly originate from the interstellar
medium (ISM) with non-solar composition, or from the separation of grains and gas.
Then, \citet{venn-lambert90} proposed that $\lambda$ Bo\"otis stars likely occur when
circumstellar gas is separated from grains and then accreted to the stars, due to the similarity
of its abundance pattern with the ISM. 
Other proposed mechanisms include the interaction of a star with a diffuse interstellar cloud
\citep{kamp-paunzen02,mg09}, where the underabundances are produced by different amounts of accreted material.
\citet{tc93} estimate that once the accretion stops, the photospheric mixing and meridional
circulation would erase this peculiar signature on a $\sim$1 Myr timescale.
By studying the distribution of $\lambda$ Bo\"otis stars on the HR diagram, \citet{murphy-paunzen17}
conclude that multiple mechanisms could result in a $\lambda$ Bo\"otis spectra, depending on the age and
environment of the star.

In addition to the mentioned scenarios, a number of works propose a possible relation between the 
$\lambda$ Bo\"otis phenomena and the presence of planets. 
For instance, \citet{gray-corbally02} suggest that planetary bodies could perturb the orbits of comets and
volatile-rich objects, likely sending them toward the star and possibly originating this pattern.
\citet{marois08} detected three planets and a debris disk by direct imaging orbiting around the bright A5-type star HR 8799.
This object was one of the first early-type stars with planets detected, also showing $\lambda$ Bo\"otis-like abundances \citep{gray-kaye99,sadakane06}.
Then, \citet{kama15} studied the chemical abundances and disk properties for a sample of pre-main-sequence Herbig Ae/Be stars.
They propose that the depletion of heavy elements observed in $\sim$33\% of these stars \citep{folsom12},
originates when Jupiter-like planets (with mass between 0.1 and 10 M$_{Jup}$) block the accretion of part of the metal-rich 
dust of the primordial disk, while gas-phase C and O continues to flow toward the central star.
The authors also suggest that main-sequence stars, with a $\lambda$ Bo\"otis fraction of $\sim$2\%,
do not host massive protoplanetary disks and then their peculiarity should disappear on a $\sim$1 Myr timescale.
For the case of main-sequence stars, \citet{jura15} proposed that this peculiar abundance pattern could also be 
directly originated from the winds of hot-Jupiters,
taking the planet as a possible source of gas relatively near to the star. However, he caution that other channels
could likely result in the $\lambda$ Bo\"otis pattern \citep[see e.g.][]{murphy-paunzen17}.
Recently, \citet{kunimoto18} simulated numerically the pre-main-sequence
evolution of stars including the effects of planet formation, and concluded that stars with
{T$_{\rm eff}$ $>$ 7000 K} may show a metallic deficit compatible with the refractory-poor $\lambda$ Bo\"otis stars.
Therefore, the possible link between planet-bearing stars and the $\lambda$ Bo\"otis chemical pattern
motivated the present study.

To date, important trends are known for the case of late-type stars with planets, such as the giant
planet-metallicity correlation \citep{santos04,santos05,fischer-valenti05,johnson10a,sousa11}.
However, early-type stars with planets are poorly studied compared to late-type stars.
This is partly because hot stars rotate rapidly and have few spectral lines, making radial velocity searches
of planets more difficult. Transit and microlensing surveys are also more sensitive to planets orbiting
low-mass stars \citep[although for different reasons, see e.g.][]{wright-gaudi13}.
In the past few years, a slowly growing number of planets orbiting early-type stars were found,
detected mainly by transits \citep[e.g. from the KELT Collaboration, ][]{pepper07} and also from direct
imaging \citep[such as the GPIES survey, ][]{nielsen19}.
This give us the opportunity to start a homogeneous study of this interesting group of stars,
by performing (to our knowledge, for the first time) a detailed chemical analysis, allowing a
comparison with the $\lambda$ Bo\"otis pattern.
Our sample includes some remarkable objects (such as HR 8799, $\beta$ Pictoris, Fomalhaut, KELT-9), 
some important prototype/standard stars for comparison ($\lambda$ Bo\"otis, Vega),
and a number of stars for which no abundance study was previously performed in the literature
(HR 4502 A, BU Psc, HD 105850, HD 110058, HD 129926, HD 153053, HD 156751, HD 188228,
HD 23281, HD 50445, HD 56537 and V435 Car).


This work is organized as follows.
In Section 2 we describe the observations and data reduction, while in Section 3 we present
the stellar parameters and chemical abundance analysis. In Section 4 we show the results and
discussion, and finally in Section 5 we highlight our main conclusions.

\section{Stellar samples and observations}

We compiled a list of early-type stars with planets taken from the Extrasolar Planets Encyclopaedia\footnote{http://exoplanet.eu/}.
These stars comprise mainly A-type and few F-type stars, and are listed in the Table \ref{tab.samples}.
Some stars listed in the mentioned catalog present companions with masses above $\sim$13 M$_\mathrm{Jup}$,
the minimum mass to burn Deuterium \citep[e.g. ][]{grossman73,saumon96}. These stars are likely orbited by brown dwarfs (BDs)
rather than planets, and some of them are also included in this work for comparison.
We also included a group of early-type stars without planets nor brown dwarfs detected, taken from the
GPIES imaging survey \citep{nielsen19}, which is mainly focused on young and nearby stars.
We note that the stars included in this work were taken from different surveys using for instance transits and imaging, for which
the comparisons should be then performed with caution.
We also take the opportunity and include in our sample the stars Vega and $\lambda$ Bo\"otis,
as standard or prototype stars for comparison. 

Overall, the sample of stars analyzed in this work consists of 37 objects, including 13 stars with planets
(detected by transits or direct imaging), 3 stars likely orbited by brown dwarfs and 21 objects without planets nor brown dwarfs
detected. 
The effective temperatures cover 6730 K $<$ T$_{\rm eff}$ $<$ 10262 K and superficial gravities between 3.60 $<$ {$\log g$} $<$ 4.37.
The complete list of stars analyzed in this work is presented in Table \ref{tab.samples}, showing
the name of the star, spectral type (taken from the SIMBAD\footnote{http://simbad.u-strasbg.fr/simbad/} database),
companion detection method (transits, imaging or RV), companion mass, type of companion (planet or brown dwarf),
infrared IR excess, source of the spectra, signal-to-noise ratio (S/N) measured at $\sim$5000 \AA, and finally the 
reference for the companion data and/or IR excess. The column labeled IR Excess was added to give information about
the possible presence of circumstellar dust orbiting around the stars.

We downloaded archive spectra for the case of HARPS, HARPS-N, HIRES, SOPHIE and ELODIE spectrographs.
General characteristics of these instruments are showed in the Table \ref{table.spectrographs},
including the resolving power, CCD detector, pixel size, telescope and approximate wavelength range.
The reduction was performed by using the Data Reduction Software (DRS) pipeline for the case of HARPS and
HARPS-N spectra\footnote{https://www.eso.org/sci/facilities/lasilla/instruments/harps/doc.html},
using the reduction package MAKEE 3 with HIRES spectra\footnote{http://www.astro.caltech.edu/~tb/makee/},
the DRS pipeline with SOPHIE spectra\footnote{http://www.obs-hp.fr/guide/sophie/sophie-eng.shtml\#drs}
and the TACOS program with ELODIE spectra\footnote{http://www.obs-hp.fr/www/guide/elodie/manuser2.html}.
The continuum normalization and other operations (such as Doppler correction and combining spectra)
were performed using Image Reduction and Analysis Facility (IRAF)\footnote{IRAF is distributed by
the National Optical Astronomical Observatories, which is operated by the Association of Universities for
Research in Astronomy, Inc., under a cooperative agreement with the National Science Foundation.}.

\begin{table}
\centering
\caption{General characteristics of the spectrographs used in this work.}
\scriptsize
\begin{tabular}{lrcclc}
\hline
Instrument & R & CCD        & Pixel  & Telescope & Approx.     \\
           &   & Detector   & size   &           & Wave. Range \\
\hline
HARPS   & 115000 &  4k x 4k &  15 $\mu$m & La Silla 3.6m & 3800 - 6800 \\
HARPS-N & 115000 &  4k x 4k &  15 $\mu$m & TNG 3.6m      & 3800 - 6800 \\
HIRES   &  67000 &  2k x 4k &  15 $\mu$m & Keck 10m      & 3750 - 9000 \\
SOPHIE  &  75000 &  4k x 2k &  15 $\mu$m & OHP 1.93m     & 3900 - 6800 \\
ELODIE  &  42000 &  1k x 1k &  24 $\mu$m & OHP 1.93m     & 3850 - 6800 \\
REOSC   &  13000 &  1k x 1k &  24 $\mu$m & CASLEO 2.1m   & 3800 - 6000 \\
\hline
\end{tabular}
\normalsize
\label{table.spectrographs}
\end{table}

We also completed the sample with observations obtained at Complejo Astr\'onomico El Leoncito (CASLEO) during the
observing runs 2018A, 2018B and 2019A. We used the \emph{Jorge Sahade} 2.15 m telescope equipped with a REOSC high-resolution
echelle spectrograph\footnote{On loan from the Institute d'Astrophysique de Liege, Belgium} and a TEK 1024x1024 CCD detector.
The REOSC spectrograph uses gratings as cross dispersers, selecting in this case a grating with 400 lines mm$^{-1}$.
We take a number of stellar spectra for each target, followed by a ThAr lamp in order to derive an appropriate pixel versus
wavelength solution. The REOSC spectra were reduced using IRAF by performing the usual operations including, for example, bias
subtraction, flat fielding, spectral order extraction and wavelength calibration. The final spectra covered the visual
range $\lambda\lambda$3800-6000, and the average S/N ratio of the spectra was around $\sim$350.

\begin{table*}
\centering
\caption{Sample of stars studied in this work.}
\scriptsize
\begin{tabular}{lccccccccl}
\hline
\hline
Star name  & Spectral & Companion & Companion               & Companion & IR     & Spectra & S/N       & Companion \&\\
           & Type     & Detection & Mass [M$_\mathrm{Jup}$] & Type      & Excess &         & @5000 \AA & IR Excess Refs.\\
           &          & Method    &                  &           &         &           &     \\
\hline
\multicolumn{3}{l}{Exoplanet host stars} \\
\hline
$\beta$ Pic    &    A6V    &    Imaging    &    9.9, 8.9    &    Planet     &    Yes    &    HARPS    &    1500    &    R1, R14, R20, R21\\
Fomalhaut    &    A4V    &    Imaging    &    $<$ 1    &    Planet     &    Yes    &    REOSC    &    490    &    R2, R20, R21 \\
KELT-9    &    B9.5 – A0    &    Transit     &    2.88    &    Planet     &    No    &    HARPS-N    &    550    &    R3, R23 \\
HD 95086    &    A8III    &    Imaging    &    2.6    &    Planet     &    Yes    &    HARPS    &    310    &    R4, R14, R24, R25 \\
HD 169142    &    F1VekA3mA3 HAeBe     &    Imaging    &    $<$ 1, $<$ 10    &    Planet     &    Yes    &    HARPS    &    210    & R5, R26 \\
HR 8799    &    F0+VkA5mA5    &    Imaging    &    8.3,  8.3,  9.2    &    Planet     &    Yes    &    ELODIE    &    370    &  R6, R14, R27, R28 \\
KELT-17    &    A2    &    Transit     &    1.31    &    Planet     &    No    &    REOSC    &    210    &  R7 \\
KELT-20    &    A0    &    Transit     &    3.5    &    Planet     &    No    &    HARPS-N    &    450    &  R8 \\
MASCARA-1    &    A8    &    Transit     &    3.7    &    Planet     &    …    &    HARPS-N    &    560    &  R9  \\
WASP-33    &    kA5hA8mF4    &    Transit     &    $<$ 4.1    &    Planet     &    …    &    HIRES    &    250    & R10 \\
HAT-P-49    &    A8    &    Transit     &    1.73    &    Planet     &    …    &    SOPHIE    &    135    & R15 \\
WASP-167    &    F1V    &    Transit     &    $<$ 8    &    Planet     &    …    &    HARPS    &    205    &   R16 \\
WASP-189    &    A4/5IV/V    &    Transit     &    1.99    &    Planet     &    …    &    HARPS    &    1105    & R17 \\
\hline
\multicolumn{3}{l}{Stars with a brown dwarf (BD) companion} \\
\hline
$\beta$ Cir    &    A3Va    &    Imaging    &    58.7    &    BD    &    Yes    &    HARPS    &    250    &  R11, R22 \\
59 Dra    &    A7 HAeBe     &    RV    &    25    &    BD    &    …    &    ELODIE    &    350    &  R12 \\
$\zeta$ Del    &    A3Va    &    …    &    50    &    BD    &    …    &    ELODIE    &    290    &  R18  \\
\hline
\multicolumn{3}{l}{Stars with no-companion detected} \\
\hline
HD 133803    &    F2IVm-2    &    …    &    …    &    ...    &    …    &    HARPS    &    300    & R14 \\
HR 4502 A    &    A0V    &    …    &    …    &    ...    &    …    &    HARPS    &    130    & R14 \\
BU Psc    &    A9V/IV    &    …    &    …    &    ...    &    …    &    HARPS    &    110    &  R14  \\
HD 29391    &    F0 IV    &    …    &    …    &    ...    &    …    &    HARPS    &    410    &  R14  \\
HD 105850    &    A1V    &    …    &    …    &    ...    &    …    &    HARPS    &    650    &  R14  \\
HD 110058    &    A0V    &    …    &    …    &    ...    &    Yes    &    HARPS    &    450    &  R14, R29 \\
HD 115820    &    A7/8V    &    …    &    …    &    ...    &    …    &    HARPS    &    250    &  R14  \\
HD 120326    &    F0V    &    …    &    …    &    ...    &    …    &    HARPS    &    200    &  R14  \\
HD 129926    &    F0VSr+G1V    &    …    &    …    &    ...    &    …    &    HARPS    &    600    &  R14  \\
HD 146624    &    A1 Va    &    …    &    …    &    ...    &    …    &    HARPS    &    620    &  R14  \\
HD 153053    &    A5IV/V    &    …    &    …    &    ...    &    Yes    &    HARPS    &    600    &  R14, R22 \\
HD 156751    &    A7II/III    &    …    &    …    &    ...    &    …    &    HARPS    &    410    &  R14  \\
HD 159492    &    A5IV/V    &    …    &    …    &    ...    &    Yes    &    HARPS    &    720    &  R14, R22  \\
HD 188228    &    A0Va    &    …    &    …    &    ...    &    Yes    &    HARPS    &    840    &  R14, R21  \\
HD 23281    &    A5III/IV    &    …    &    …    &    ...    &    …    &    HARPS    &    470    &  R14  \\
HD 50445    &    A3V    &    …    &    …    &    ...    &    …    &    HARPS    &    680    &  R14  \\
HD 56537    &    A4IV    &    …    &    …    &    ...    &    Yes    &    HARPS    &    620    &  R14, R20  \\
HD 88955    &    A2Va    &    …    &    …    &    ...    &    Yes    &    HARPS    &    590    &  R14, R20  \\
V435 Car    &    A5V    &    …    &    …    &    ...    &    Yes    &    HARPS    &    200    &  R14  \\
$\lambda$ Bo\"otis    &    A0Va lBoo    &    …    &    …    &    …    &    Yes    &    ELODIE    &    525    & R20, R21 \\
Vega    &    A0Va    &    …    &    …    &    …    &    Yes    &    SOPHIE    &    1230    & R19, R20, R21 \\
\hline
\end{tabular}
\tablebib{
Companions data: R1 \citep{lagrange19}, R2 \citep{kalas13}, R3 \citep{borsa19}, R4 \citep{derosa16}, 
R5 \citep{fedele17}, R6 \citep{wang18}, R7 \citep{zhou16}, R8 \citep{lund17}, R9 \citep{talens17}, R10 \citep{cc10}, 
R11 \citep{smith15}, R12 \citep{galland06}, R13 \citep{aller13}, R14 \citep{nielsen19}, R15 \citep{bieryla14},
R16 \citep{temple17}, R17 \citep{lendl20}, R18 \citep{derosa14}, R19 \citep{matra20},
R20 \citep{rieke05}, R21 \citep{su06}, R22 \citep{morales11}, R23 \citep{gaudi17}, R24 \citep{moor13},
R25 \citep{su17}, R26 \citep{reggiani14}, R27 \citep{su09}, R28 \citep{matthews14}, R29 \citep{esposito20}
}
\normalsize
\label{tab.samples}
\end{table*}

\begin{figure}
\centering
\includegraphics[width=8cm]{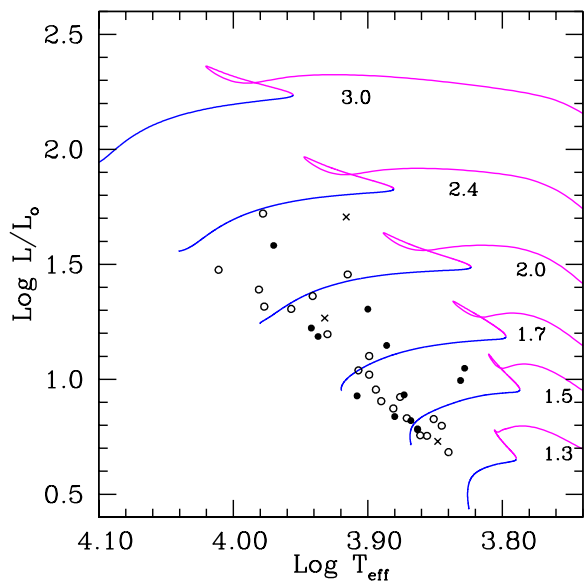}
\caption{Effective temperature versus luminosity for stars with planets, without planets and
with a BD companion (filled circles, empty circles and crosses).
Evolutionary tracks for stars of different masses are showed in blue/magenta
for main-sequence/evolved stars. The numbers are expressed in solar masses.}
\label{fig.hr}%
\end{figure}

We present in the Fig. \ref{fig.hr} a theoretical HR diagram ($\log$ T$_{\rm eff}$ versus $\log$ L/L$_{\sun}$)
for the stars in our sample.
The luminosity L was evaluated from the visual magnitude corrected by interstellar reddening according to the 
extinction maps of \citet{schlegel98},\footnote{https://irsa.ipac.caltech.edu/applications/DUST/} following the procedure of \citet{bilir08}.
We used Gaia DR2 parallaxes \citep[][]{gaia18} and bolometric corrections interpolated in the tables of \citet{flower96}.
Stars with planets, without planets and with a BD companion are showed with filled circles, 
empty circles and crosses, respectively. We also show PARSEC evolutionary tracks \citep{bressan12} for stars with different masses.
Blue and magenta lines correspond to main-sequence and evolved phases.

\section{Stellar parameters and abundance analysis}

\subsection{Effective temperature and gravity}

\begin{figure}
\centering
\includegraphics[width=8cm]{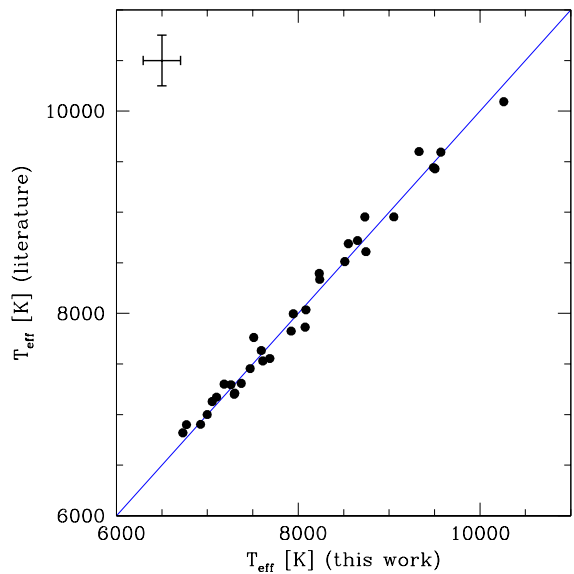}
\caption{Effective temperature T$_{\rm eff}$ derived in this work versus literature data.
Average dispersion bars are showed in the upper left corner of the panel.}
\label{fig.teff}%
\end{figure}

Effective temperature T$_{\rm eff}$ and superficial gravity {$\log g$} were first estimated by using the
Str\"omgren uvby$\beta$ mean photometry of \citet{hauck-mermilliod98} for most stars in our sample
or by taking previously published results.
We used the program TempLogG \citep{kaiser06} together with the calibration of \citet{napi93} and
derredenned colors according to \citet{domingo-figueras99}, in order to derive the fundamental parameters.
Then, these values were refined (when neccesary and/or possible) by enforcing 
excitation and ionization balances of the iron lines. The same strategy was previously
applied in the literature \citep[e.g. ][]{acke04,saffe-levato14}.
The values derived in this way are listed in the Table \ref{table.params}, with an average dispersion of
$\sim$175 K and $\sim$0.15 dex for T$_{\rm eff}$ and {$\log g$}.
In the Fig. \ref{fig.teff} we compare T$_{\rm eff}$ values with those collected from literature 
\citep{erspamer-north03,lepine03,paunzen06,masana06,zorec-royer12,derosa16,zhou16,kah16,gray17,lund17,talens17,borsa19}, 
showing a general agreement. Average dispersion bars are showed in the upper left corner of the panel.

\subsection{Rotational velocities $v\sin i$}

\begin{figure}
\centering
\includegraphics[width=8cm]{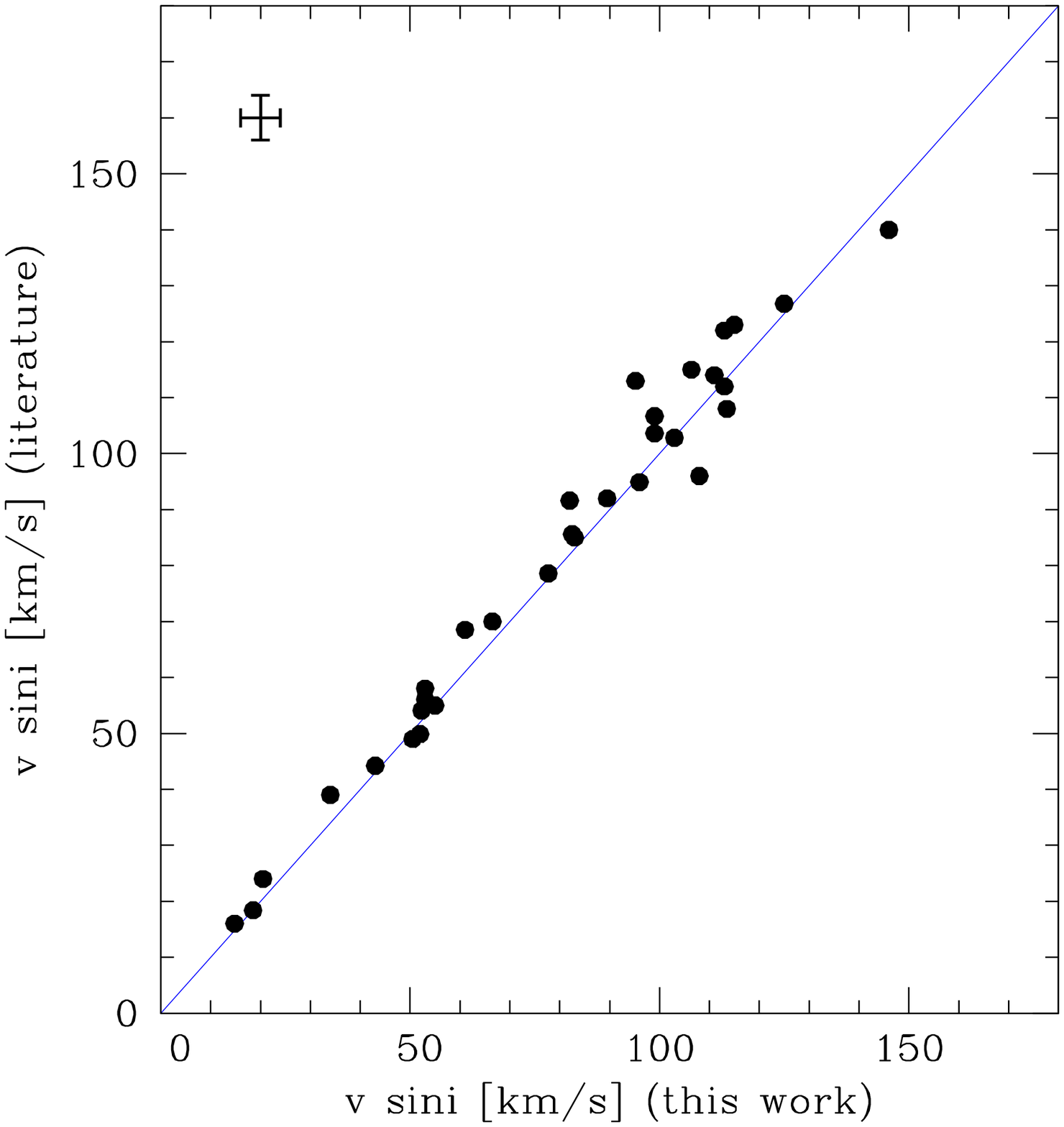}
\caption{Projected rotational velocity $v\sin i$ derived in this work versus literature data.
Average dispersion bars are showed in the upper left corner of the panel.}
\label{fig.vsini}%
\end{figure}

Projected rotational velocities $v\sin i$ were first estimated by fitting the observed profile of the line \ion{Mg}{II} 4481.23 \AA\
and then refined by fitting most \ion{Fe}{I} and \ion{Fe}{II} lines in the spectra. Synthetic spectra were calculated
using the program SYNTHE \citep{kurucz-avrett81} together with ATLAS9 \citep{kurucz93} model atmospheres,
and then convolved with a rotational profile (using the Kurucz's command \textit{rotate}) and 
with an instrumental profile for each spectrograph (using the command \textit{broaden}).
Rotational velocities were varied for each line, adopting as final value the average of different lines. 
The resulting $v\sin i$ values are showed in the 4th column of Table \ref{table.params},
covering between 14.8 and 153.5 km s$^{-1}$ for the stars in our sample.
The average dispersion in $v\sin i$ resulted $\sim$2.2 km s$^{-1}$.
In the Fig. \ref{fig.vsini} we compare $v\sin i$ values derived in this work with those from literature,
showing a general agreement. Average dispersion bars are showed in the upper left corner of the panel.

\subsection{Microturbulence velocity}

Microturbulence velocity v$_\mathrm{micro}$ is commonly used as a free parameter in abundance analysis.
Several studies have found that microturbulence appears to vary systematically with T$_{\rm eff}$ for early-type stars
\citep{chaffee70,nissen81,coupry-burkhart92,gray01,takeda08,gebran14}.
They showed that v$_\mathrm{micro}$ increases with increasing T$_{\rm eff}$, peaking around mid-A type ($\sim$8500 K),
before falling away to almost $\sim$0 km s$^{-1}$ for B-type stars.
In this work, we adopted the formula derived by \citet{gebran14} in order to estimate v$_\mathrm{micro}$ as a
function of T$_{\rm eff}$, which is valid for $\sim$6000 K $<$ T$_{\rm eff}$ $<$ $\sim$10000 K. 
\citet{gebran14} showed that the formula has an uncertainty of $\sim$25 $\%$,
which we adopted as the error in the v$_\mathrm{micro}$ values. 
The effect of this relatively large uncertainty (and uncertainties of other fundamental parameters)
on the abundance values is estimated in the next sections.

\begin{table*}
\centering
\caption{Fundamental parameters derived for the stars in this work.}
\begin{tabular}{lrrrr}
\hline
Star & T$_{\rm eff}$ [K]& $\log g$ [dex]& v$_\mathrm{micro}$ [km s$^{-1}$]& $v\sin i$ [km s$^{-1}$]\\
\hline
$\beta$ Pic             &   8084  $\pm$ 130  &   4.22 $\pm$  0.13  &   3.31 $\pm$  0.83  &  113.0 $\pm$  1.3   \\ 
Fomalhaut               &   8745  $\pm$ 195  &   4.17 $\pm$  0.10  &   2.95 $\pm$  0.74  &   82.0 $\pm$  3.2   \\ 
KELT-9                  &   9329  $\pm$ 118  &   4.00 $\pm$  0.14  &   2.27 $\pm$  0.57  &  113.5 $\pm$  4.6   \\ 
HD 95086                &   7593  $\pm$ 122  &   4.02 $\pm$  0.14  &   3.09 $\pm$  0.77  &   30.0 $\pm$  1.4   \\ 
HD 169142               &   7296  $\pm$ 365  &   4.20 $\pm$  0.25  &   2.75 $\pm$  0.69  &   55.0 $\pm$  0.8   \\ 
HR 8799                 &   7301  $\pm$ 190  &   4.12 $\pm$  0.23  &   2.76 $\pm$  0.69  &   50.5 $\pm$  1.0   \\ 
KELT-17                 &   7471  $\pm$ 210  &   4.20 $\pm$  0.14  &   2.50 $\pm$  0.63  &   43.0 $\pm$  0.6   \\ 
KELT-20                 &   8652  $\pm$ 160  &   4.11 $\pm$  0.20  &   3.03 $\pm$  0.76  &  111.0 $\pm$  3.1   \\ 
MASCARA-1               &   7687  $\pm$ 238  &   4.09 $\pm$  0.14  &   3.17 $\pm$  0.79  &   99.0 $\pm$  3.9   \\ 
WASP-33                 &   7373  $\pm$ 164  &   4.14 $\pm$  0.20  &   2.86 $\pm$  0.71  &   82.5 $\pm$  2.0   \\ 
HD 133803               &   6998  $\pm$ 193  &   4.08 $\pm$  0.23  &   2.29 $\pm$  0.57  &   89.5 $\pm$  1.3   \\ 
$\beta$ Cir             &   8552  $\pm$ 297  &   3.94 $\pm$  0.18  &   3.12 $\pm$  0.78  &   61.0 $\pm$  0.8   \\ 
59 Dra                  &   7053  $\pm$ 194  &   4.18 $\pm$  0.21  &   2.38 $\pm$  0.60  &   53.0 $\pm$  1.1   \\ 
HR 4502 A               &   9569  $\pm$ 253  &   4.00 $\pm$  0.16  &   1.96 $\pm$  0.49  &   18.5 $\pm$  0.7   \\ 
BU Psc                  &   7185  $\pm$ 160  &   4.22 $\pm$  0.14  &   2.59 $\pm$  0.65  &   53.0 $\pm$  2.6   \\ 
HD 29391                &   7259  $\pm$ 167  &   4.12 $\pm$  0.20  &   2.70 $\pm$  0.68  &   70.0 $\pm$  1.7   \\ 
HD 105850               &   9052  $\pm$ 167  &   4.08 $\pm$  0.08  &   2.61 $\pm$  0.65  &  125.0 $\pm$  1.5   \\ 
HD 110058               &   7839  $\pm$ 202  &   4.00 $\pm$  0.14  &   3.26 $\pm$  0.81  &  153.5 $\pm$  2.9   \\ 
HD 115820               &   7610  $\pm$ 135  &   4.30 $\pm$  0.15  &   3.11 $\pm$  0.78  &   92.2 $\pm$  1.8   \\ 
HD 120326               &   6925  $\pm$ 143  &   4.37 $\pm$  0.11  &   2.17 $\pm$  0.54  &   66.5 $\pm$  1.5   \\ 
HD 129926               &   7101  $\pm$ 167  &   4.18 $\pm$  0.20  &   2.46 $\pm$  0.62  &  113.0 $\pm$  2.1   \\ 
HD 146624               &   9489  $\pm$ 184  &   4.13 $\pm$  0.07  &   2.06 $\pm$  0.52  &   34.0 $\pm$  0.5   \\ 
HD 153053               &   7916  $\pm$ 129  &   4.03 $\pm$  0.14  &   3.29 $\pm$  0.82  &  103.0 $\pm$  1.6   \\ 
HD 156751               &   7432  $\pm$ 175  &   3.92 $\pm$  0.21  &   2.93 $\pm$  0.73  &  100.0 $\pm$  4.3   \\ 
HD 159492               &   8076  $\pm$ 128  &   4.24 $\pm$  0.13  &   3.31 $\pm$  0.83  &   52.3 $\pm$  1.2   \\ 
HD 188228               &   10262 $\pm$ 172  &   4.29 $\pm$  0.06  &   1.17 $\pm$  0.29  &   83.0 $\pm$  3.5   \\ 
HD 23281                &   7761  $\pm$ 135  &   4.18 $\pm$  0.15  &   3.22 $\pm$  0.80  &   77.7 $\pm$  2.4   \\ 
HD 50445                &   7922  $\pm$ 117  &   4.03 $\pm$  0.14  &   3.29 $\pm$  0.82  &   96.0 $\pm$  1.9   \\ 
HD 56537                &   8231  $\pm$ 122  &   3.70 $\pm$  0.12  &   3.29 $\pm$  0.82  &  146.0 $\pm$  4.3   \\ 
HD 88955                &   8733  $\pm$ 154  &   3.76 $\pm$  0.10  &   2.96 $\pm$  0.74  &   99.0 $\pm$  3.0   \\ 
V435 Car                &   7510  $\pm$ 165  &   4.21 $\pm$  0.14  &   3.01 $\pm$  0.75  &  106.4 $\pm$  1.9   \\ 
HAT-P-49                &   6730  $\pm$ 234  &   4.02 $\pm$  0.14  &   1.82 $\pm$  0.46  &   14.8 $\pm$  0.2   \\ 
$\lambda$ Bo\"otis      &   8512  $\pm$ 144  &   3.95 $\pm$  0.06  &   3.15 $\pm$  0.79  &  115.0 $\pm$  3.1   \\ 
Vega                    &   9505  $\pm$ 188  &   3.95 $\pm$  0.19  &   2.04 $\pm$  0.51  &   20.5 $\pm$  0.7   \\ 
WASP-167                &   6770  $\pm$ 210  &   4.05 $\pm$  0.14  &   1.90 $\pm$  0.47  &   52.0 $\pm$  2.0   \\ 
WASP-189                &   7946  $\pm$ 136  &   3.85 $\pm$  0.12  &   3.30 $\pm$  0.82  &   95.2 $\pm$  3.3   \\ 
$\zeta$ Del             &   8236  $\pm$ 124  &   3.60 $\pm$  0.12  &   3.29 $\pm$  0.82  &  108.0 $\pm$  3.6   \\ 
\hline
\end{tabular}
\normalsize
\label{table.params}
\end{table*}

\subsection{Abundance analyses}

We applied an iterative procedure to determine the chemical abundances for the stars in our sample.
We started by computing an ATLAS12 \citep{kurucz93} model atmosphere for the adopted
T$_{\rm eff}$, $\log g$ and v$_\mathrm{micro}$ values. For this initial model, we used solar metallicity values
taken from \citet{asplund09}. The abundances were determined by fitting a synthetic spectra
to the different lines using the program SYNTHE \citep{kurucz93}.
With the new abundance values, we derived a new model atmosphere and started the process again.
If neccessary, T$_{\rm eff}$ and $\log g$ were refined to achieve the balance of Fe I and Fe II lines.
In each step, opacities were calculated for an arbitrary composition and v$_\mathrm{micro}$ using the opacity
sampling (OS) method, similar to previous works \citep{saffe18,saffe19,saffe20}.
In this case, two runs of ATLAS12 are used \citep[see e.g.][]{castelli05,saffe18}: the first for a 
preselection of important lines and the second for the final calculation of the model structure.
In this way, abundances are consistently derived using specific opacities rather than solar-scaled values.
An estimation of the differences obtained when using these two approaches for the case of solar-type
stars, can be seen in \citet{saffe18}.

We compared observed and synthetic spectra using a $\chi^{2}$ function, calculated as the 
quadratic sum of the differences between both spectra. The abundances were varied in steps of 0.01 dex
until reach a minimum in $\chi^{2}$, similar to \citet{saffe-levato14}. 
The fits were also verified by eye inspection, using intervals spanning 10 \AA\ around the lines of interest.
Abundances derived in this way are presented in the Table \ref{table.abunds1} of the Appendix,
showing the average and dispersion for the stars in our sample.
The values are showed using the square bracket notation, which denotes abundances
relative to the Sun, that is [N/H]=log(N/H)$_{Star}$ $-$ log(N/H)$_{Sun}$,
where log(N/H)$_{Star}$ and log(N/H)$_{Sun}$ are abundance values for
the star and for the Sun, the later taken from \citet{asplund09}.

\begin{figure*}
\centering
\includegraphics[width=16cm]{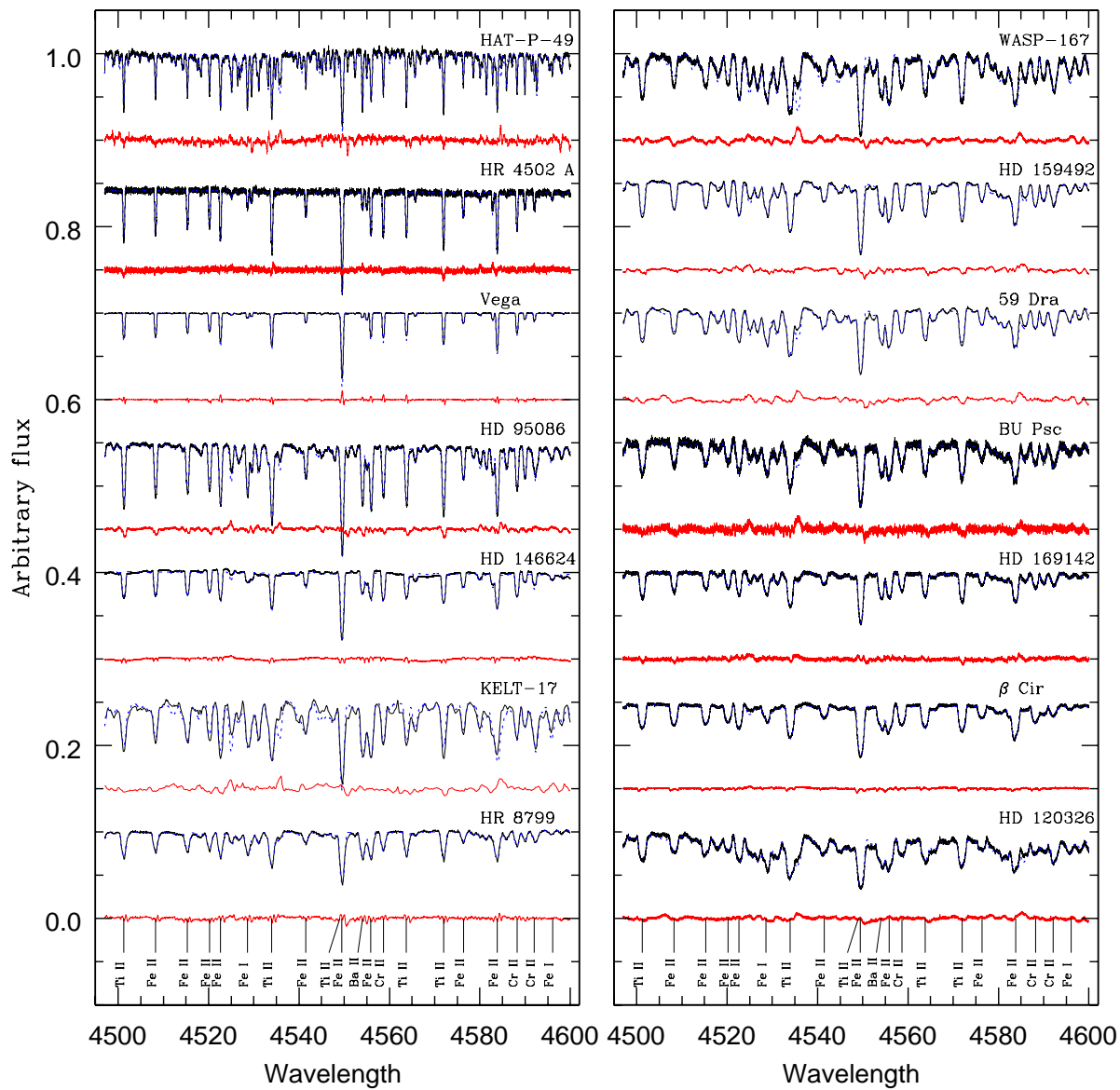}
\caption{Observed, synthetic and difference spectra (black, blue dotted and red lines) 
for the stars in our sample, sorted by $v\sin i$.}
\label{fig.range1}%
\end{figure*}

\begin{figure*}
\centering
\includegraphics[width=16cm]{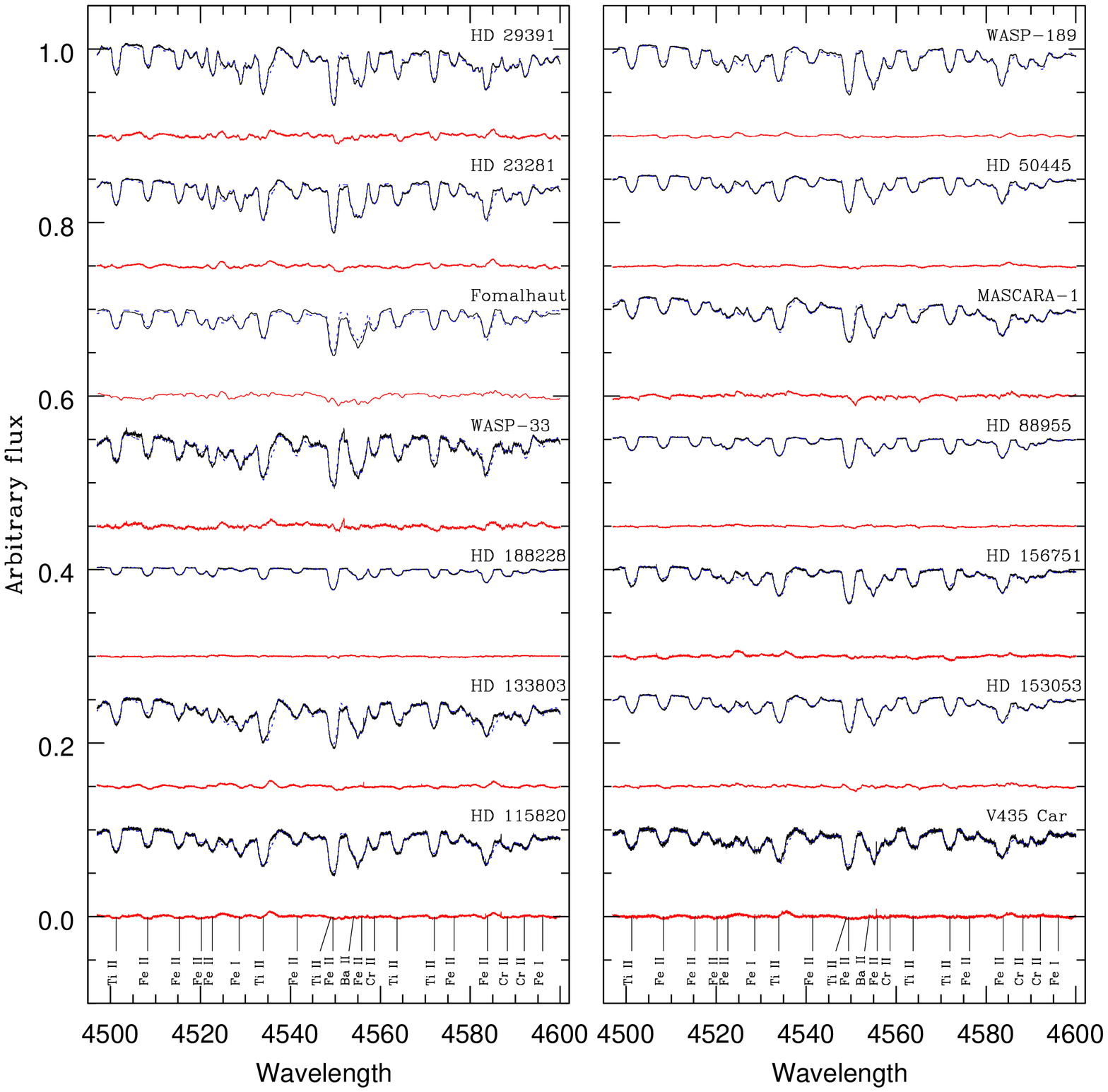}
\caption{Observed, synthetic and difference spectra (black, blue dotted and red lines) 
for the stars in our sample, sorted by $v\sin i$.}
\label{fig.range2}%
\end{figure*}

\begin{figure*}
\centering
\includegraphics[width=16cm]{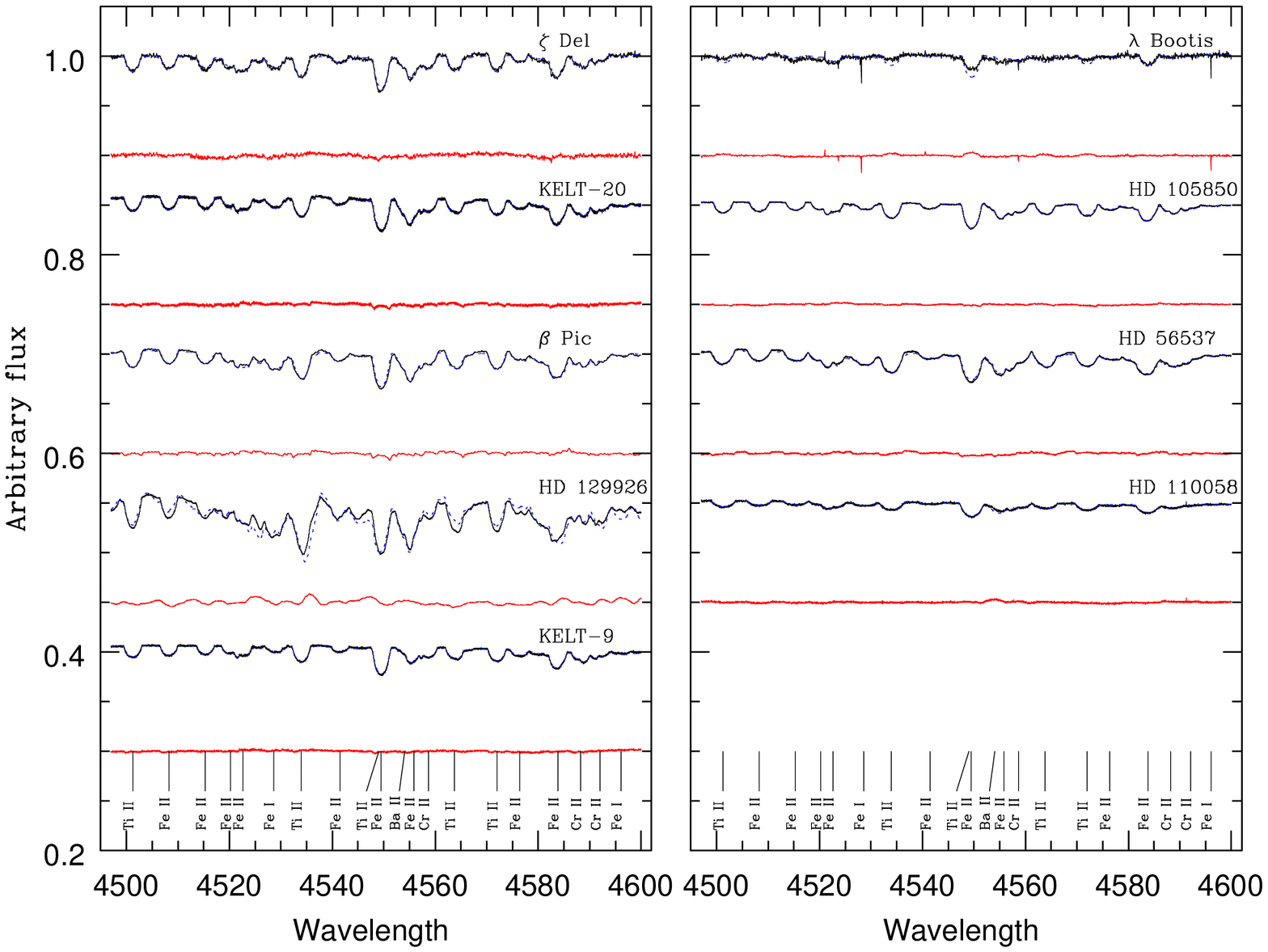}
\caption{Observed, synthetic and difference spectra (black, blue dotted and red lines) 
for the stars in our sample, sorted by $v\sin i$.}
\label{fig.range3}%
\end{figure*}

Chemical abundances were derived for 23 different species, including
\ion{C}{I}, \ion{C}{II}, \ion{O}{I}, \ion{Mg}{I}, \ion{Mg}{II},
\ion{Al}{I}, \ion{Al}{II}, \ion{Si}{II}, \ion{Ca}{I}, \ion{Ca}{II}, 
\ion{Sc}{II}, \ion{Ti}{II}, \ion{Cr}{II}, \ion{Mn}{I}, \ion{Mn}{II}, 
\ion{Fe}{I}, \ion{Fe}{II}, \ion{Ni}{II}, \ion{Zn}{I}, \ion{Sr}{II},
\ion{Y}{II}, \ion{Zr}{II} and \ion{Ba}{II}. 
The atomic line list and laboratory data used in this work are basically those described in \cite{castelli-hubrig04},
updated with specialized references as described in Sect. 7 of \citet{gonzalez14}.
In Figs. \ref{fig.range1} to \ref{fig.range3} we present an example of observed, synthetic and difference spectra
(black, blue dotted and red lines) for the stars in our sample. The stars are sorted in the plot by increasing $v\sin i$.
There is a good agreement between modeling and observations for the lines of
different chemical species. 

\subsection{Uncertainty of abundance values}

In general, the uncertainty in the abundance values have different sources.
First, we estimated the measurement error e$_{1}$ from the line-to-line dispersion
as $\sigma/\sqrt{n}$ , where $\sigma$ is the standard deviation of the line-by-line
abundances and n is the number of lines. For elements with only one line, we adopted
for $\sigma$ the standard deviation of the iron lines.
Then, we determined the contribution to the abundance error due to the uncertainty 
in stellar parameters. We modified T$_{\rm eff}$ and $\log g$ by their uncertainties 
and recalculated the abundances, obtaining the corresponding differences e$_{2}$ and e$_{3}$.
We also wanted to estimate the contribution to the error due to the 
uncertainty in v$_\mathrm{micro}$, which is not always included in the error calculation.
Similarly, we modified v$_\mathrm{micro}$ by its uncertainty and obtained the difference e$_{4}$
by recalculating the abundances.
Finally, the total error was estimated as the quadratic sum of e$_{1}$, e$_{2}$, e$_{3}$ and e$_{4}$.
These values are presented in the Table \ref{table.abunds1} of the Appendix, in order to determine their
contribution to the total error.

\subsection{NLTE effects}

The basic difference between LTE and NLTE is the behavior of atomic level populations.
LTE allows a relatively simple calculation using a Saha-Boltzman distribution, while in NLTE
the level populations are affected by the radiation field and should be determined by
kinetic equilibrium \citep[see e.g.][]{kubat14}.
Departures from LTE are more pronounced in stars with high temperature and with low gravity
and metallicity.  For instance, the reduction of surface gravity results in a decreased
efficiency of collisions with electrons and hydrogen atoms, reducing the thermalizing
effect, which leads to stronger NLTE effects \citep[e.g. ][]{gratton99}.
Subsequently, some particular species and transitions should be taken with caution.

\citet{rentzsch96} estimated NLTE departures for \ion{C}{I} lines in early-type stars (up to $\sim$0.2 dex
for lines with equivalent widths W$_{eq}\sim$100 m\AA, and lower departures with decreasing W$_{eq}$). Weak and intense
\ion{C}{II} lines could also show pronounced NLTE effects \citep{przybilla11}, except for \ion{C}{II} 5145 {\AA} and
other members of the multiplet.
For \ion{O}{I}, NLTE effects are expected, especially in the near-IR triplet \ion{O}{I} 7771 {\AA} and the
other lines of the same multiplet \citep[e.g.][]{sitnova13,przybilla11}.
For the case of Mg, \citet{przybilla01} found that the intense line \ion{Mg}{II} 4481 systematically
yields notably higher abundances due to NLTE effects (between 0.2 - 0.8 dex for early-type stars). 
More recently, \citet{alexeeva18} studied the formation of Mg lines under LTE and NLTE.
For stars with 7000 K $<$ T$_{\rm eff}$ $<$ 17500 K, they recommend to use the \ion{Mg}{II} lines 
3848.21 {\AA}, 4427.99 {\AA}, 4384.64 {\AA} and 4390.57 {\AA} even at the LTE assumption. However, these lines are not always
available in our spectra. For stars with 7000 K $<$ T$_{\rm eff}$ $<$ 8000 K, they showed that the
line 4702.99 and the \ion{Mg}{I} b triplet could be safely used in LTE. 
For each star, they also estimate that the average difference between \ion{Mg}{I} and \ion{Mg}{II}
diminish from $\sim$0.23 dex in LTE to $\sim$0.09 dex in NLTE. This could explain, at least in part, the
higher abundances  of \ion{Mg}{II} compared to \ion{Mg}{I} observed in some stars.
However, the authors also caution that the difference \ion{Mg}{II} $-$ \ion{Mg}{I} even in NLTE
could amount up to $\sim$0.24 dex for metal-poor stars, for a reason that require future investigation.
For the case of \ion{Ti}{II}, \citet{sitnova16} showed some small NLTE departures (for instance, up to
$\sim$0.03 dex for the line 5336.79) and originating possible discrepancies between \ion{Ti}{I} and \ion{Ti}{II}.

\subsection{Comparison with literature}

\begin{figure}
\centering
\includegraphics[width=8cm]{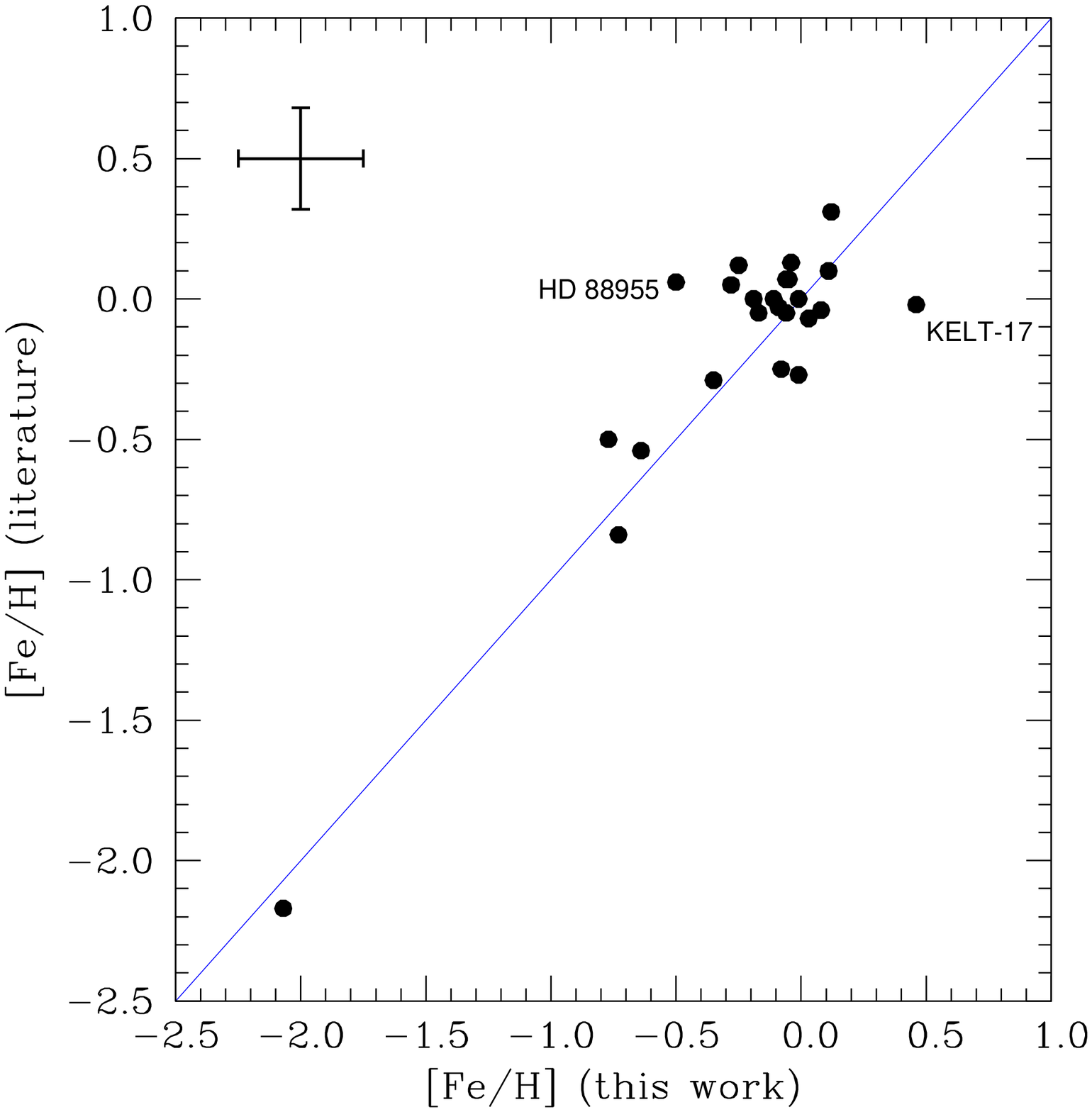}
\caption{Metallicity values derived in this work ([Fe/H]) versus literature data.
Average dispersion bars are showed in the upper left corner of the panel.
The stars HD 88955 and KELT-17 are identified in the plot (see text for more details).}
\label{fig.metal1}%
\end{figure}

We present in the Fig. \ref{fig.metal1} a comparison of metallicity values derived in this work
and from literature. We collected these values from different sources 
\citep{andrievsky02,erspamer-north03,gray06,sadakane06,cc10,yoon10,prugniel11,folsom12,moor13,bieryla14,zhou16,gaudi17,luck17,lund17,talens17,temple17,
anderson18,bochanski18,arentsen19,cheng19}.
The Fig. \ref{fig.metal1} shows a good agreement with literature data in general for most stars.
The point in the lower corner of the plot corresponds to the star $\lambda$ Bo\"otis.

The largest differences in metallicity corresponds to the stars KELT-17 and HD 88955, with [Fe/H] differences
of 0.48 and 0.56 dex, respectively. Both stars are identified in the Fig. \ref{fig.metal1}.
Given the relatively large differences, we briefly explore the possible source of discrepancies.
\citet{zhou16} discovered a giant planet around KELT-17 and adopted a final [Fe/H] value of -0.018$\pm$0.073 dex by
iterating through a global analysis of the transit, constrained by a transit-derived stellar density and stellar isochrone models.
The authors also note that, when using only spectroscopic data (a single interval centered around the Mg b lines),
the resulting metallicity could vary between -0.10$\pm$0.08 dex and +0.25$\pm$0.08 dex (by fixing the $\log g$ value). 
In this way, the global solution adopted by \citet{zhou16} depends on several parameters and models,
not specifically designed to derive abundances.
On the other hand, \citet{saffe20} showed that the chemical pattern of KELT-17 closely resembles those of Am stars
(for many species and not only iron), using a process different than those used by \citet{zhou16}.

The other star with a notable difference in metallicity is HD 88955. \citet{erspamer-north03} derived for this object
[Fe/H] $=$ 0.06 dex using an automatic procedure with an ELODIE (R$\sim$42000) spectra.
They used a spectral synthesis program (modified by the authors) in order to accept ATLAS9 model atmospheres instead of ATLAS5
models as in their original version.
On the other hand, for this star we obtained [Fe/H] $=$ -0.50$\pm$0.14 dex using a HARPS (R$\sim$115000) spectra together with ATLAS12 models.
The difference is possibly due to the adopted stellar parameters (difference of 220 K in T$_{\rm eff}$ and 0.37 dex in $\log g$)
and laboratory data for the lines. Unfortunately, \citet{erspamer-north03} do not report a [Fe/H] error for this star, 
which difficults a later comparison.

\section{Discussion}

In the present section, we discuss different aspects related to the metallic content of the early-type stars with planets.
We start by searching $\lambda$ Bo\"otis stars in our sample and studying the possible relation between hot-Jupiter planets and $\lambda$ Bo\"otis stars.
Then, we search for other possible chemically peculiar stars in our sample.
Finally, we discuss the possible implications of our results within the context of the planet formation models.

\subsection{$\lambda$ Bo\"otis stars in our sample}

\begin{figure*}
\centering
\includegraphics[width=8cm]{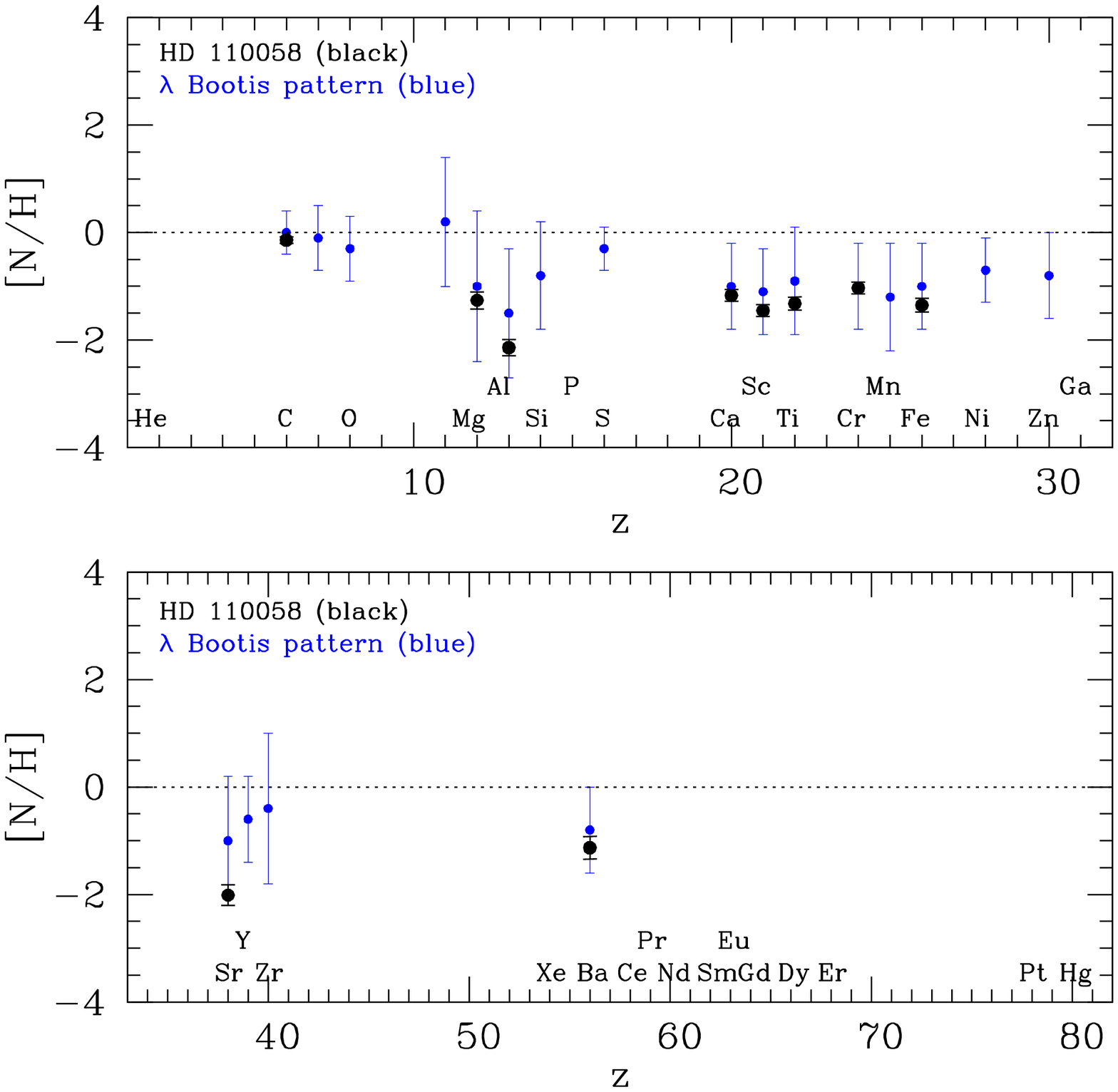}
\includegraphics[width=8cm]{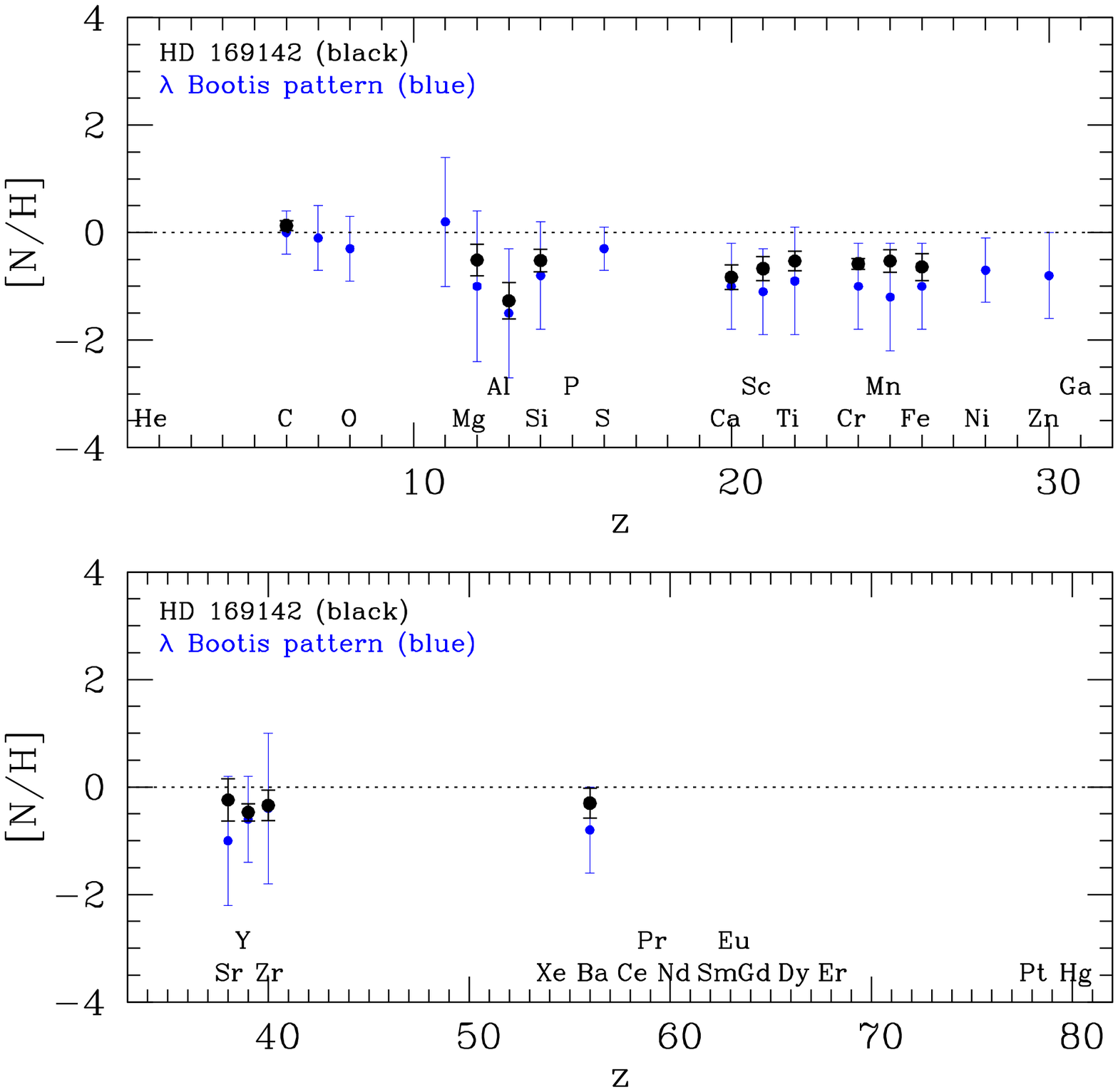}
\includegraphics[width=8cm]{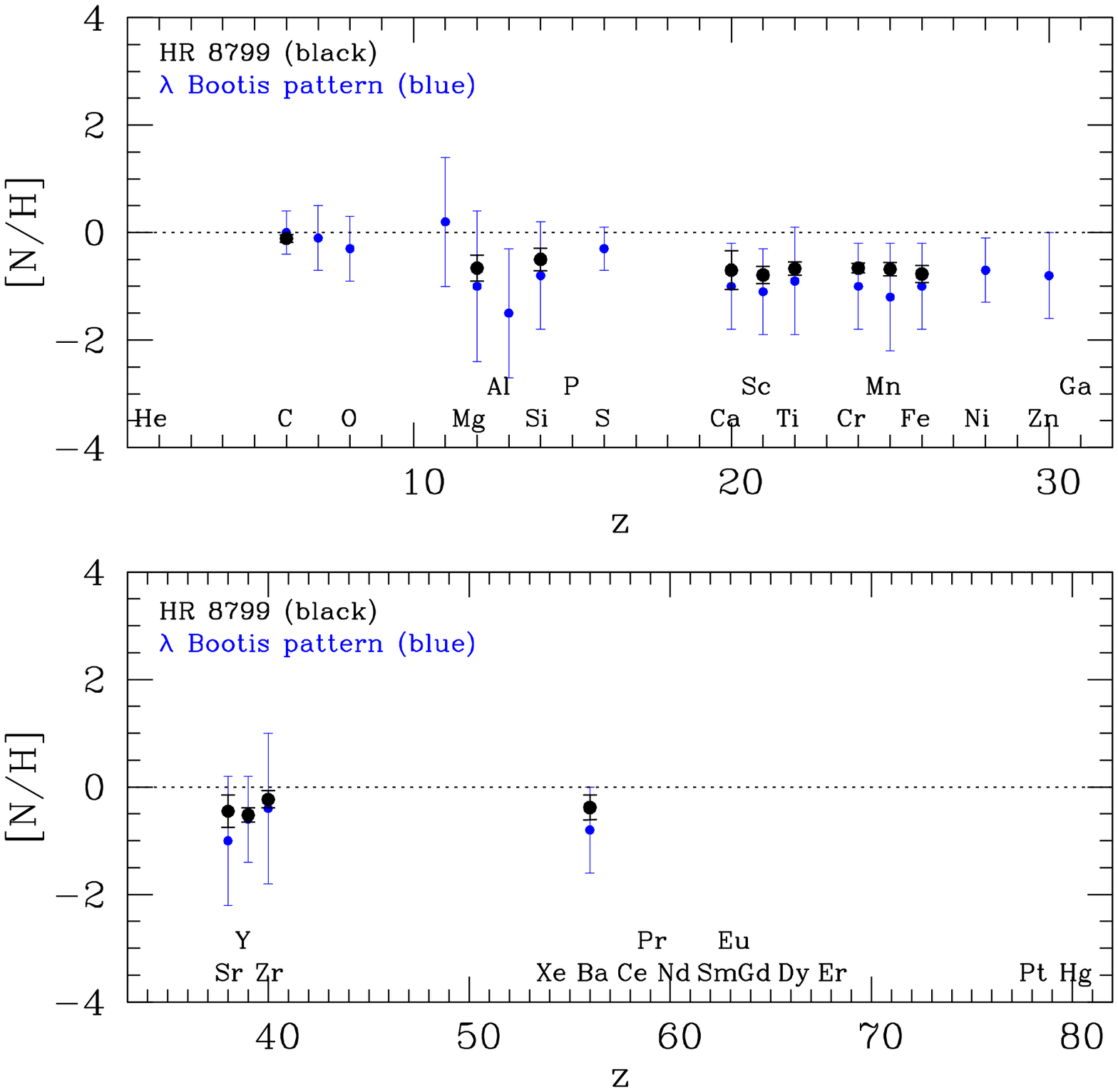}
\includegraphics[width=8cm]{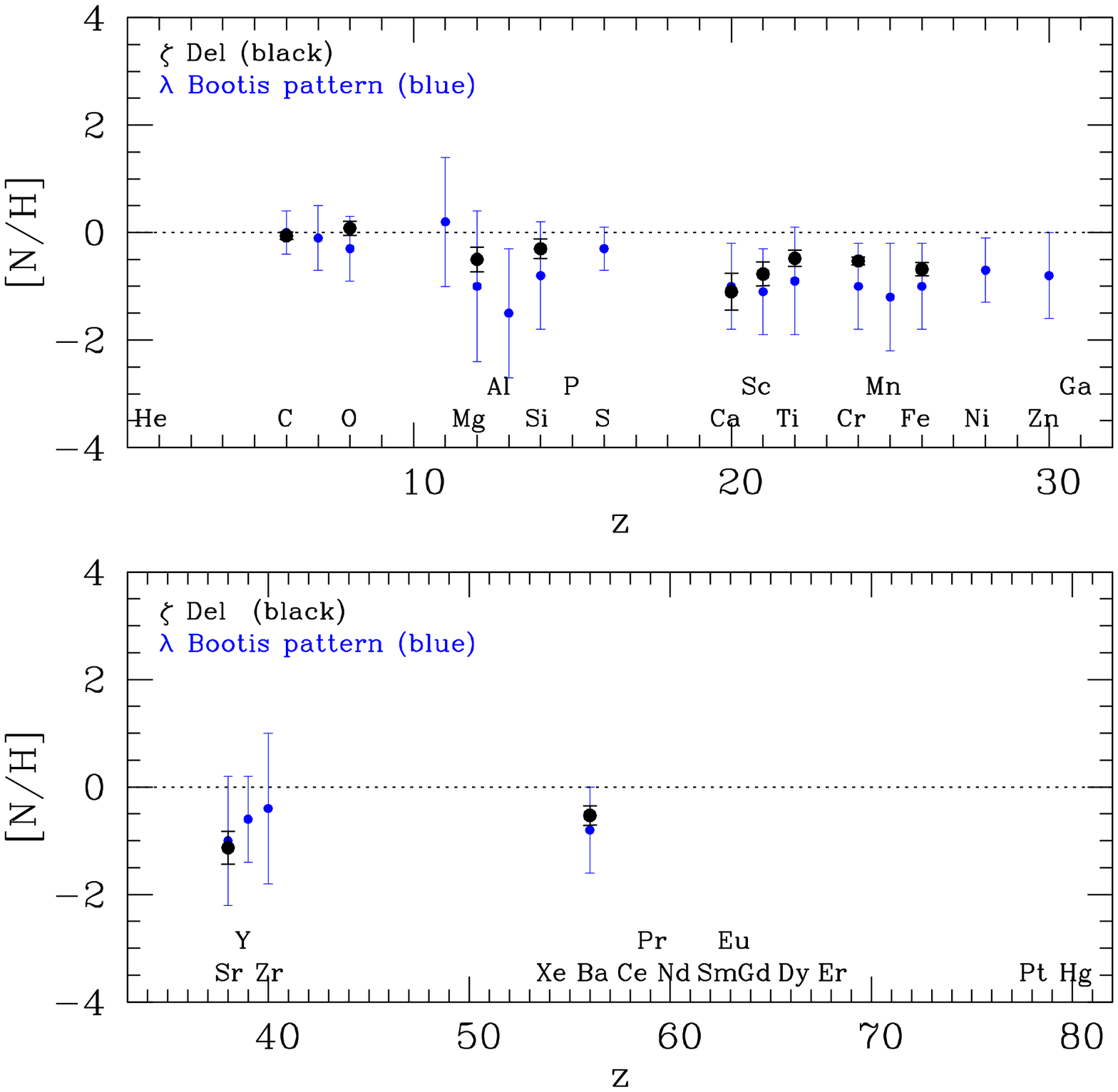}
\caption{Abundances of four early-type stars (black), compared to the average pattern of $\lambda$ Bo\"otis stars (blue).
We show two panels for each star, corresponding to elements with {z$<$32} and {z$>$32}.}
\label{fig.lamboo.stars}%
\end{figure*}

We compared the abundances of our sample stars with the average pattern of $\lambda$ Bo\"otis stars.
As explained in the Introduction, $\lambda$ Bo\"otis stars are early-type objects showing underabundances ($\sim$1-2 dex)
of iron-peak elements and near-solar abundances of \ion{C}{}, \ion{N}{}, \ion{O}{} and \ion{S}{} \citep[e.g. ][]{kamp01,heiter02,gray17}.
A representative chemical pattern was derived by using the average of 12 $\lambda$ Bo\"otis stars taken from \citet{heiter02}.
As result of this comparison, we have identified four (or five) stars with the $\lambda$ Bo\"otis pattern:
HD 110058, HD 169142, HR 8799 and $\zeta$ Del (the star $\lambda$ Bo\"otis itself would be the 5th object).
We present in the Fig. \ref{fig.lamboo.stars} the abundances of these four stars, compared to the average pattern
of $\lambda$ Bo\"otis stars. The vertical bars in the average $\lambda$ Bo\"otis pattern correspond to the standard deviation
of the different stars, as derived by \citet{heiter02}.
The figure shows two panels for each star, corresponding to elements with atomic number 
{z$<$32} and {z$>$32}. In general, these stars present near-solar values of C and O, together with subsolar values of the
other metals. An individual description for each star can be read in the Appendix B.
HD 169142 was also previously identified as a $\lambda$ Bo\"otis object \citep{folsom12,gray17},
as well as HR 8799 \citep{gray-kaye99,sadakane06} and $\lambda$ Bo\"otis \citep{venn-lambert90,paunzen99,cheng19}.

Most of the $\lambda$ Bo\"otis stars identified here present evidence of circumstellar material.
HR 8799 and HD 169142 are orbited by dusty disks and planets detected by direct imaging \citep{marois08,fedele17}.
Up to now, there is no planet detected around HD 110058, although presents an IR excess indicative of a dusty disk \citep{nielsen19,esposito20}.
It is also interesting to note that $\zeta$ Del is orbited by a brown dwarf detected by direct imaging \citep{derosa14},
with a projected separation of 13.51$\pm$0.08 arcsec, or 912$\pm$15 AU at the distance of $\zeta$ Del. 
There is also evidence of IR excess around $\zeta$ Del likely related to the presence of dust \citep{nielsen19,esposito20}.
To our knowledge, $\zeta$ Del is the first $\lambda$ Bo\"otis star orbited by a brown dwarf,
which could be used as a laboratory to test stellar formation scenarios.
Interestingly, our sample includes other two stars orbited by brown dwarfs (59 Dra and $\beta$ Cir), although not
showing $\lambda$ Bo\"otis characteristics.

\begin{figure*}
\centering
\includegraphics[width=8cm]{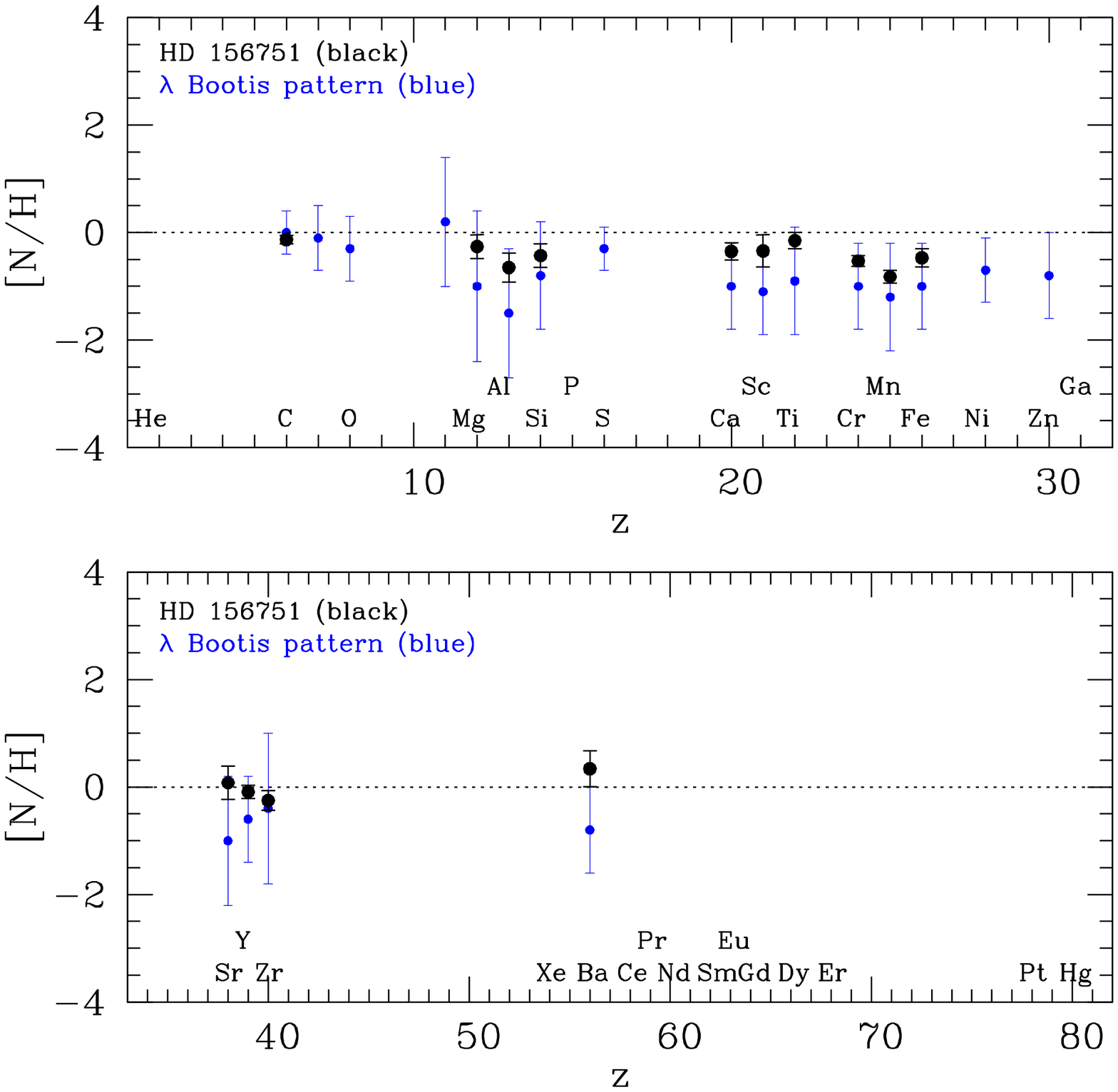}
\includegraphics[width=8cm]{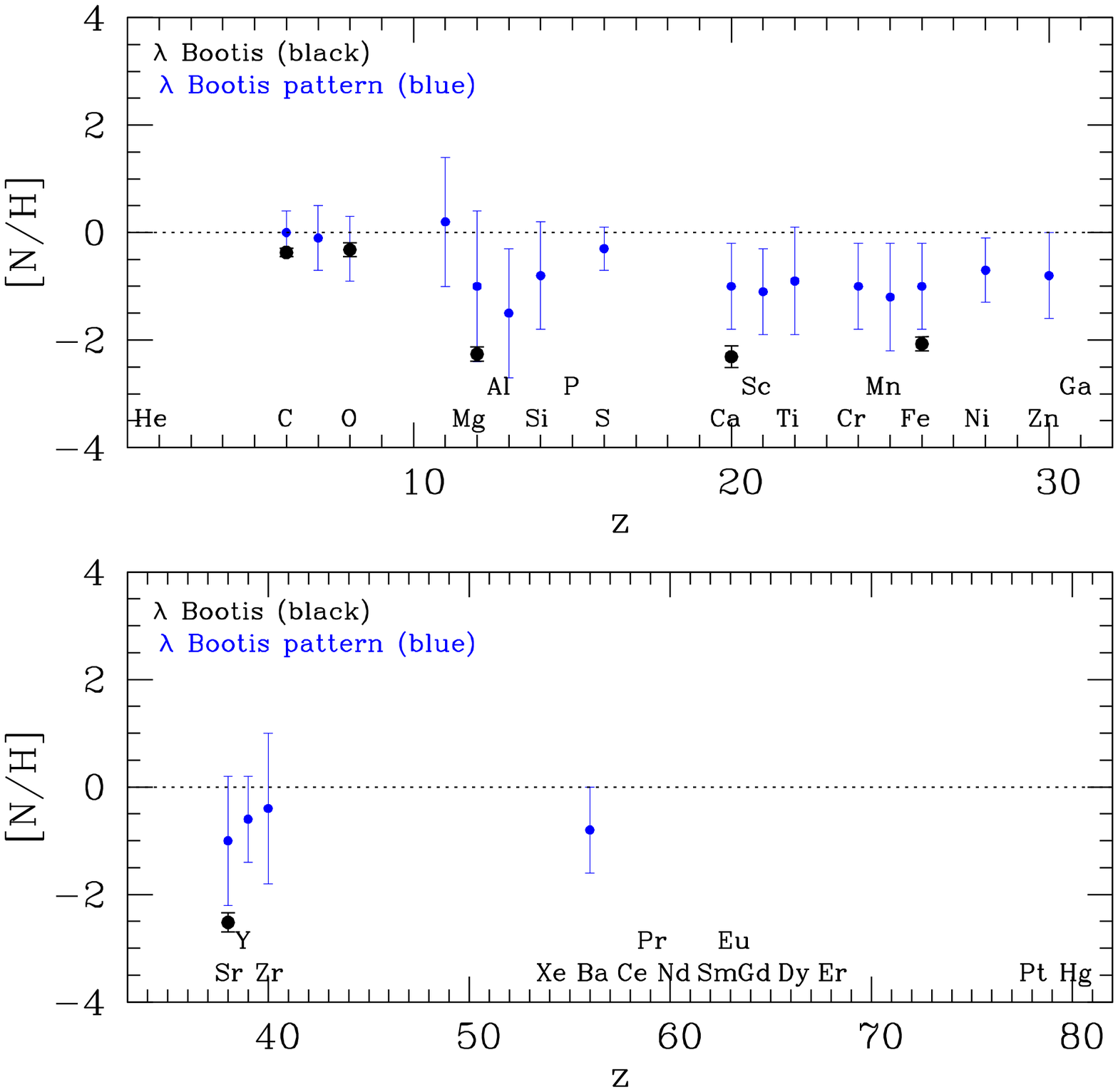}
\caption{Abundances of the stars HD 156751 and $\lambda$ Bo\"otis (black), compared to the average pattern
of $\lambda$ Bo\"otis stars (blue). We show two panels for each star, corresponding to elements with {z$<$32}
and {z$>$32}.}
\label{fig.mild-lamboo.stars}%
\end{figure*}

We also present in the Fig. \ref{fig.mild-lamboo.stars} the abundances of HD 156751, which shows a less clear
$\lambda$ Bo\"otis signature (their $\sim$solar values of Sr, Y and Ba are not typical of $\lambda$ Bo\"otis stars).
The same Fig. \ref{fig.mild-lamboo.stars} presents the star $\lambda$ Bo\"otis itself, which shows rather extreme characteristics for its class
(a very low metallic content). Up to now, there is no planet detected around HD 156751 nor $\lambda$ Bo\"otis
\citep{nielsen19}, while there is IR excess detected around $\lambda$ Bo\"otis indicative of a dusty disk
\citep{rieke05,su06}.

In our sample of early-type stars, there is no obvious relation between the presence of planets
(13 stars) and the $\lambda$ Bo\"otis pattern (4 objects, or 5 if we include the star $\lambda$ Bo\"otis).
\citet{kama15} proposed that the $\lambda$ Bo\"otis observed in $\sim$33\% of pre-main-sequence
Herbig AeBe stars \citep{folsom12}, originates when Jupiter-like planets (with mass between 0.1 and 10 M$_{Jup}$)
block the accretion of dust from the primordial disk.
Our sample includes the star HD 169142, which have been classified as a young pre-main-sequence object and
also as a $\lambda$ Bo\"otis star \citep{folsom12}. This object is included in our sample and shows the mentioned
chemical peculiarity (see Fig. \ref{fig.lamboo.stars}).
Although not strictly a pre-main-sequence object, HR 8799 is also a young star \citep[age of 42$\pm$5 Myr, ][]{nielsen19}
which shows the $\lambda$ Bo\"otis signature (see Fig. \ref{fig.lamboo.stars}).
Then, the $\lambda$ Bo\"otis signature that we observe in these two stars, support in principle the scenario proposed
by \citet{kama15}.

It is worthwhile to mention that $\beta$ Pictoris and HD 95086 are also young stars (ages $<\sim$50 Myr)
hosting giant planets \citep{lagrange19,derosa16}, but not showing a clear $\lambda$ Bo\"otis signature.
$\beta$ Pictoris is sligthly metal-poor (including C), while Ca and Ba show solar values.
Its $\lambda$ Bo\"otis classification was initially suggested based on its evolutionary status and the presence 
of a circumstellar disk \citep{king-patten92}. However, this peculiar classification was then ruled out
by different authors, using an optical spectral \citep{holweger97} and also from the ratio of UV lines \citep{cheng16}.
In this work, we do not identify the star $\beta$ Pictoris with a $\lambda$ Bo\"otis pattern.
The other young star (HD 95086) presents mostly solar or very slightly subsolar abundances, different than 
average $\lambda$ Bo\"otis stars.
However, the fact that these young stars ($\beta$ Pictoris and HD 95086) do not display the $\lambda$ Bo\"otis
peculiarity, do not rule out the scenario proposed by \citet{kama15}.
$\beta$ Pictoris and HD 95086 are young stars although not strictly accreting pre-main-sequence objects.
\citet{kama15} propose that for main-sequence stars without a massive protoplanetary disk
(that is, after the accretion of volatile-rich gas), the peculiarity should disappear on a short timescale ($\sim$1 Myr),
as estimated by \citet{tc93}. Following the scenario of \citet{kama15}, it is possible that these young stars showed
the $\lambda$ Bo\"otis peculiarity in the past, which was then erased few Myrs ago.

\subsection{$\lambda$ Bo\"otis stars and hot-Jupiter planets}

Hot-Jupiter planets present short orbital periods {($<$ 10 d)} and large planetary masses {($>$ 0.1 M$_{Jup}$)},
that is, they are gas giants orbiting very close to their stars \citep[e.g. ][]{wang15}.
Recently, \citet{jura15} proposed that $\lambda$ Bo\"otis stars could be originated by accreting volatile-rich gas
from the winds of hot-Jupiter planets, rather than from the interaction with a molecular cloud.
In our sample, there are some stars orbited by hot-Jupiter planets, all detected by the transits technique:
WASP-33, WASP-167, WASP-189, KELT-9, KELT-17, KELT-20, MASCARA-1 and HAT-P-49.
We present in the Figures \ref{fig.hot-jup1} and \ref{fig.hot-jup2} of the Appendix, the abundances of these
hot-Jupiter hosts, compared to the average pattern of $\lambda$ Bo\"otis stars.
We do not recognize a clear $\lambda$ Bo\"otis pattern in these objects.
Most of them show approximately solar abundances; the possible exceptions are WASP-167 showing a metal-rich spectra,
and KELT-17 which is an Am star \citep{saffe20}.
KELT-20 presents subsolar metallic abundances, being likely closer to the $\lambda$ Bo\"otis pattern.
However, this star still presents differences with this class ([C/H]=-0.42$\pm$0.11 is subsolar, while Si and Ba present suprasolar values),
which rule out its $\lambda$ Bo\"otis nature.

In principle, the abundances derived for these stars do not support the accretion scenario from the winds of hot-Jupiters.
This is possibly due to a number of reasons, as explained by \citet{jura15}.
For example, the hot-Jupiter composition is assumed to be nearly solar; the elements in the flow should be efficiently separated
(this depends on the viscosity of the gas); the amount of planetary gas expelled (and then accreted onto the star)
should be enough to produce the effect; and finally, the flow from the planet should not be magnetically funneled
(this perhaps avoid elemental separations). If these conditions are not met, the $\lambda$ Bo\"otis pattern will likely not appear
\citep{jura15}. The author caution that other channels could also result in a $\lambda$ Bo\"otis pattern.
In particular, \citet{murphy-paunzen17} conclude that multiple mechanisms could possibly produce a $\lambda$ Bo\"otis spectra,
depending on the age and environment of the star.

\subsection{Other chemically peculiar stars in our sample}

\begin{figure}
\centering
\includegraphics[width=8cm]{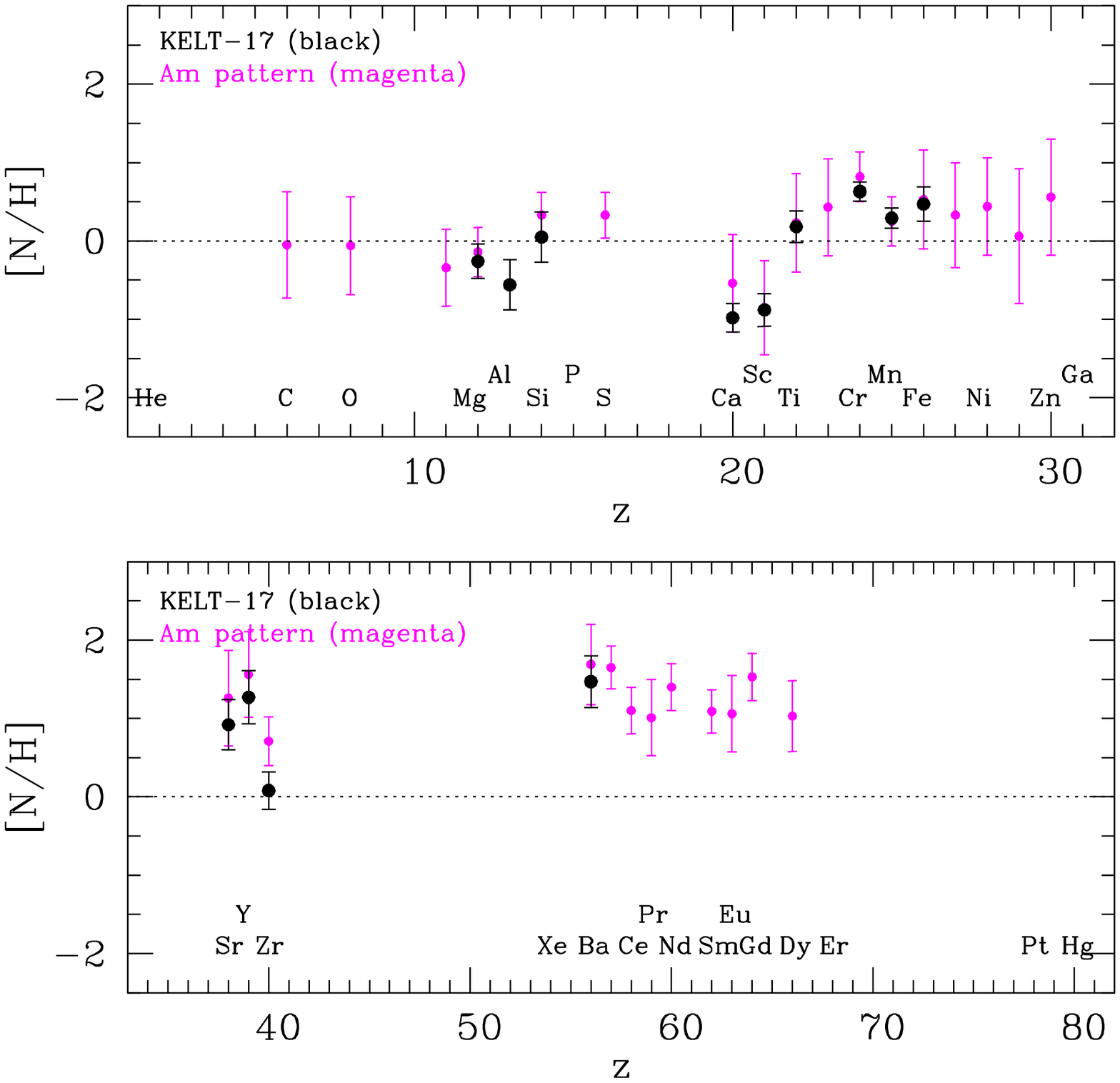}
\caption{Abundances of the exoplanet host star KELT-17 (black), compared to the average pattern
of Am stars (magenta). The two panels correspond to elements with {z$<$32} and {z$>$32}.}
\label{fig.kelt-17}%
\end{figure}

We also compared the abundances of the early-type stars with those of chemically peculiar Am, ApSi and HgMn stars.
In this case, the origin of their peculiar abundances is commonly attributed to diffussion processes
\citep[e.g. ][]{michaud70,michaud76,michaud83,vauclair78,richer00}, taking place in the stable atmospheres of 
slowly rotating A-type stars. In principle, there is no direct relation between these peculiar patterns and the presence of planets.
After inspecting the chemical patterns of our early-type stars, we only found one clear Am object: the exoplanet host star KELT-17 \citep{saffe20}.
In general, Am stars present overabundances of most heavy elements in their spectra, particularly \ion{Fe}{} and \ion{Ni}{}, 
together with underabundances of \ion{Ca}{II} and \ion{Sc}{II} \citep[see e.g. the work of ][ and references therein]{catanzaro19}.
We compared the abundances of KELT-17 with an average pattern of 62 Am stars recently determined by \citet{catanzaro19}
and found a good agreement, as we can see in the Fig. \ref{fig.kelt-17}.
KELT-17 is the first exoplanet host whose chemical pattern was identified as an Am star \citep{saffe20},
being an early result of this study.
Although not included in the present sample, other planet bearing stars with a possible Am pattern include
KELT-19 \citep{siverd17} and KELT-26 A \citep{rm19}.

In general, by inspecting the abundances of the planet host stars in our sample,
we found a number of stars showing mostly solar values 
(Fomalhaut, KELT-9, MASCARA-1, WASP-33, HAT-P-49 and WASP-189).
Other objects show a $\lambda$ Bo\"otis signature (such as the young stars HD 169142 and HR 8799),
while other planet hosts show a subsolar or slightly subsolar metallic content ($\beta$ Pictoris, HD 95086 and KELT-20).
In addition, we also found a chemically peculiar Am star (KELT-17) and a metal-rich star (WASP-167).
Then, no single chemical pattern could account for the complete group of planet bearing stars.

\subsection{Early-type stars and models of planet formation}

In this section we briefly review planet formation models and the possible relation to our observations.
The two main models of planet formation are the Core Accretion (CA) and the Gravitational
Instability (GI). In the CA model, the accretion of dust particles and planetesimals could result
in a solid core of few M$_{\oplus}$, forming low-mass planets and cores of giant planets.  
If the core reach a critical mass of $\sim$5-10 M$_{\oplus}$ before the dissipation of gas disk,
then they could undergo a runaway accretion of gas and form giant planets
\citep[e.g.][]{safronov69,pollack96,ida-lin04a,alibert05,alibert11,mordasini12a}.
In the GI model, a massive and cold protoplanetary disk fragments into clumps which then
cool and contract to form giant planets \citep[e.g.][]{kuiper51,boss98,boss02,boss17}.
In the last years, CA models become the dominant planet formation theory for solar-mass stars,
matching observed features such as the abundance of Neptune and Jupiter-mass planets \citep[e.g. ][]{udry07}
and the planet-metallicity correlation \citep[][]{ida-lin04a,mordasini12a}.
On the other hand, initial GI models cannot provide an explanation of the planet-metallicity
correlation\footnote{But note that recent GI works seem to find the correlation with the inclusion
of peeble accretion \citep{nayakshin15}}.
Both CA and GI models have received considerable improvements in the last years
(such as planet migration, peeble accretion, etc.) and are still under developement
\citep[see e.g. the reviews of ][]{lissauer-stevenson07,durisen07,helled14,rm20}.

It has been claimed that, for solar-type stars, CA models could face some difficulties
(in principle) explaining the presence of giant planets around metal-poor stars, or massive planets at long radial distances
\citep[see e.g.][]{helled14}. In particular, HR 8799 is a metal-poor A5 star included in our sample, hosting
three giant planets orbiting beyond 10 AU \citep{marois08}. Then, some authors proposed that the planets
orbiting HR 8799 are likely formed by GI, given that GI models can take place at large radii and in low metal environments
\citep{marois08,dodson09,meru-bate10}. Other works cast some doubts about the planet formation
around HR 8799, showing that CA models are also possible \citep{currie11} or even a combination
of GI and CA \citep{marois10}.
For the case of $\beta$ Pictoris and HD 169142, also having giant planets at long distances,
CA models seem to be favored \citep{gravity20,nowak20,perez19}.
However, an important word of caution about these works is in order.
Some of the stars mentioned (HR 8799 and HD 169142) display (superficial) metal-poor abundances,
showing in fact a $\lambda$ Bo\"otis pattern (see for example Fig. \ref{fig.lamboo.stars}).
As previously mentioned, the most accepted idea about the origin of this peculiar signature, 
suppose a solar-like composition for the original molecular cloud where the stars born, and then
some kind of selective accretion to obtain a $\lambda$ Bo\"otis pattern.
In this way, it would be not entirely appropriate to assume a metal-poor natal environment for stars like
HR 8799, as assumed by some works to support a GI planet formation \citep[e.g. ][]{meru-bate10}.
This fact was early noted by \citet{paunzen14}, in their comparison
of $\lambda$ Bo\"otis stars and Population II type stars.
Numerical simulations of planet formation around $\lambda$ Bo\"otis stars should assume a solar-like
composition (rather than a metal-poor natal environment), and this could have important consequences
for the subsequent results. 

We prefer to be more cautious about the origin of planets orbiting around HR 8799 and similar stars,
and explore subsequent predictions of CA models, even including a low metallicity natal environment.
\citet{alibert11} studied the formation of planets under the CA assumption, specially for the case
of stars with different masses (0.5, 1.0 and 2.0 M$_{\sun}$).
They found that the metallicity effect of planet formation is weakly dependent on
stellar mass, that is, stars hosting giant planets resulted, on average, more metal rich
than stars without planets.
However, they showed that the metallicity effect depends also on the mass of the disk.
If the disk mass scales with the stellar mass, the effect of metallicity decreases
as the mass of the primary increases. Then, the minimum metallicity required to form a
massive planet is correspondingly lower for massive stars than lower mass stars.
\citet{mordasini12a} also suggest a "compensation effect", where giant planet cores
could form even at low metallicities and large distances but compensated by high disk masses.
They caution, however, that low metallicities cannot be compensated by high gas masses ad infinitum,
at least if higher mass disks have an ice line farther out due to stronger viscous dissipation.
In other words, it seems possible for stars having massive gas disks to form giant planets though CA,
without the necessity of higher metallicities.

\section{Conclusions}

In the present work, we performed a detailed abundance determination for a number
of early-type stars with and without planets, by fitting high-resolution stellar spectra
with a synthetic spectra.
We compared the complete chemical pattern of the sample with those of $\lambda$ Bo\"otis
as well as with other chemically peculiar stars.
Then, the main results of this study are as follows:

-We have found four $\lambda$ Bo\"otis stars in our sample, two of which
present planets (HR 8799 and HD 169142), one without planets (HD 110058),
and the first $\lambda$ Bo\"otis star orbited by a brown dwarf ($\zeta$ Del).
This last interesting pair composed by a $\lambda$ Bo\"otis star + brown dwarf, 
could help to test stellar formation scenarios.

-We find no unique chemical pattern for (early-type) planet-bearing stars.
Within this group, we found $\lambda$ Bo\"otis stars (HD 8799 and HD 169142),
a chemically peculiar Am star (KELT-17), a number of stars showing mostly
solar abundaces, and one metal-rich object (WASP-167).


-The $\lambda$ Bo\"otis signature that we observe in the Herbig AeBe star HD 169142
and in the young star HR 8799, support in principle the scenario proposed by \citet{kama15}.
They suggest that the presence of giant planets in very young stars possibly block the dust
from protostellar disks and allow the accretion of volatile-rich gas, resulting in a $\lambda$ Bo\"otis pattern.

-The abundances derived in this work for different hot-Jupiter exoplanet host stars do not support,
in principle, the accretion from hot-Jupiters winds proposed to explain the origin of $\lambda$ Bo\"otis stars.
We suggest that other mechanisms should account for the presence of main-sequence $\lambda$ Bo\"otis stars.

-It was previously suggested that gravitational instability could account for the formation of planets
around low metallicity stars like HR 8799. However, we caution that 
it seems also possible for stars having massive gas disks
to form giant planets though core accretion. 

The interesting initial findings that we found here encourage us to continue investigating 
early-type stars and the possible relation between planets and $\lambda$ Bo\"otis stars.
We suggest to increase the number of stars studied in order to improve the statistical significance
of the results.

\begin{acknowledgements}
We thank the referee Dr. Ernst Paunzen for constructive comments that improved the paper.
CS acknowledge the project grants CICITCA E1134 and PICT 2017-2294.
The authors wish to recognize and acknowledge the very significant cultural role
and reverence that the summit of Mauna Kea has always had within the indigenous
Hawaiian community.  We are most fortunate to have the opportunity to conduct
observations from this mountain. 
The authors also thank Dr. Robert Kurucz for making their codes available to us.
\end{acknowledgements}

\begin{appendix}

\section{Detailed chemical abundances}

We present in this section individual abundances for each star,
showing average$\pm$total error and the uncertainties e1, ..., e4 for the chemical species
(see Table \ref{table.abunds1}).

\begin{table*}
\centering
\caption{Abundances derived in this work. For each specie, we present average$\pm$total error and the uncertainties e1, ..., e4.}
\scriptsize
\begin{tabular}{lccccccccccccccc}
\hline
Specie & [N/H] & e$_{1}$ & e$_{2}$ & e$_{3}$ & e$_{4}$ & [N/H] & e$_{1}$ & e$_{2}$ & e$_{3}$ & e$_{4}$ & [N/H] & e$_{1}$ & e$_{2}$ & e$_{3}$ & e$_{4}$ \\
\hline
 & \multicolumn{5}{c}{HD 135379} & \multicolumn{5}{c}{$\beta$ Pictoris} & \multicolumn{5}{c}{HD 216956} \\
\hline
\ion{C}{I} & -0.49 $\pm$ 0.17 & 0.04 & 0.16 & 0.02 & 0.02 & -0.20 $\pm$ 0.11 & 0.09 & 0.02 & 0.04 & 0.02 & -0.43 $\pm$ 0.19 & 0.17 & 0.09 & 0.01 & 0.02 \\
\ion{Mg}{I} & -0.16 $\pm$ 0.38 & 0.13 & 0.25 & 0.06 & 0.25 & -0.22 $\pm$ 0.20 & 0.07 & 0.05 & 0.03 & 0.18 & 0.02 $\pm$ 0.28 & 0.14 & 0.15 & 0.05 & 0.18 \\
\ion{Mg}{II} & -0.19 $\pm$ 0.10 & 0.08 & 0.03 & 0.04 & 0.04 & -0.01 $\pm$ 0.19 & 0.13 & 0.06 & 0.01 & 0.12 & 0.06 $\pm$ 0.15 & 0.14 & 0.02 & 0.01 & 0.05 \\
\ion{Al}{I} & -0.63 $\pm$ 0.28 & 0.05 & 0.21 & 0.01 & 0.17 & -0.84 $\pm$ 0.21 & 0.11 & 0.02 & 0.01 & 0.18 & -0.32 $\pm$ 0.28 & 0.14 & 0.18 & 0.01 & 0.17 \\
\ion{Si}{II} & -0.24 $\pm$ 0.14 & 0.06 & 0.10 & 0.06 & 0.05 & -0.14 $\pm$ 0.24 & 0.19 & 0.08 & 0.04 & 0.11 & -0.20 $\pm$ 0.14 & 0.09 & 0.07 & 0.03 & 0.07 \\
\ion{Ca}{I} & -0.02 $\pm$ 0.48 & 0.08 & 0.35 & 0.06 & 0.32 & 0.11 $\pm$ 0.38 & 0.13 & 0.13 & 0.01 & 0.33 &   & & & & \\
\ion{Ca}{II} & -0.14 $\pm$ 0.24 & 0.08 & 0.23 & 0.02 & 0.01 & 0.10 $\pm$ 0.14 & 0.13 & 0.03 & 0.02 & 0.01 & -0.10 $\pm$ 0.18 & 0.14 & 0.12 & 0.02 & 0.01 \\
\ion{Sc}{II} & 0.07 $\pm$ 0.35 & 0.06 & 0.24 & 0.04 & 0.24 & -0.14 $\pm$ 0.25 & 0.07 & 0.07 & 0.08 & 0.22 & -0.06 $\pm$ 0.32 & 0.10 & 0.19 & 0.04 & 0.23 \\
\ion{Ti}{II} & -0.16 $\pm$ 0.15 & 0.02 & 0.13 & 0.04 & 0.06 & -0.08 $\pm$ 0.19 & 0.03 & 0.03 & 0.05 & 0.18 & 0.13 $\pm$ 0.17 & 0.03 & 0.10 & 0.02 & 0.13 \\
\ion{Cr}{II} & -0.23 $\pm$ 0.11 & 0.02 & 0.08 & 0.04 & 0.05 & -0.18 $\pm$ 0.10 & 0.05 & 0.02 & 0.04 & 0.08 & 0.23 $\pm$ 0.08 & 0.05 & 0.05 & 0.02 & 0.04 \\
\ion{Mn}{I} & -0.39 $\pm$ 0.29 & 0.10 & 0.27 & 0.04 & 0.01 & -0.23 $\pm$ 0.13 & 0.08 & 0.08 & 0.02 & 0.06 & -0.27 $\pm$ 0.29 & 0.08 & 0.28 & 0.02 & 0.03 \\
\ion{Fe}{I} & -0.25 $\pm$ 0.26 & 0.01 & 0.25 & 0.03 & 0.07 & -0.28 $\pm$ 0.14 & 0.02 & 0.06 & 0.02 & 0.12 & 0.12 $\pm$ 0.19 & 0.02 & 0.15 & 0.02 & 0.11 \\
\ion{Fe}{II} & -0.19 $\pm$ 0.16 & 0.02 & 0.12 & 0.04 & 0.10 & -0.26 $\pm$ 0.14 & 0.04 & 0.03 & 0.04 & 0.12 & 0.11 $\pm$ 0.15 & 0.04 & 0.07 & 0.02 & 0.13 \\
\ion{Ni}{II} & -0.11 $\pm$ 0.22 & 0.05 & 0.19 & 0.02 & 0.10 &   & & & & & 0.46 $\pm$ 0.16 & 0.09 & 0.13 & 0.02 & 0.04 \\
\ion{Sr}{II} & 0.28 $\pm$ 0.48 & 0.07 & 0.32 & 0.03 & 0.35 & -0.47 $\pm$ 0.36 & 0.09 & 0.11 & 0.07 & 0.32 & 0.53 $\pm$ 0.34 & 0.08 & 0.17 & 0.03 & 0.28 \\
\ion{Y}{II} & 0.30 $\pm$ 0.20 & 0.05 & 0.19 & 0.04 & 0.03 & -0.01 $\pm$ 0.09 & 0.08 & 0.02 & 0.04 & 0.01 & 0.59 $\pm$ 0.16 & 0.07 & 0.14 & 0.03 & 0.03 \\
\ion{Zr}{II} & 0.37 $\pm$ 0.14 & 0.05 & 0.12 & 0.06 & 0.02 & 0.50 $\pm$ 0.11 & 0.09 & 0.01 & 0.02 & 0.04 & 0.57 $\pm$ 0.18 & 0.14 & 0.12 & 0.01 & 0.03 \\
\ion{Ba}{II} & 0.30 $\pm$ 0.26 & 0.04 & 0.25 & 0.01 & 0.07 & -0.04 $\pm$ 0.17 & 0.11 & 0.07 & 0.02 & 0.11 & 1.22 $\pm$ 0.34 & 0.15 & 0.15 & 0.01 & 0.27 \\
\hline
 & \multicolumn{5}{c}{HD 180777} & \multicolumn{5}{c}{HD 195689} & \multicolumn{5}{c}{HD 95086} \\
\hline
\ion{C}{I} & -0.21 $\pm$ 0.10 & 0.06 & 0.03 & 0.05 & 0.05 & -0.13 $\pm$ 0.09 & 0.05 & 0.07 & 0.03 & 0.01 & -0.12 $\pm$ 0.07 & 0.06 & 0.03 & 0.03 & 0.02 \\
\ion{O}{I} &   & & & & & -0.08 $\pm$ 0.13 & 0.13 & 0.02 & 0.01 & 0.01 &   & & & & \\
\ion{Mg}{I} & -0.21 $\pm$ 0.20 & 0.06 & 0.13 & 0.11 & 0.08 & 0.04 $\pm$ 0.18 & 0.08 & 0.09 & 0.07 & 0.12 & 0.10 $\pm$ 0.23 & 0.06 & 0.07 & 0.04 & 0.21 \\
\ion{Mg}{II} & 0.01 $\pm$ 0.20 & 0.10 & 0.11 & 0.03 & 0.13 & 0.12 $\pm$ 0.11 & 0.10 & 0.01 & 0.02 & 0.05 & 0.24 $\pm$ 0.18 & 0.11 & 0.05 & 0.02 & 0.13 \\
\ion{Al}{I} &   & & & & & -0.61 $\pm$ 0.12 & 0.06 & 0.09 & 0.04 & 0.05 & -0.45 $\pm$ 0.27 & 0.08 & 0.07 & 0.01 & 0.25 \\
\ion{Si}{II} & 0.20 $\pm$ 0.18 & 0.10 & 0.12 & 0.08 & 0.01 & 0.08 $\pm$ 0.07 & 0.05 & 0.02 & 0.03 & 0.04 & -0.05 $\pm$ 0.12 & 0.07 & 0.08 & 0.05 & 0.04 \\
\ion{Ca}{I} & -0.21 $\pm$ 0.32 & 0.10 & 0.22 & 0.09 & 0.19 & -0.09 $\pm$ 0.38 & 0.08 & 0.26 & 0.10 & 0.24 & 0.08 $\pm$ 0.30 & 0.11 & 0.17 & 0.04 & 0.22 \\
\ion{Ca}{II} &   & & & & & 0.04 $\pm$ 0.13 & 0.08 & 0.09 & 0.04 & 0.01 & -0.18 $\pm$ 0.12 & 0.11 & 0.05 & 0.01 & 0.01 \\
\ion{Sc}{II} & 0.03 $\pm$ 0.37 & 0.09 & 0.09 & 0.09 & 0.34 & -0.14 $\pm$ 0.17 & 0.08 & 0.07 & 0.01 & 0.13 & -0.12 $\pm$ 0.26 & 0.06 & 0.11 & 0.07 & 0.22 \\
\ion{Ti}{II} & -0.08 $\pm$ 0.17 & 0.03 & 0.06 & 0.06 & 0.14 & 0.10 $\pm$ 0.16 & 0.02 & 0.06 & 0.03 & 0.15 & -0.07 $\pm$ 0.10 & 0.02 & 0.04 & 0.04 & 0.08 \\
\ion{Cr}{II} & -0.12 $\pm$ 0.12 & 0.03 & 0.07 & 0.06 & 0.07 & 0.00 $\pm$ 0.07 & 0.03 & 0.03 & 0.02 & 0.06 & -0.05 $\pm$ 0.09 & 0.03 & 0.04 & 0.04 & 0.07 \\
\ion{Mn}{I} & -0.29 $\pm$ 0.14 & 0.06 & 0.11 & 0.02 & 0.07 &   & & & & & -0.18 $\pm$ 0.10 & 0.05 & 0.06 & 0.01 & 0.07 \\
\ion{Fe}{I} & -0.19 $\pm$ 0.22 & 0.02 & 0.13 & 0.03 & 0.18 & -0.09 $\pm$ 0.11 & 0.02 & 0.09 & 0.04 & 0.04 & -0.08 $\pm$ 0.18 & 0.01 & 0.09 & 0.01 & 0.15 \\
\ion{Fe}{II} & -0.16 $\pm$ 0.14 & 0.02 & 0.07 & 0.02 & 0.12 & -0.08 $\pm$ 0.14 & 0.02 & 0.03 & 0.03 & 0.13 & -0.05 $\pm$ 0.16 & 0.02 & 0.04 & 0.03 & 0.15 \\
\ion{Ni}{II} &   & & & & & -0.01 $\pm$ 0.10 & 0.08 & 0.03 & 0.01 & 0.06 &   & & & & \\
\ion{Sr}{II} & 0.19 $\pm$ 0.27 & 0.07 & 0.16 & 0.01 & 0.22 & -0.54 $\pm$ 0.21 & 0.08 & 0.14 & 0.01 & 0.14 & 0.35 $\pm$ 0.33 & 0.08 & 0.14 & 0.03 & 0.29 \\
\ion{Y}{II} & 0.01 $\pm$ 0.14 & 0.09 & 0.06 & 0.08 & 0.07 &   & & & & & -0.07 $\pm$ 0.09 & 0.05 & 0.03 & 0.04 & 0.05 \\
\ion{Zr}{II} & 0.00 $\pm$ 0.14 & 0.07 & 0.04 & 0.07 & 0.09 &   & & & & & -0.20 $\pm$ 0.12 & 0.06 & 0.08 & 0.04 & 0.05 \\
\ion{Ba}{II} & 0.25 $\pm$ 0.29 & 0.10 & 0.11 & 0.03 & 0.26 &   & & & & & 0.25 $\pm$ 0.27 & 0.16 & 0.07 & 0.03 & 0.20 \\
\hline
 & \multicolumn{5}{c}{HD 169142} & \multicolumn{5}{c}{HD 218396} & \multicolumn{5}{c}{KELT-17} \\
\hline
\ion{C}{I} & 0.13 $\pm$ 0.09 & 0.06 & 0.02 & 0.06 & 0.01 & -0.11 $\pm$ 0.07 & 0.04 & 0.01 & 0.05 & 0.01 & -4.09 $\pm$ 0.17 & 0.16 & 0.02 & 0.01 & 0.02 \\
\ion{Mg}{I} & -0.51 $\pm$ 0.29 & 0.08 & 0.22 & 0.09 & 0.15 & -0.66 $\pm$ 0.24 & 0.10 & 0.13 & 0.07 & 0.17 & -0.26 $\pm$ 0.22 & 0.09 & 0.11 & 0.04 & 0.16 \\
\ion{Mg}{II} & -0.31 $\pm$ 0.25 & 0.12 & 0.17 & 0.06 & 0.12 & -0.48 $\pm$ 0.18 & 0.08 & 0.09 & 0.05 & 0.12 & -0.02 $\pm$ 0.24 & 0.16 & 0.10 & 0.01 & 0.14 \\
\ion{Al}{I} & -1.27 $\pm$ 0.34 & 0.08 & 0.28 & 0.05 & 0.16 &   & & & & & -0.56 $\pm$ 0.32 & 0.16 & 0.18 & 0.03 & 0.20 \\
\ion{Si}{II} & -0.52 $\pm$ 0.21 & 0.06 & 0.19 & 0.08 & 0.05 & -0.50 $\pm$ 0.21 & 0.06 & 0.19 & 0.07 & 0.04 & 0.05 $\pm$ 0.32 & 0.31 & 0.09 & 0.01 & 0.02 \\
\ion{Ca}{I} &   & & & & & -0.80 $\pm$ 0.36 & 0.08 & 0.24 & 0.05 & 0.25 &   & & & & \\
\ion{Ca}{II} & -0.83 $\pm$ 0.23 & 0.12 & 0.20 & 0.01 & 0.01 &   & & & & & -0.98 $\pm$ 0.18 & 0.16 & 0.08 & 0.01 & 0.02 \\
\ion{Sc}{II} & -0.67 $\pm$ 0.22 & 0.06 & 0.16 & 0.10 & 0.11 & -0.79 $\pm$ 0.16 & 0.04 & 0.09 & 0.09 & 0.11 & -0.88 $\pm$ 0.21 & 0.12 & 0.04 & 0.07 & 0.15 \\
\ion{Ti}{II} & -0.53 $\pm$ 0.18 & 0.03 & 0.12 & 0.08 & 0.11 & -0.67 $\pm$ 0.12 & 0.02 & 0.06 & 0.07 & 0.08 & 0.18 $\pm$ 0.20 & 0.05 & 0.06 & 0.05 & 0.18 \\
\ion{Cr}{II} & -0.58 $\pm$ 0.10 & 0.03 & 0.06 & 0.07 & 0.04 & -0.66 $\pm$ 0.09 & 0.02 & 0.04 & 0.06 & 0.04 & 0.63 $\pm$ 0.12 & 0.05 & 0.05 & 0.03 & 0.09 \\
\ion{Mn}{I} & -0.53 $\pm$ 0.21 & 0.05 & 0.20 & 0.02 & 0.05 & -0.68 $\pm$ 0.12 & 0.03 & 0.11 & 0.03 & 0.02 & 0.29 $\pm$ 0.13 & 0.08 & 0.06 & 0.01 & 0.09 \\
\ion{Fe}{I} & -0.64 $\pm$ 0.25 & 0.02 & 0.22 & 0.02 & 0.11 & -0.77 $\pm$ 0.16 & 0.01 & 0.12 & 0.02 & 0.10 & 0.47 $\pm$ 0.22 & 0.03 & 0.11 & 0.02 & 0.19 \\
\ion{Fe}{II} & -0.67 $\pm$ 0.19 & 0.05 & 0.04 & 0.07 & 0.16 & -0.70 $\pm$ 0.15 & 0.02 & 0.04 & 0.06 & 0.13 & 0.51 $\pm$ 0.17 & 0.03 & 0.07 & 0.02 & 0.15 \\
\ion{Sr}{II} & -0.24 $\pm$ 0.39 & 0.08 & 0.28 & 0.03 & 0.26 & -0.45 $\pm$ 0.30 & 0.06 & 0.14 & 0.05 & 0.25 & 0.92 $\pm$ 0.32 & 0.16 & 0.16 & 0.02 & 0.22 \\
\ion{Y}{II} & -0.47 $\pm$ 0.16 & 0.06 & 0.13 & 0.08 & 0.01 & -0.52 $\pm$ 0.13 & 0.08 & 0.08 & 0.08 & 0.01 & 1.27 $\pm$ 0.34 & 0.10 & 0.09 & 0.08 & 0.30 \\
\ion{Zr}{II} & -0.34 $\pm$ 0.28 & 0.18 & 0.20 & 0.08 & 0.02 & -0.23 $\pm$ 0.16 & 0.13 & 0.06 & 0.06 & 0.02 & 0.08 $\pm$ 0.24 & 0.16 & 0.05 & 0.07 & 0.15 \\
\ion{Ba}{II} & -0.30 $\pm$ 0.28 & 0.11 & 0.21 & 0.04 & 0.15 & -0.38 $\pm$ 0.23 & 0.10 & 0.12 & 0.03 & 0.17 & 1.47 $\pm$ 0.33 & 0.16 & 0.11 & 0.04 & 0.26 \\
\hline
 & \multicolumn{5}{c}{HD 185603} & \multicolumn{5}{c}{MASCARA-1} & \multicolumn{5}{c}{WASP-33} \\
\hline
\ion{C}{I} & -0.42 $\pm$ 0.10 & 0.07 & 0.06 & 0.01 & 0.02 & -0.32 $\pm$ 0.08 & 0.07 & 0.02 & 0.02 & 0.04 & -0.24 $\pm$ 0.16 & 0.15 & 0.04 & 0.03 & 0.02 \\
\ion{O}{I} &   & & & & &   & & & & & 0.58 $\pm$ 0.23 & 0.11 & 0.03 & 0.01 & 0.20 \\
\ion{Mg}{I} & -0.52 $\pm$ 0.21 & 0.08 & 0.12 & 0.07 & 0.14 & -0.33 $\pm$ 0.24 & 0.06 & 0.08 & 0.03 & 0.22 & -0.11 $\pm$ 0.21 & 0.09 & 0.11 & 0.09 & 0.12 \\
\ion{Mg}{II} & -0.33 $\pm$ 0.11 & 0.07 & 0.04 & 0.02 & 0.06 & -0.12 $\pm$ 0.21 & 0.09 & 0.14 & 0.02 & 0.13 & 0.06 $\pm$ 0.21 & 0.15 & 0.04 & 0.02 & 0.13 \\
\ion{Al}{I} & -1.11 $\pm$ 0.16 & 0.07 & 0.13 & 0.01 & 0.05 & -0.34 $\pm$ 0.25 & 0.09 & 0.06 & 0.01 & 0.23 & -0.64 $\pm$ 0.25 & 0.11 & 0.15 & 0.03 & 0.16 \\
\ion{Si}{II} & -0.07 $\pm$ 0.13 & 0.08 & 0.05 & 0.08 & 0.04 & 0.07 $\pm$ 0.17 & 0.09 & 0.14 & 0.05 & 0.02 & 0.11 $\pm$ 0.20 & 0.15 & 0.10 & 0.07 & 0.01 \\
\ion{Ca}{I} & -0.59 $\pm$ 0.39 & 0.07 & 0.28 & 0.07 & 0.25 & 0.01 $\pm$ 0.37 & 0.09 & 0.20 & 0.04 & 0.29 & -0.01 $\pm$ 0.39 & 0.15 & 0.23 & 0.07 & 0.26 \\
\ion{Ca}{II} & -0.45 $\pm$ 0.14 & 0.07 & 0.12 & 0.02 & 0.01 & 0.13 $\pm$ 0.10 & 0.09 & 0.04 & 0.01 & 0.01 & -0.20 $\pm$ 0.17 & 0.15 & 0.08 & 0.01 & 0.01 \\
\ion{Sc}{II} & -0.71 $\pm$ 0.24 & 0.07 & 0.20 & 0.10 & 0.06 & 0.00 $\pm$ 0.28 & 0.05 & 0.09 & 0.06 & 0.25 & 0.02 $\pm$ 0.36 & 0.08 & 0.12 & 0.10 & 0.32 \\
\ion{Ti}{II} & -0.25 $\pm$ 0.18 & 0.02 & 0.09 & 0.05 & 0.15 & -0.04 $\pm$ 0.13 & 0.02 & 0.02 & 0.04 & 0.12 & 0.08 $\pm$ 0.16 & 0.04 & 0.04 & 0.06 & 0.13 \\
\ion{Cr}{II} & -0.21 $\pm$ 0.10 & 0.02 & 0.03 & 0.04 & 0.08 & -0.09 $\pm$ 0.12 & 0.05 & 0.06 & 0.04 & 0.08 & 0.13 $\pm$ 0.11 & 0.06 & 0.06 & 0.03 & 0.08 \\
\ion{Mn}{I} & -0.26 $\pm$ 0.29 & 0.17 & 0.23 & 0.08 & 0.01 & -0.07 $\pm$ 0.11 & 0.09 & 0.06 & 0.02 & 0.03 & 0.05 $\pm$ 0.11 & 0.08 & 0.06 & 0.02 & 0.05 \\
\ion{Fe}{I} & -0.35 $\pm$ 0.15 & 0.02 & 0.13 & 0.03 & 0.06 & -0.11 $\pm$ 0.15 & 0.02 & 0.06 & 0.01 & 0.14 & 0.11 $\pm$ 0.23 & 0.04 & 0.12 & 0.02 & 0.20 \\
\ion{Fe}{II} & -0.30 $\pm$ 0.15 & 0.02 & 0.06 & 0.04 & 0.13 & -0.10 $\pm$ 0.13 & 0.02 & 0.02 & 0.03 & 0.13 & 0.13 $\pm$ 0.16 & 0.04 & 0.04 & 0.02 & 0.15 \\
\ion{Ni}{II} & -0.15 $\pm$ 0.15 & 0.07 & 0.08 & 0.01 & 0.10 &   & & & & &   & & & & \\
\ion{Sr}{II} & -0.81 $\pm$ 0.32 & 0.07 & 0.20 & 0.04 & 0.24 & 0.49 $\pm$ 0.33 & 0.14 & 0.15 & 0.02 & 0.26 & 0.45 $\pm$ 0.36 & 0.11 & 0.16 & 0.02 & 0.31 \\
\ion{Y}{II} &   & & & & & 0.54 $\pm$ 0.11 & 0.06 & 0.02 & 0.06 & 0.08 & 0.33 $\pm$ 0.15 & 0.13 & 0.03 & 0.05 & 0.05 \\
\ion{Zr}{II} &   & & & & & 0.30 $\pm$ 0.15 & 0.06 & 0.11 & 0.06 & 0.05 & -0.20 $\pm$ 0.18 & 0.15 & 0.08 & 0.03 & 0.05 \\
\ion{Ba}{II} & 0.43 $\pm$ 0.20 & 0.08 & 0.15 & 0.01 & 0.11 & 0.58 $\pm$ 0.29 & 0.10 & 0.09 & 0.02 & 0.26 & 0.94 $\pm$ 0.26 & 0.10 & 0.08 & 0.03 & 0.22 \\
\hline
\end{tabular}
\normalsize
\label{table.abunds1}
\end{table*}
 
\setcounter{table}{0}
\begin{table*}
\centering
\caption{Continued. Abundances derived in this work. For each specie, we present average$\pm$total error and the uncertainties e1, ..., e4.}
\scriptsize
\begin{tabular}{lccccccccccccccc}
\hline
Specie & [N/H] & e$_{1}$ & e$_{2}$ & e$_{3}$ & e$_{4}$ & [N/H] & e$_{1}$ & e$_{2}$ & e$_{3}$ & e$_{4}$ & [N/H] & e$_{1}$ & e$_{2}$ & e$_{3}$ & e$_{4}$ \\
\hline
 & \multicolumn{5}{c}{HR 4502 A} & \multicolumn{5}{c}{BU Psc} & \multicolumn{5}{c}{HD 29391} \\
\hline
\ion{C}{I} & -0.15 $\pm$ 0.14 & 0.04 & 0.13 & 0.03 & 0.01 & -0.20 $\pm$ 0.07 & 0.05 & 0.01 & 0.03 & 0.02 & -0.22 $\pm$ 0.09 & 0.07 & 0.03 & 0.04 & 0.04 \\
\ion{O}{I} & -0.25 $\pm$ 0.07 & 0.07 & 0.01 & 0.02 & 0.01 &   & & & & &   & & & & \\
\ion{Mg}{I} & 0.18 $\pm$ 0.28 & 0.15 & 0.20 & 0.07 & 0.11 & -0.29 $\pm$ 0.14 & 0.06 & 0.06 & 0.07 & 0.09 & -0.13 $\pm$ 0.20 & 0.08 & 0.12 & 0.09 & 0.10 \\
\ion{Mg}{II} & 0.04 $\pm$ 0.12 & 0.11 & 0.02 & 0.04 & 0.01 & -0.08 $\pm$ 0.17 & 0.11 & 0.04 & 0.02 & 0.13 & 0.04 $\pm$ 0.20 & 0.13 & 0.06 & 0.02 & 0.13 \\
\ion{Al}{I} & -0.35 $\pm$ 0.21 & 0.06 & 0.19 & 0.02 & 0.06 & -0.90 $\pm$ 0.18 & 0.07 & 0.05 & 0.02 & 0.15 & -0.72 $\pm$ 0.21 & 0.09 & 0.11 & 0.03 & 0.15 \\
\ion{Al}{II} & -0.21 $\pm$ 0.12 & 0.09 & 0.07 & 0.05 & 0.01 &   & & & & &   & & & & \\
\ion{Si}{II} & -0.15 $\pm$ 0.07 & 0.04 & 0.03 & 0.04 & 0.04 & -0.05 $\pm$ 0.12 & 0.09 & 0.06 & 0.05 & 0.01 & -0.04 $\pm$ 0.27 & 0.24 & 0.10 & 0.07 & 0.01 \\
\ion{Ca}{I} & 0.22 $\pm$ 0.38 & 0.09 & 0.29 & 0.08 & 0.22 & -0.20 $\pm$ 0.28 & 0.11 & 0.10 & 0.06 & 0.23 & -0.09 $\pm$ 0.37 & 0.13 & 0.23 & 0.08 & 0.24 \\
\ion{Ca}{II} & 0.00 $\pm$ 0.23 & 0.09 & 0.21 & 0.06 & 0.01 & -0.34 $\pm$ 0.11 & 0.07 & 0.06 & 0.01 & 0.05 & -0.09 $\pm$ 0.18 & 0.13 & 0.11 & 0.03 & 0.01 \\
\ion{Sc}{II} & -0.11 $\pm$ 0.16 & 0.04 & 0.15 & 0.02 & 0.03 & 0.01 $\pm$ 0.32 & 0.08 & 0.04 & 0.06 & 0.31 & 0.05 $\pm$ 0.34 & 0.07 & 0.10 & 0.10 & 0.30 \\
\ion{Ti}{II} & 0.01 $\pm$ 0.12 & 0.01 & 0.12 & 0.02 & 0.04 & -0.10 $\pm$ 0.14 & 0.03 & 0.02 & 0.04 & 0.13 & -0.02 $\pm$ 0.15 & 0.03 & 0.08 & 0.06 & 0.12 \\
\ion{Cr}{II} & -0.07 $\pm$ 0.08 & 0.02 & 0.07 & 0.03 & 0.02 & -0.20 $\pm$ 0.07 & 0.03 & 0.02 & 0.04 & 0.04 & -0.08 $\pm$ 0.10 & 0.04 & 0.04 & 0.05 & 0.07 \\
\ion{Mn}{I} & -0.22 $\pm$ 0.25 & 0.05 & 0.24 & 0.02 & 0.01 & -0.34 $\pm$ 0.07 & 0.04 & 0.04 & 0.01 & 0.05 & -0.24 $\pm$ 0.11 & 0.05 & 0.09 & 0.01 & 0.04 \\
\ion{Fe}{I} & -0.04 $\pm$ 0.18 & 0.01 & 0.18 & 0.04 & 0.02 & -0.31 $\pm$ 0.18 & 0.01 & 0.06 & 0.01 & 0.17 &  & & & & \\
\ion{Fe}{II} & -0.11 $\pm$ 0.10 & 0.02 & 0.07 & 0.03 & 0.07 & -0.25 $\pm$ 0.11 & 0.03 & 0.04 & 0.02 & 0.10 & -0.06 $\pm$ 0.10 & 0.03 & 0.06 & 0.03 & 0.07 \\
\ion{Ni}{II} & -0.22 $\pm$ 0.08 & 0.06 & 0.05 & 0.02 & 0.01 &   & & & & &   & & & & \\
\ion{Sr}{II} & -0.08 $\pm$ 0.33 & 0.06 & 0.25 & 0.01 & 0.22 & 0.23 $\pm$ 0.28 & 0.07 & 0.05 & 0.01 & 0.27 & 0.13 $\pm$ 0.34 & 0.09 & 0.16 & 0.03 & 0.28 \\
\ion{Y}{II} & 0.01 $\pm$ 0.19 & 0.06 & 0.18 & 0.01 & 0.01 & -0.04 $\pm$ 0.11 & 0.09 & 0.03 & 0.05 & 0.05 & 0.03 $\pm$ 0.13 & 0.07 & 0.06 & 0.07 & 0.05 \\
\ion{Zr}{II} & 0.22 $\pm$ 0.13 & 0.03 & 0.13 & 0.03 & 0.01 & 0.15 $\pm$ 0.21 & 0.20 & 0.02 & 0.04 & 0.06 & -0.16 $\pm$ 0.17 & 0.13 & 0.08 & 0.04 & 0.05 \\
\ion{Ba}{II} & 0.29 $\pm$ 0.26 & 0.12 & 0.23 & 0.04 & 0.01 & 0.19 $\pm$ 0.28 & 0.11 & 0.04 & 0.02 & 0.25 & 0.20 $\pm$ 0.25 & 0.06 & 0.09 & 0.03 & 0.22 \\
\hline
 & \multicolumn{5}{c}{HD 105850} & \multicolumn{5}{c}{HD 110058} & \multicolumn{5}{c}{HD 115820} \\
\hline
\ion{C}{I} & -0.41 $\pm$ 0.13 & 0.08 & 0.10 & 0.01 & 0.02 & -0.14 $\pm$ 0.06 & 0.04 & 0.02 & 0.03 & 0.02 & -0.13 $\pm$ 0.08 & 0.07 & 0.03 & 0.03 & 0.02 \\
\ion{Mg}{I} & -0.27 $\pm$ 0.19 & 0.10 & 0.11 & 0.03 & 0.12 & -1.26 $\pm$ 0.16 & 0.09 & 0.06 & 0.02 & 0.12 & -0.04 $\pm$ 0.18 & 0.08 & 0.08 & 0.06 & 0.12 \\
\ion{Mg}{II} & -0.07 $\pm$ 0.14 & 0.12 & 0.04 & 0.01 & 0.07 & -1.01 $\pm$ 0.21 & 0.09 & 0.14 & 0.05 & 0.11 & 0.20 $\pm$ 0.19 & 0.14 & 0.05 & 0.01 & 0.11 \\
\ion{Al}{I} & -0.80 $\pm$ 0.13 & 0.06 & 0.10 & 0.01 & 0.07 & -2.14 $\pm$ 0.15 & 0.09 & 0.12 & 0.01 & 0.01 & -0.46 $\pm$ 0.22 & 0.10 & 0.08 & 0.03 & 0.18 \\
\ion{Si}{II} & -0.45 $\pm$ 0.12 & 0.08 & 0.04 & 0.03 & 0.07 &   & & & & & 0.21 $\pm$ 0.21 & 0.10 & 0.07 & 0.04 & 0.17 \\
\ion{Ca}{I} & -0.07 $\pm$ 0.39 & 0.09 & 0.26 & 0.05 & 0.27 & -1.32 $\pm$ 0.36 & 0.09 & 0.20 & 0.01 & 0.29 & 0.08 $\pm$ 0.33 & 0.14 & 0.18 & 0.06 & 0.23 \\
\ion{Ca}{II} & -0.15 $\pm$ 0.14 & 0.09 & 0.11 & 0.01 & 0.01 & -1.17 $\pm$ 0.11 & 0.09 & 0.06 & 0.01 & 0.01 & -0.07 $\pm$ 0.15 & 0.14 & 0.05 & 0.01 & 0.01 \\
\ion{Sc}{II} & -0.21 $\pm$ 0.17 & 0.07 & 0.15 & 0.01 & 0.06 & -1.45 $\pm$ 0.11 & 0.09 & 0.02 & 0.04 & 0.03 & 0.12 $\pm$ 0.31 & 0.10 & 0.11 & 0.06 & 0.26 \\
\ion{Ti}{II} & -0.09 $\pm$ 0.18 & 0.03 & 0.09 & 0.02 & 0.15 & -1.32 $\pm$ 0.12 & 0.04 & 0.06 & 0.05 & 0.09 & 0.20 $\pm$ 0.17 & 0.04 & 0.03 & 0.05 & 0.15 \\
\ion{Cr}{II} & -0.27 $\pm$ 0.07 & 0.03 & 0.04 & 0.02 & 0.04 & -1.03 $\pm$ 0.11 & 0.10 & 0.05 & 0.01 & 0.02 & 0.05 $\pm$ 0.13 & 0.04 & 0.04 & 0.03 & 0.12 \\
\ion{Mn}{I} &   & & & & &   & & & & & -0.07 $\pm$ 0.11 & 0.05 & 0.06 & 0.02 & 0.07 \\
\ion{Fe}{I} & -0.29 $\pm$ 0.14 & 0.02 & 0.12 & 0.02 & 0.06 & -1.35 $\pm$ 0.13 & 0.03 & 0.08 & 0.01 & 0.09 & -0.06 $\pm$ 0.18 & 0.02 & 0.09 & 0.02 & 0.15 \\
\ion{Fe}{II} & -0.24 $\pm$ 0.13 & 0.02 & 0.05 & 0.02 & 0.12 & -1.28 $\pm$ 0.08 & 0.03 & 0.02 & 0.04 & 0.06 & -0.02 $\pm$ 0.09 & 0.04 & 0.05 & 0.02 & 0.06 \\
\ion{Ni}{II} & -0.22 $\pm$ 0.18 & 0.09 & 0.06 & 0.01 & 0.14 &   & & & & &   & & & & \\
\ion{Sr}{II} & -0.59 $\pm$ 0.29 & 0.08 & 0.19 & 0.01 & 0.21 & -2.01 $\pm$ 0.19 & 0.13 & 0.11 & 0.04 & 0.08 & 0.09 $\pm$ 0.33 & 0.15 & 0.15 & 0.04 & 0.25 \\
\ion{Y}{II} &   & & & & &   & & & & & 0.17 $\pm$ 0.11 & 0.09 & 0.03 & 0.05 & 0.02 \\
\ion{Zr}{II} & -0.06 $\pm$ 0.10 & 0.09 & 0.03 & 0.01 & 0.04 &   & & & & & -0.13 $\pm$ 0.17 & 0.14 & 0.09 & 0.02 & 0.01 \\
\ion{Ba}{II} & -0.01 $\pm$ 0.25 & 0.20 & 0.15 & 0.02 & 0.04 & -1.13 $\pm$ 0.21 & 0.16 & 0.13 & 0.03 & 0.04 & 0.49 $\pm$ 0.25 & 0.07 & 0.08 & 0.02 & 0.23 \\
\hline
 & \multicolumn{5}{c}{HD 120326} & \multicolumn{5}{c}{HD 129926} & \multicolumn{5}{c}{HD 133803} \\
\hline
\ion{C}{I} & -0.17 $\pm$ 0.09 & 0.08 & 0.01 & 0.03 & 0.02 & -0.09 $\pm$ 0.19 & 0.16 & 0.02 & 0.08 & 0.07 & -0.07 $\pm$ 0.10 & 0.08 & 0.01 & 0.05 & 0.03 \\
\ion{Mg}{I} & -0.23 $\pm$ 0.15 & 0.08 & 0.11 & 0.06 & 0.05 & 0.09 $\pm$ 0.20 & 0.10 & 0.12 & 0.10 & 0.06 & -0.03 $\pm$ 0.21 & 0.10 & 0.14 & 0.12 & 0.05 \\
\ion{Mg}{II} & -0.02 $\pm$ 0.20 & 0.14 & 0.07 & 0.02 & 0.13 & 0.38 $\pm$ 0.23 & 0.18 & 0.03 & 0.02 & 0.13 & 0.29 $\pm$ 0.22 & 0.17 & 0.07 & 0.04 & 0.12 \\
\ion{Al}{I} & -0.51 $\pm$ 0.14 & 0.10 & 0.09 & 0.04 & 0.04 & -0.23 $\pm$ 0.20 & 0.13 & 0.09 & 0.07 & 0.11 & -0.23 $\pm$ 0.22 & 0.12 & 0.14 & 0.10 & 0.07 \\
\ion{Si}{II} & 0.32 $\pm$ 0.13 & 0.10 & 0.07 & 0.03 & 0.01 & 0.53 $\pm$ 0.19 & 0.13 & 0.08 & 0.07 & 0.09 & 0.27 $\pm$ 0.25 & 0.15 & 0.10 & 0.06 & 0.17 \\
\ion{Ca}{I} &  0.01 $\pm$ 0.25 & 0.14 & 0.17 & 0.05 & 0.11 & 0.41 $\pm$ 0.31 & 0.18 & 0.18 & 0.09 & 0.16 & 0.03 $\pm$ 0.36 & 0.17 & 0.24 & 0.10 & 0.19 \\
\ion{Ca}{II} & 0.20 $\pm$ 0.17 & 0.14 & 0.09 & 0.06 & 0.02 & 0.52 $\pm$ 0.25 & 0.18 & 0.11 & 0.14 & 0.02 & 0.32 $\pm$ 0.32 & 0.17 & 0.12 & 0.24 & 0.01 \\
\ion{Sc}{II} & 0.06 $\pm$ 0.28 & 0.17 & 0.07 & 0.04 & 0.21 & 0.95 $\pm$ 0.33 & 0.09 & 0.14 & 0.07 & 0.28 & 0.22 $\pm$ 0.31 & 0.10 & 0.15 & 0.11 & 0.23 \\
\ion{Ti}{II} & -0.05 $\pm$ 0.14 & 0.03 & 0.05 & 0.02 & 0.13 & 0.99 $\pm$ 0.21 & 0.04 & 0.05 & 0.04 & 0.20 & 0.10 $\pm$ 0.15 & 0.04 & 0.05 & 0.08 & 0.11 \\
\ion{Cr}{II} & -0.11 $\pm$ 0.08 & 0.05 & 0.04 & 0.02 & 0.03 & 0.58 $\pm$ 0.18 & 0.14 & 0.04 & 0.04 & 0.09 & 0.19 $\pm$ 0.11 & 0.06 & 0.07 & 0.05 & 0.05 \\
\ion{Mn}{I} & -0.33 $\pm$ 0.14 & 0.05 & 0.09 & 0.01 & 0.10 &   & & & & & 0.00 $\pm$ 0.14 & 0.07 & 0.05 & 0.05 & 0.10 \\
\ion{Fe}{I} & -0.17 $\pm$ 0.20 & 0.02 & 0.11 & 0.02 & 0.17 & 0.39 $\pm$ 0.29 & 0.06 & 0.12 & 0.06 & 0.25 & 0.03 $\pm$ 0.24 & 0.03 & 0.14 & 0.04 & 0.19 \\
\ion{Fe}{II} & -0.11 $\pm$ 0.10 & 0.04 & 0.06 & 0.01 & 0.07 & 0.38 $\pm$ 0.12 & 0.11 & 0.03 & 0.01 & 0.04 & 0.02 $\pm$ 0.15 & 0.06 & 0.08 & 0.02 & 0.11 \\
\ion{Sr}{II} & 0.12 $\pm$ 0.22 & 0.10 & 0.12 & 0.01 & 0.15 & 0.62 $\pm$ 0.27 & 0.13 & 0.15 & 0.04 & 0.18 & 0.16 $\pm$ 0.33 & 0.12 & 0.19 & 0.02 & 0.24 \\
\ion{Y}{II} & 0.07 $\pm$ 0.14 & 0.09 & 0.05 & 0.04 & 0.08 & 1.10$\pm$ 0.25 & 0.18 & 0.03 & 0.04 & 0.16 & 0.10 $\pm$ 0.21 & 0.13 & 0.09 & 0.12 & 0.05 \\
\ion{Zr}{II} & -0.02 $\pm$ 0.20 & 0.17 & 0.07 & 0.04 & 0.07 &   & & & & & -0.27 $\pm$ 0.21 & 0.17 & 0.09 & 0.01 & 0.08 \\
\ion{Ba}{II} & 0.16 $\pm$ 0.23 & 0.07 & 0.07 & 0.01 & 0.21 & 1.27 $\pm$ 0.28 & 0.21 & 0.08 & 0.01 & 0.17 & 0.52 $\pm$ 0.27 & 0.13 & 0.09 & 0.03 & 0.23 \\
\hline
 & \multicolumn{5}{c}{HD 146624} & \multicolumn{5}{c}{HD 153053} & \multicolumn{5}{c}{HD 156751} \\
\hline
\ion{C}{I} & -0.42 $\pm$ 0.11 & 0.05 & 0.10 & 0.01 & 0.01 & -0.18 $\pm$ 0.06 & 0.05 & 0.01 & 0.02 & 0.02 & -0.13 $\pm$ 0.08 & 0.07 & 0.02 & 0.04 & 0.01 \\
\ion{O}{I} & -0.39 $\pm$ 0.13 & 0.12 & 0.02 & 0.01 & 0.02 &   & & & & &   & & & & \\
\ion{Mg}{I} & -0.10 $\pm$ 0.20 & 0.07 & 0.14 & 0.03 & 0.12 & -0.30 $\pm$ 0.23 & 0.09 & 0.04 & 0.03 & 0.20 & -0.26 $\pm$ 0.22 & 0.09 & 0.10 & 0.08 & 0.15 \\
\ion{Mg}{II} & -0.01 $\pm$ 0.08 & 0.08 & 0.01 & 0.02 & 0.01 & -0.11 $\pm$ 0.17 & 0.11 & 0.06 & 0.05 & 0.11 & -0.01 $\pm$ 0.20 & 0.14 & 0.07 & 0.04 & 0.12 \\
\ion{Al}{I} & -0.48 $\pm$ 0.18 & 0.07 & 0.16 & 0.01 & 0.06 & -0.69 $\pm$ 0.24 & 0.14 & 0.05 & 0.02 & 0.20 & -0.65 $\pm$ 0.27 & 0.10 & 0.07 & 0.04 & 0.24 \\
\ion{Si}{II} & -0.05 $\pm$ 0.08 & 0.05 & 0.03 & 0.02 & 0.05 & -0.35 $\pm$ 0.18 & 0.11 & 0.05 & 0.04 & 0.13 & -0.43 $\pm$ 0.22 & 0.09 & 0.09 & 0.08 & 0.17 \\
\ion{Ca}{I} & -0.23 $\pm$ 0.27 & 0.10 & 0.20 & 0.03 & 0.15 & -0.04 $\pm$ 0.33 & 0.11 & 0.13 & 0.01 & 0.28 & -0.03 $\pm$ 0.36 & 0.14 & 0.23 & 0.05 & 0.24 \\
\ion{Ca}{II} & -0.23 $\pm$ 0.16 & 0.10 & 0.13 & 0.02 & 0.01 & -0.02 $\pm$ 0.11 & 0.11 & 0.02 & 0.03 & 0.01 & -0.35 $\pm$ 0.16 & 0.14 & 0.08 & 0.02 & 0.01 \\
\ion{Sc}{II} & -0.34 $\pm$ 0.12 & 0.06 & 0.10 & 0.02 & 0.02 & -0.22 $\pm$ 0.28 & 0.06 & 0.06 & 0.08 & 0.26 & -0.34 $\pm$ 0.30 & 0.11 & 0.14 & 0.10 & 0.22 \\
\ion{Ti}{II} & 0.01 $\pm$ 0.09 & 0.02 & 0.08 & 0.01 & 0.04 & -0.17 $\pm$ 0.10 & 0.02 & 0.02 & 0.05 & 0.09 & -0.15 $\pm$ 0.15 & 0.04 & 0.05 & 0.07 & 0.12 \\
\ion{Cr}{II} & 0.15 $\pm$ 0.07 & 0.02 & 0.05 & 0.01 & 0.04 & -0.28 $\pm$ 0.08 & 0.04 & 0.02 & 0.03 & 0.07 & -0.53 $\pm$ 0.10 & 0.05 & 0.05 & 0.04 & 0.05 \\
\ion{Mn}{I} & -0.07 $\pm$ 0.18 & 0.08 & 0.16 & 0.01 & 0.01 & -0.21 $\pm$ 0.10 & 0.05 & 0.07 & 0.03 & 0.05 & -0.82 $\pm$ 0.12 & 0.07 & 0.07 & 0.04 & 0.05 \\
\ion{Fe}{I} & -0.01 $\pm$ 0.14 & 0.01 & 0.13 & 0.02 & 0.03 & -0.29 $\pm$ 0.14 & 0.02 & 0.06 & 0.01 & 0.13 & -0.47 $\pm$ 0.17 & 0.03 & 0.11 & 0.02 & 0.13 \\
\ion{Fe}{II} & 0.01 $\pm$ 0.11 & 0.02 & 0.06 & 0.02 & 0.09 & -0.22 $\pm$ 0.15 & 0.02 & 0.03 & 0.04 & 0.14 & -0.45 $\pm$ 0.09 & 0.04 & 0.04 & 0.04 & 0.07 \\
\ion{Ni}{II} & 0.16 $\pm$ 0.08 & 0.07 & 0.04 & 0.01 & 0.01 &   & & & & &   & & & & \\
\ion{Sr}{II} & -0.11 $\pm$ 0.30 & 0.07 & 0.18 & 0.01 & 0.23 & 0.25 $\pm$ 0.32 & 0.07 & 0.10 & 0.05 & 0.29 & 0.08 $\pm$ 0.31 & 0.10 & 0.18 & 0.02 & 0.23 \\
\ion{Y}{II} & 0.00 $\pm$ 0.18 & 0.06 & 0.17 & 0.01 & 0.01 & 0.34 $\pm$ 0.10 & 0.06 & 0.03 & 0.06 & 0.05 & -0.09 $\pm$ 0.12 & 0.07 & 0.05 & 0.08 & 0.03 \\
\ion{Zr}{II} & 0.21 $\pm$ 0.16 & 0.12 & 0.11 & 0.01 & 0.01 & 0.50 $\pm$ 0.13 & 0.11 & 0.02 & 0.04 & 0.04 & -0.25 $\pm$ 0.18 & 0.14 & 0.11 & 0.04 & 0.01 \\
\ion{Ba}{II} & 1.01 $\pm$ 0.19 & 0.06 & 0.17 & 0.02 & 0.07 & 0.35 $\pm$ 0.22 & 0.12 & 0.05 & 0.03 & 0.18 & 0.34 $\pm$ 0.33 & 0.10 & 0.11 & 0.02 & 0.30 \\
\hline
\end{tabular}
\normalsize
\label{table.abunds2}
\end{table*}

\setcounter{table}{0}
\begin{table*}
\centering
\caption{Continued. Abundances derived in this work. For each specie, we present average$\pm$total error and the uncertainties e1, ..., e4.}
\scriptsize
\begin{tabular}{lccccccccccccccc}
\hline
Specie & [N/H] & e$_{1}$ & e$_{2}$ & e$_{3}$ & e$_{4}$ & [N/H] & e$_{1}$ & e$_{2}$ & e$_{3}$ & e$_{4}$ & [N/H] & e$_{1}$ & e$_{2}$ & e$_{3}$ & e$_{4}$ \\
\hline
 & \multicolumn{5}{c}{HD 159492} & \multicolumn{5}{c}{HD 188228} & \multicolumn{5}{c}{HD 23281} \\
\hline
\ion{C}{I} & -0.24 $\pm$ 0.06 & 0.05 & 0.03 & 0.02 & 0.02 & -0.55 $\pm$ 0.16 & 0.13 & 0.08 & 0.02 & 0.02 & -0.30 $\pm$ 0.07 & 0.06 & 0.03 & 0.02 & 0.02 \\
\ion{Mg}{I} & -0.06 $\pm$ 0.23 & 0.08 & 0.03 & 0.03 & 0.21 & 0.02 $\pm$ 0.13 & 0.07 & 0.10 & 0.03 & 0.02 & -0.13 $\pm$ 0.20 & 0.08 & 0.06 & 0.02 & 0.17 \\
\ion{Mg}{II} & 0.16 $\pm$ 0.16 & 0.09 & 0.08 & 0.03 & 0.10 & 0.02 $\pm$ 0.10 & 0.10 & 0.03 & 0.01 & 0.01 & 0.10 $\pm$ 0.17 & 0.09 & 0.07 & 0.05 & 0.11 \\
\ion{Al}{I} & -0.61 $\pm$ 0.25 & 0.08 & 0.05 & 0.01 & 0.24 & -0.36 $\pm$ 0.18 & 0.08 & 0.16 & 0.01 & 0.01 & -0.46 $\pm$ 0.24 & 0.07 & 0.10 & 0.01 & 0.21 \\
\ion{Si}{II} & -0.06 $\pm$ 0.14 & 0.11 & 0.06 & 0.04 & 0.04 & -0.13 $\pm$ 0.09 & 0.08 & 0.02 & 0.02 & 0.03 & -0.07 $\pm$ 0.17 & 0.14 & 0.06 & 0.05 & 0.05 \\
\ion{Ca}{I} & 0.07 $\pm$ 0.31 & 0.09 & 0.10 & 0.01 & 0.28 & -0.19 $\pm$ 0.26 & 0.08 & 0.24 & 0.04 & 0.02 & -0.34 $\pm$ 0.33 & 0.09 & 0.19 & 0.04 & 0.25 \\
\ion{Ca}{II} & 0.23 $\pm$ 0.10 & 0.09 & 0.02 & 0.03 & 0.01 & -0.22 $\pm$ 0.12 & 0.08 & 0.08 & 0.03 & 0.01 & -0.26 $\pm$ 0.10 & 0.09 & 0.03 & 0.03 & 0.01 \\
\ion{Sc}{II} & 0.09 $\pm$ 0.24 & 0.05 & 0.03 & 0.07 & 0.23 & -0.56 $\pm$ 0.11 & 0.08 & 0.08 & 0.01 & 0.01 & -0.41 $\pm$ 0.26 & 0.05 & 0.07 & 0.12 & 0.21 \\
\ion{Ti}{II} & 0.07 $\pm$ 0.13 & 0.03 & 0.03 & 0.04 & 0.12 & -0.03 $\pm$ 0.11 & 0.02 & 0.08 & 0.01 & 0.07 & 0.00 $\pm$ 0.14 & 0.02 & 0.04 & 0.06 & 0.12 \\
\ion{Cr}{II} & -0.06 $\pm$ 0.08 & 0.03 & 0.01 & 0.04 & 0.07 & 0.12 $\pm$ 0.06 & 0.03 & 0.04 & 0.01 & 0.04 & 0.20 $\pm$ 0.07 & 0.03 & 0.03 & 0.04 & 0.04 \\
\ion{Mn}{I} & -0.11 $\pm$ 0.09 & 0.06 & 0.06 & 0.01 & 0.02 &   & & & & & -0.10 $\pm$ 0.11 & 0.05 & 0.05 & 0.01 & 0.09 \\
\ion{Fe}{I} & -0.04 $\pm$ 0.16 & 0.01 & 0.06 & 0.01 & 0.15 & 0.03 $\pm$ 0.11 & 0.02 & 0.10 & 0.02 & 0.02 & 0.08 $\pm$ 0.18 & 0.02 & 0.07 & 0.01 & 0.16 \\
\ion{Fe}{II} & -0.02 $\pm$ 0.11 & 0.02 & 0.02 & 0.03 & 0.10 & 0.02 $\pm$ 0.08 & 0.02 & 0.04 & 0.02 & 0.06 & 0.11 $\pm$ 0.10 & 0.03 & 0.04 & 0.04 & 0.08 \\
\ion{Ni}{II} &   & & & & & 0.09 $\pm$ 0.09 & 0.08 & 0.04 & 0.01 & 0.01 &   & & & & \\
\ion{Sr}{II} & 0.23 $\pm$ 0.31 & 0.06 & 0.06 & 0.04 & 0.29 & -0.32 $\pm$ 0.16 & 0.06 & 0.14 & 0.01 & 0.06 & 0.48 $\pm$ 0.28 & 0.07 & 0.12 & 0.02 & 0.24 \\
\ion{Y}{II} & 0.32 $\pm$ 0.09 & 0.06 & 0.04 & 0.04 & 0.04 &   & & & & & 0.48 $\pm$ 0.10 & 0.06 & 0.04 & 0.02 & 0.08 \\
\ion{Zr}{II} & 0.33 $\pm$ 0.08 & 0.07 & 0.01 & 0.03 & 0.03 &   & & & & & 0.14 $\pm$ 0.18 & 0.07 & 0.08 & 0.03 & 0.15 \\
\ion{Ba}{II} & 0.39 $\pm$ 0.21 & 0.10 & 0.07 & 0.02 & 0.18 & 0.99 $\pm$ 0.18 & 0.11 & 0.14 & 0.02 & 0.01 & 0.92 $\pm$ 0.22 & 0.06 & 0.07 & 0.02 & 0.20 \\
\hline
 & \multicolumn{5}{c}{HD 50445} & \multicolumn{5}{c}{HD 56537} & \multicolumn{5}{c}{HD 88955} \\
\hline
\ion{C}{I} & -0.22 $\pm$ 0.05 & 0.04 & 0.01 & 0.02 & 0.02 & -0.42 $\pm$ 0.13 & 0.11 & 0.07 & 0.02 & 0.01 & -0.61 $\pm$ 0.11 & 0.07 & 0.09 & 0.02 & 0.02 \\
\ion{Mg}{I} & -0.25 $\pm$ 0.24 & 0.09 & 0.04 & 0.02 & 0.21 & -0.39 $\pm$ 0.21 & 0.07 & 0.07 & 0.03 & 0.18 & -0.37 $\pm$ 0.21 & 0.09 & 0.11 & 0.04 & 0.14 \\
\ion{Mg}{II} & -0.09 $\pm$ 0.16 & 0.11 & 0.05 & 0.03 & 0.11 & -0.09 $\pm$ 0.13 & 0.11 & 0.03 & 0.01 & 0.07 & -0.09 $\pm$ 0.14 & 0.12 & 0.02 & 0.01 & 0.07 \\
\ion{Al}{I} & -0.69 $\pm$ 0.26 & 0.10 & 0.07 & 0.01 & 0.23 & -1.23 $\pm$ 0.14 & 0.08 & 0.06 & 0.01 & 0.11 & -0.97 $\pm$ 0.16 & 0.07 & 0.11 & 0.03 & 0.09 \\
\ion{Si}{II} & -0.34 $\pm$ 0.15 & 0.08 & 0.06 & 0.05 & 0.09 & -0.28 $\pm$ 0.16 & 0.09 & 0.08 & 0.04 & 0.10 & -0.34 $\pm$ 0.11 & 0.09 & 0.03 & 0.04 & 0.05 \\
\ion{Ca}{I} & -0.23 $\pm$ 0.28 & 0.08 & 0.11 & 0.02 & 0.24 &   & & & & & -0.39 $\pm$ 0.42 & 0.09 & 0.32 & 0.05 & 0.24 \\
\ion{Ca}{II} & -0.25 $\pm$ 0.09 & 0.08 & 0.02 & 0.03 & 0.01 & -0.11 $\pm$ 0.12 & 0.11 & 0.06 & 0.02 & 0.01 & -0.28 $\pm$ 0.15 & 0.09 & 0.11 & 0.01 & 0.01 \\
\ion{Sc}{II} & -0.53 $\pm$ 0.21 & 0.04 & 0.05 & 0.08 & 0.19 & -0.27 $\pm$ 0.21 & 0.05 & 0.11 & 0.06 & 0.17 & -0.39 $\pm$ 0.19 & 0.05 & 0.14 & 0.03 & 0.12 \\
\ion{Ti}{II} & -0.41 $\pm$ 0.19 & 0.03 & 0.02 & 0.05 & 0.19 & -0.16 $\pm$ 0.19 & 0.03 & 0.05 & 0.04 & 0.18 & -0.19 $\pm$ 0.19 & 0.02 & 0.08 & 0.02 & 0.17 \\
\ion{Cr}{II} & -0.33 $\pm$ 0.10 & 0.03 & 0.02 & 0.04 & 0.09 & -0.37 $\pm$ 0.11 & 0.05 & 0.04 & 0.03 & 0.08 & -0.40 $\pm$ 0.08 & 0.04 & 0.04 & 0.02 & 0.06 \\
\ion{Mn}{I} & -0.26 $\pm$ 0.12 & 0.10 & 0.06 & 0.01 & 0.04 & -0.28 $\pm$ 0.16 & 0.07 & 0.14 & 0.03 & 0.02 &   & & & & \\
\ion{Fe}{I} & -0.31 $\pm$ 0.13 & 0.01 & 0.05 & 0.01 & 0.12 & -0.51 $\pm$ 0.13 & 0.02 & 0.10 & 0.01 & 0.08 & -0.50 $\pm$ 0.14 & 0.02 & 0.13 & 0.02 & 0.04 \\
\ion{Fe}{II} & -0.28 $\pm$ 0.16 & 0.02 & 0.01 & 0.05 & 0.16 & -0.43 $\pm$ 0.16 & 0.03 & 0.03 & 0.03 & 0.15 & -0.45 $\pm$ 0.13 & 0.02 & 0.06 & 0.02 & 0.12 \\
\ion{Ni}{II} &   & & & & &   & & & & & -0.31 $\pm$ 0.14 & 0.09 & 0.07 & 0.01 & 0.08 \\
\ion{Sr}{II} & 0.03 $\pm$ 0.26 & 0.07 & 0.09 & 0.06 & 0.23 & -0.72 $\pm$ 0.30 & 0.07 & 0.16 & 0.04 & 0.24 & -0.58 $\pm$ 0.31 & 0.07 & 0.18 & 0.01 & 0.24 \\
\ion{Y}{II} & 0.14 $\pm$ 0.10 & 0.07 & 0.03 & 0.05 & 0.03 & -0.07 $\pm$ 0.10 & 0.07 & 0.06 & 0.04 & 0.02 & -0.08 $\pm$ 0.12 & 0.07 & 0.10 & 0.02 & 0.01 \\
\ion{Zr}{II} & 0.04 $\pm$ 0.11 & 0.08 & 0.05 & 0.05 & 0.01 & -0.21 $\pm$ 0.12 & 0.11 & 0.03 & 0.03 & 0.05 & 0.03 $\pm$ 0.12 & 0.09 & 0.07 & 0.02 & 0.01 \\
\ion{Ba}{II} & 0.25 $\pm$ 0.19 & 0.06 & 0.05 & 0.02 & 0.17 & -0.22 $\pm$ 0.19 & 0.15 & 0.11 & 0.01 & 0.05 & -0.10 $\pm$ 0.18 & 0.12 & 0.14 & 0.01 & 0.02 \\
\hline
 & \multicolumn{5}{c}{ V435 Car } & \multicolumn{5}{c}{ HAT-P-49 } & \multicolumn{5}{c}{ $\lambda$ Bo\"otis } \\
\hline
\ion{C}{I} & -0.07 $\pm$ 0.08 & 0.07 & 0.02 & 0.01 & 0.02 & 0.04 $\pm$ 0.06 & 0.04 & 0.02 & 0.04 & 0.01 & -0.37 $\pm$ 0.08 & 0.06 & 0.05 & 0.01 & 0.01 \\
\ion{O}{I} &   & & & & &   & & & & & -0.32 $\pm$ 0.13 & 0.13 & 0.01 & 0.02 & 0.01 \\
\ion{Mg}{I} & -0.09 $\pm$ 0.20 & 0.08 & 0.09 & 0.11 & 0.11 & -0.09 $\pm$ 0.12 & 0.07 & 0.06 & 0.07 & 0.03 & -2.26 $\pm$ 0.13 & 0.09 & 0.09 & 0.01 & 0.04 \\
\ion{Mg}{II} & 0.32 $\pm$ 0.20 & 0.14 & 0.08 & 0.08 & 0.09 & 0.25 $\pm$ 0.19 & 0.10 & 0.08 & 0.01 & 0.14 & -2.07 $\pm$ 0.13 & 0.13 & 0.02 & 0.01 & 0.02 \\
\ion{Al}{I} & -0.60 $\pm$ 0.22 & 0.10 & 0.06 & 0.06 & 0.18 & -0.53 $\pm$ 0.15 & 0.10 & 0.08 & 0.03 & 0.07 &   & & & & \\
\ion{Si}{II} & 0.07 $\pm$ 0.32 & 0.25 & 0.06 & 0.02 & 0.19 &   & & & & &   & & & & \\
\ion{Ca}{I} & -0.06 $\pm$ 0.38 & 0.14 & 0.21 & 0.14 & 0.24 & -0.08 $\pm$ 0.15 & 0.10 & 0.08 & 0.07 & 0.05 & -2.31 $\pm$ 0.20 & 0.13 & 0.15 & 0.01 & 0.02 \\
\ion{Ca}{II} & -0.08 $\pm$ 0.15 & 0.14 & 0.06 & 0.01 & 0.01 &   & & & & &   & & & & \\
\ion{Sc}{II} & -0.06 $\pm$ 0.30 & 0.07 & 0.12 & 0.04 & 0.26 & 0.24 $\pm$ 0.32 & 0.27 & 0.02 & 0.04 & 0.18 &   & & & & \\
\ion{Ti}{II} & 0.12 $\pm$ 0.15 & 0.04 & 0.04 & 0.03 & 0.14 & 0.11 $\pm$ 0.12 & 0.02 & 0.02 & 0.04 & 0.11 &   & & & & \\
\ion{Cr}{II} & 0.08 $\pm$ 0.12 & 0.06 & 0.03 & 0.01 & 0.10 & 0.07 $\pm$ 0.07 & 0.02 & 0.01 & 0.04 & 0.05 &   & & & & \\
\ion{Mn}{I} & -0.22 $\pm$ 0.22 & 0.08 & 0.10 & 0.03 & 0.17 & -0.10 $\pm$ 0.08 & 0.02 & 0.04 & 0.01 & 0.06 &   & & & & \\
\ion{Fe}{I} & -0.15 $\pm$ 0.18 & 0.03 & 0.09 & 0.05 & 0.15 & -0.05 $\pm$ 0.16 & 0.01 & 0.06 & 0.02 & 0.15 & -2.07 $\pm$ 0.13 & 0.06 & 0.12 & 0.01 & 0.02 \\
\ion{Fe}{II} & -0.11 $\pm$ 0.09 & 0.04 & 0.05 & 0.02 & 0.06 & 0.00 $\pm$ 0.10 & 0.02 & 0.04 & 0.02 & 0.09 & -2.04 $\pm$ 0.10 & 0.08 & 0.05 & 0.02 & 0.02 \\
\ion{Ni}{II} &   & & & & & 0.15 $\pm$ 0.13 & 0.07 & 0.06 & 0.04 & 0.08 &   & & & & \\
\ion{Zn}{I} &   & & & & & -0.32 $\pm$ 0.10 & 0.07 & 0.04 & 0.01 & 0.07 &   & & & & \\
\ion{Sr}{II} & -0.39 $\pm$ 0.33 & 0.11 & 0.20 & 0.02 & 0.24 & 0.31 $\pm$ 0.11 & 0.07 & 0.06 & 0.01 & 0.07 & -2.52 $\pm$ 0.18 & 0.14 & 0.12 & 0.01 & 0.01 \\
\ion{Y}{II} & 0.07 $\pm$ 0.09 & 0.08 & 0.03 & 0.03 & 0.01 & 0.06 $\pm$ 0.11 & 0.05 & 0.02 & 0.05 & 0.08 &   & & & & \\
\ion{Zr}{II} & -0.26 $\pm$ 0.18 & 0.14 & 0.10 & 0.04 & 0.01 & 0.13 $\pm$ 0.08 & 0.04 & 0.02 & 0.04 & 0.04 &   & & & & \\
\ion{Ba}{II} & 0.21 $\pm$ 0.28 & 0.10 & 0.09 & 0.03 & 0.24 & 0.53 $\pm$ 0.24 & 0.09 & 0.04 & 0.01 & 0.22 &   & & & & \\
\hline
& \multicolumn{5}{c}{ Vega } & \multicolumn{5}{c}{ WASP-167 } & \multicolumn{5}{c}{ WASP-189 } \\
\hline
\ion{C}{I} & -0.33 $\pm$ 0.11 & 0.03 & 0.10 & 0.04 & 0.01 & 0.09 $\pm$ 0.12 & 0.09 & 0.02 & 0.04 & 0.07 & -0.40 $\pm$ 0.11 & 0.09 & 0.04 & 0.01 & 0.04 \\
\ion{O}{I} & -0.36 $\pm$ 0.10 & 0.09 & 0.01 & 0.04 & 0.01 &   & & & & &   & & & & \\
\ion{Mg}{I} & -0.57 $\pm$ 0.21 & 0.07 & 0.16 & 0.06 & 0.11 & -0.02 $\pm$ 0.13 & 0.09 & 0.06 & 0.07 & 0.03 & 0.06 $\pm$ 0.24 & 0.10 & 0.06 & 0.04 & 0.21 \\
\ion{Mg}{II} & -0.53 $\pm$ 0.07 & 0.05 & 0.02 & 0.04 & 0.03 & 0.37 $\pm$ 0.20 & 0.16 & 0.06 & 0.01 & 0.11 & 0.19 $\pm$ 0.24 & 0.21 & 0.04 & 0.02 & 0.11 \\
\ion{Al}{I} & -1.13 $\pm$ 0.17 & 0.04 & 0.16 & 0.04 & 0.02 & -0.43 $\pm$ 0.18 & 0.16 & 0.06 & 0.03 & 0.04 & -0.19 $\pm$ 0.34 & 0.29 & 0.10 & 0.01 & 0.15 \\
\ion{Si}{II} & -0.61 $\pm$ 0.10 & 0.04 & 0.03 & 0.09 & 0.02 &   & & & & & 0.08 $\pm$ 0.22 & 0.20 & 0.02 & 0.04 & 0.08 \\
\ion{Ca}{I} & -0.87 $\pm$ 0.25 & 0.06 & 0.22 & 0.09 & 0.04 &  0.08 $\pm$ 0.21 & 0.16 & 0.08 & 0.07 & 0.09 & 0.04 $\pm$ 0.33 & 0.15 & 0.17 & 0.01 & 0.24 \\
\ion{Ca}{II} & -0.76 $\pm$ 0.19 & 0.06 & 0.16 & 0.08 & 0.03 & 0.32 $\pm$ 0.18 & 0.16 & 0.04 & 0.08 & 0.01 & -0.14 $\pm$ 0.15 & 0.15 & 0.03 & 0.01 & 0.01 \\
\ion{Sc}{II} & -0.98 $\pm$ 0.14 & 0.06 & 0.12 & 0.02 & 0.03 & 0.17 $\pm$ 0.22 & 0.16 & 0.02 & 0.06 & 0.13 & 0.05 $\pm$ 0.28 & 0.16 & 0.05 & 0.05 & 0.22 \\
\ion{Ti}{II} & -0.65 $\pm$ 0.10 & 0.01 & 0.09 & 0.03 & 0.02 & 0.31 $\pm$ 0.16 & 0.04 & 0.04 & 0.02 & 0.15 & 0.06 $\pm$ 0.12 & 0.03 & 0.02 & 0.04 & 0.11 \\
\ion{Cr}{II} & -0.57 $\pm$ 0.07 & 0.02 & 0.05 & 0.03 & 0.01 & 0.27 $\pm$ 0.09 & 0.04 & 0.02 & 0.04 & 0.06 & 0.12 $\pm$ 0.09 & 0.05 & 0.03 & 0.03 & 0.06 \\
\ion{Mn}{I} & -0.78 $\pm$ 0.20 & 0.04 & 0.20 & 0.03 & 0.01 & 0.23 $\pm$ 0.13 & 0.07 & 0.04 & 0.02 & 0.10 & 0.11 $\pm$ 0.11 & 0.07 & 0.04 & 0.01 & 0.08 \\
\ion{Fe}{I} & -0.73 $\pm$ 0.15 & 0.01 & 0.15 & 0.05 & 0.01 & 0.08 $\pm$ 0.15 & 0.03 & 0.06 & 0.03 & 0.13 & -0.01 $\pm$ 0.15 & 0.03 & 0.07 & 0.01 & 0.13 \\
\ion{Fe}{II} & -0.72 $\pm$ 0.08 & 0.01 & 0.06 & 0.04 & 0.03 & 0.03 $\pm$ 0.15 & 0.07 & 0.04 & 0.01 & 0.13 & 0.05 $\pm$ 0.13 & 0.04 & 0.06 & 0.02 & 0.11 \\
\ion{Ni}{II} & -0.73 $\pm$ 0.06 & 0.04 & 0.03 & 0.04 & 0.01 &   & & & & & 0.59 $\pm$ 0.24 & 0.10 & 0.04 & 0.03 & 0.22 \\
\ion{Zn}{I} &   & & & & & -0.17 $\pm$ 0.14 & 0.11 & 0.03 & 0.02 & 0.07 &   & & & & \\
\ion{Sr}{II} & -1.15 $\pm$ 0.20 & 0.06 & 0.18 & 0.03 & 0.04 & 0.44 $\pm$ 0.20 & 0.16 & 0.08 & 0.01 & 0.09 & 0.58 $\pm$ 0.26 & 0.09 & 0.09 & 0.04 & 0.22 \\
\ion{Y}{II} &   & & & & & 0.22 $\pm$ 0.22 & 0.13 & 0.04 & 0.05 & 0.16 & 0.92 $\pm$ 0.17 & 0.09 & 0.04 & 0.05 & 0.14 \\
\ion{Zr}{II} & -0.59 $\pm$ 0.14 & 0.08 & 0.11 & 0.03 & 0.01 & 0.31 $\pm$ 0.21 & 0.16 & 0.02 & 0.01 & 0.13 & 0.61 $\pm$ 0.17 & 0.10 & 0.05 & 0.02 & 0.13 \\
\ion{Ba}{II} & -0.48 $\pm$ 0.22 & 0.14 & 0.16 & 0.05 & 0.02 & 0.58 $\pm$ 0.20 & 0.13 & 0.04 & 0.01 & 0.14 & 0.89 $\pm$ 0.23 & 0.11 & 0.06 & 0.02 & 0.19 \\
\hline
\end{tabular}
\normalsize
\label{table.abunds3}
\end{table*}

\setcounter{table}{0}
\begin{table*}
\centering
\caption{Continued. Abundances derived in this work. For each specie, we present average$\pm$total error and the uncertainties e1, ..., e4.}
\scriptsize
\begin{tabular}{lccccccccccccccc}
\hline
Specie & [N/H] & e$_{1}$ & e$_{2}$ & e$_{3}$ & e$_{4}$ \\
\hline
 & \multicolumn{5}{c}{ $\zeta$ Del } & \multicolumn{5}{c}{} & \multicolumn{5}{c}{} \\
\hline
\ion{C}{I} & -0.06 $\pm$ 0.07 & 0.06 & 0.04 & 0.01 & 0.01 &  &  \\
\ion{O}{I} & 0.08 $\pm$ 0.13 & 0.13 & 0.03 & 0.02 & 0.01 &  &  \\
\ion{Mg}{I} & -0.50 $\pm$ 0.23 & 0.06 & 0.07 & 0.04 & 0.21 &  &  \\
\ion{Mg}{II} & -0.38 $\pm$ 0.15 & 0.09 & 0.04 & 0.01 & 0.11 &  &  \\
\ion{Si}{II} & -0.30 $\pm$ 0.18 & 0.14 & 0.07 & 0.07 & 0.07 &  &  \\
\ion{Ca}{I} & -1.10 $\pm$ 0.34 & 0.11 & 0.23 & 0.02 & 0.22 &  &  \\
\ion{Sc}{II} & -0.77 $\pm$ 0.22 & 0.11 & 0.09 & 0.02 & 0.16 &  &  \\
\ion{Ti}{II} & -0.48 $\pm$ 0.15 & 0.03 & 0.05 & 0.03 & 0.14 &  &  \\
\ion{Cr}{II} & -0.53 $\pm$ 0.07 & 0.04 & 0.03 & 0.02 & 0.05 &  &  \\
\ion{Fe}{I} & -0.68 $\pm$ 0.12 & 0.02 & 0.10 & 0.02 & 0.07 &  &  \\
\ion{Fe}{II} & -0.59 $\pm$ 0.13 & 0.03 & 0.04 & 0.03 & 0.12 &  &  \\
\ion{Sr}{II} & -1.13 $\pm$ 0.31 & 0.16 & 0.13 & 0.03 & 0.23 &  &  \\
\ion{Ba}{II} & -0.53 $\pm$ 0.18 & 0.14 & 0.11 & 0.01 & 0.01 &  &  \\
\hline
\end{tabular}
\normalsize
\label{table.abunds4}
\end{table*}

\section{Comments about individual stars}

In this section we briefly comment the chemical features that we observe on individual stars.

59 Dra: This object presents mostly solar values. 

$\beta$ Cir: This star presents some slightly subsolar abundances (such as [Fe/H]=-0.25$\pm$0.26 dex) but also suprasolar values
             of s-process elements (Sr, Y, Zr and Ba by $\sim$+0.3 dex). In particular, C is clearly subsolar (-0.49$\pm$0.17 dex). 
             We do not identify a $\lambda$ Bo\"otis pattern.

$\beta$ Pic: This star presents some sligthly subsolar values of some species (for example [Fe/H]=-0.28$\pm$0.14 dex) but also solar values
             of Mg, Ca, Sc, Ti, Y and Ba. C resulted slightly subsolar (-0.20$\sim$0.11 dex). 
             Then, we do not detect clear $\lambda$ Bo\"otis signature. 
             Its $\lambda$ Bo\"otis classification was initially suggested based on its evolutionary status and the presence 
             of a circumstellar disk \citep{king-patten92}. However, its $\lambda$ Bo\"otis nature was then ruled out by using optical
             spectra \citep{holweger97} and also using the ratio of UV lines \citep{cheng16}.
           
BU Psc: This star resulted sligthly subsolar in some elements (Al, Cr, Fe) although solar in other species (Si, Sc, Ti, Y, Ba).

Fomalhaut: This star shows mostly solar values (such as [Fe/H]=0.12$\pm$0.19 dex), and slightly enriched in Ni and s-process
          elements (Sr, Y, Zr and Ba). We do not observe $\lambda$ Bo\"otis-like characteristics.

HD 29391: Most metals in this star show solar abundances.

HD 105850: This object shows some slightly depleted elements (Cr, Fe) and some solar species (Ca, Sc, Ti, Zr, Ba).
           We do not identify a $\lambda$ Bo\"otis pattern.
           
HD 110058: We observe a clear $\lambda$ Bo\"otis pattern. Most metals show very low abundances (between 1-2 dex),
           while C is almost solar ([C/H]=-0.14$\pm$0.06 dex), in agreement with most $\lambda$ Bo\"otis stars.

HD 115820: Most elements in this object show solar abundances.

HD 120326: Mostly solar values. Some elements are very slightly subsolar (Mn, Fe) while other show solar values (Ca, Sc, Ti, Cr).

HD 129926: \citet{corbally84} classified this star as a close binary (F0 V + G1 V), with V magnitudes of 5.10 and 7.12
           and a separation of $\sim$8.2 arcsec. He reported strong Sr and \ion{Fe}{II} for the primary, and strong
           lines for the secondary. 
           The HARPS data present a metal-rich spectra, showing similarities but also differences to Am and ApSi stars.
           For instance, Cr, Fe, and Sr show lower values than ApSi stars, while Ca and Sc show higher values than Am stars.

HD 133803: This star shows solar (or slightly suprasolar) abundances in most elements.

HD 146624: This star presents solar abundances (for example Mg, Si, Ti, Fe). C and O both shows subsolar values.

HD 153053: Some species present slightly subsolar values (such as [Fe/H]=-0.29$\pm$0.14 dex), while s-process elements show
           slightly suprasolar abundances (Sr, Y, Zr and Ba by $\sim$+0.3 dex).

HD 156751: This star would classify as mild-$\lambda$ Bo\"otis, if we only inspect those elements with atomic number z$<$28,
           showing subsolar values of metals (for example [Fe/H]=-0.47$\pm$0.17 dex) and near-solar C (-0.13$\pm$0.08 dex). 
           However, we also measured solar values for some s-process elements (Sr, Y and Ba), being somewhat higher than
           average $\lambda$ Bo\"otis stars. 
           
HD 159492: This object presents mostly solar abundances (such as Mg, Si, Sc, Ti, Cr, Fe).

HD 169142: This star presents a clear $\lambda$ Bo\"otis pattern. Subsolar abundances of most metals (between 0.50-0.75 dex) together
           with a solar abundance of C ([C/H]=0.13$\pm$0.09 dex).
           This star was previously identified as a $\lambda$ Bo\"otis object \citep{folsom12,gray17}, although for some authors
           their $\lambda$ Bo\"otis pattern is less clear \citep{murphy15}. Interestingly, \citet{gray17} suggest a likely spectral variation
           for this star, proposing a follow up to see if its $\lambda$ Bo\"otis nature could also vary.

HD 188228: Mostly solar values (Mg, Si, Ti, Cr, Fe), with subsolar C and suprasolar Ba.

HD 23281: Mostly solar values (Mg, Si, Ti, Fe), with slightly subsolar C and suprasolar values of s-process elements (Sr, Y, Zr and Ba). 

HD 50445: This star presents some subsolar abundances ([Fe/H]=-0.31$\pm$0.13 dex). However, it also shows solar
          values of Mg, Sr, Y, Zr and Ba unlike most $\lambda$ Bo\"otis stars.

HD 56537: This object shows some subsolar values (for example [Fe/H]=-0.51$\pm$0.13 dex).
          However, C also shows a low value ([C/H]=-0.42$\pm$0.13 dex)), while Ca, Ti, Y and Ba show almost solar values,
          different of most $\lambda$ Bo\"otis stars.

HD 88955: This star presents some subsolar abundances (such as [Fe/H]=-0.50$\pm$0.14 dex) but also a low value for C (-0.61$\pm$0.11 dex).
          However some s-process species show solar values (Y, Zr, Ba). Then, is different of most $\lambda$ Bo\"otis stars.

HD 95086: This object presents solar values of most metals.

HR 4502 A: This star presents mostly solar abundances.

HR 8799: This star presents a $\lambda$ Bo\"otis pattern. Subsolar metallic abundances (between 0.50-0.75 dex) together
         with a near-solar C ([C/H]=-0.11$\pm$0.07 dex). Its $\lambda$ Bo\"otis nature was also previously reported in the
          literature \citep{gray-kaye99,sadakane06}.

KELT-17: This object shows a chemically peculiar Am pattern. KELT-17 presents subsolar values of Ca and Sc, while other metals
         show overabundances (Ti, Cr, Mn, Fe, Sr, Y, Ba). The Am nature of this star was recently reported by \citet{saffe20}.
           
KELT-20: This star presents mostly subsolar values (for example [Fe/H]=-0.35$\pm$0.15 dex). 
        However, it also shows a low C abundance ([C/H]=-0.42$\pm$0.10 dex), also solar or suprasolar values of Si, Ni and Ba,
        different of average $\lambda$ Bo\"otis stars.

KELT-9: This star presents solar abundances for most metals, with slightly subsolar values for Al and Sr.

MASCARA-1: This object shows mostly solar values in general, with suprasolar abundances of Sr, Y, Zr and Ba.

V435 Car: This star presents solar values of most elements.

WASP-33:  This star shows mostly solar abundances, together with a suprasolar O abundance.
         We caution however that, for this star, the O abundance was derived using the \ion{O}{I} near-IR triplet at $\sim$7771 {\AA},
         which suffer of NLTE effects \citep[e.g.][]{sitnova13,przybilla11}.
         In this star, some s-process elements (Sr, Y and Ba) also show suprasolar values.

HAT-P-49: This object presents mostly solar values, with slightly suprasolar abundances of Sc, Sr and Ba.

$\lambda$ Bo\"otis: This object is more metal-poor than the average pattern of $\lambda$ Bo\"otis stars.
          In addition, this star shows C and O slightly subsolar rather than solar, 
          in agreement with previous works \citep{venn-lambert90,paunzen99,cheng19}.
          Then, we would consider this object as a rather extreme case of the $\lambda$ Bo\"otis class.
          
Vega:     This star presents subsolar abundances of most metals and slightly subsolar values of O (-0.36$\pm$0.10 dex): these
          features are, in principle, similar to other $\lambda$ Bo\"otis stars. However, C presents a lower abundance ([C/H]=-0.33$\pm$0.11 dex)
          than most $\lambda$ Bo\"otis stars. 
          A $\lambda$ Bo\"otis nature for this star was previously suggested \citep{yoon10}, however this class was then ruled
          out by \citet{cheng16}, in agreement with the present work.
           
WASP-167: This star presents a slightly metal-rich abundance pattern. However, it is different of Am stars (showing higher Ca, Sc and
          lower Cr, Sr, Y, Zr, Ba), different of ApSi stars (showing lower Ti, Cr, Mn, Fe) and different of HgMn stars
          (lower Mn and no Hg observed).

\begin{figure*}
\centering
\includegraphics[width=8cm]{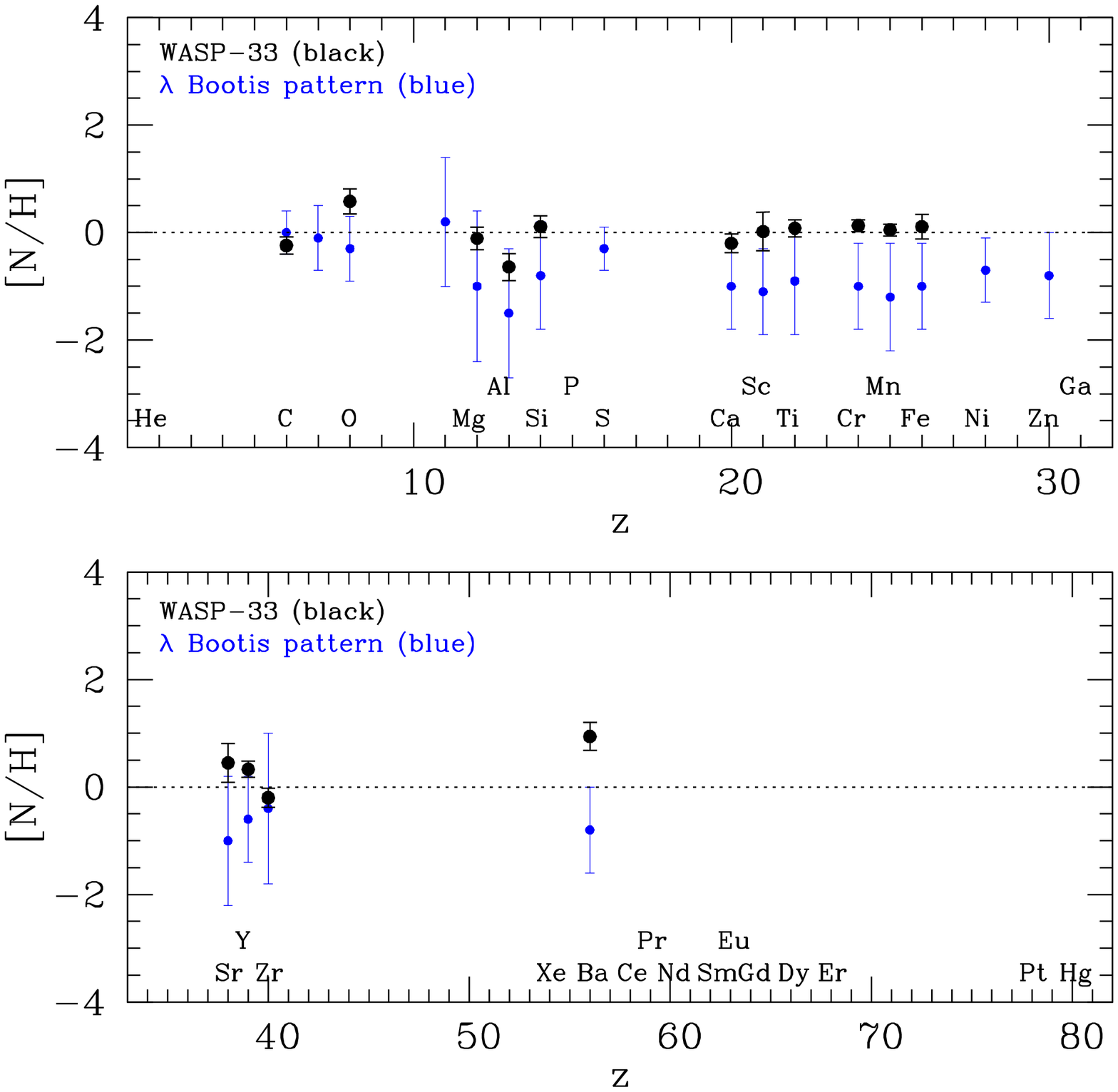}
\includegraphics[width=8cm]{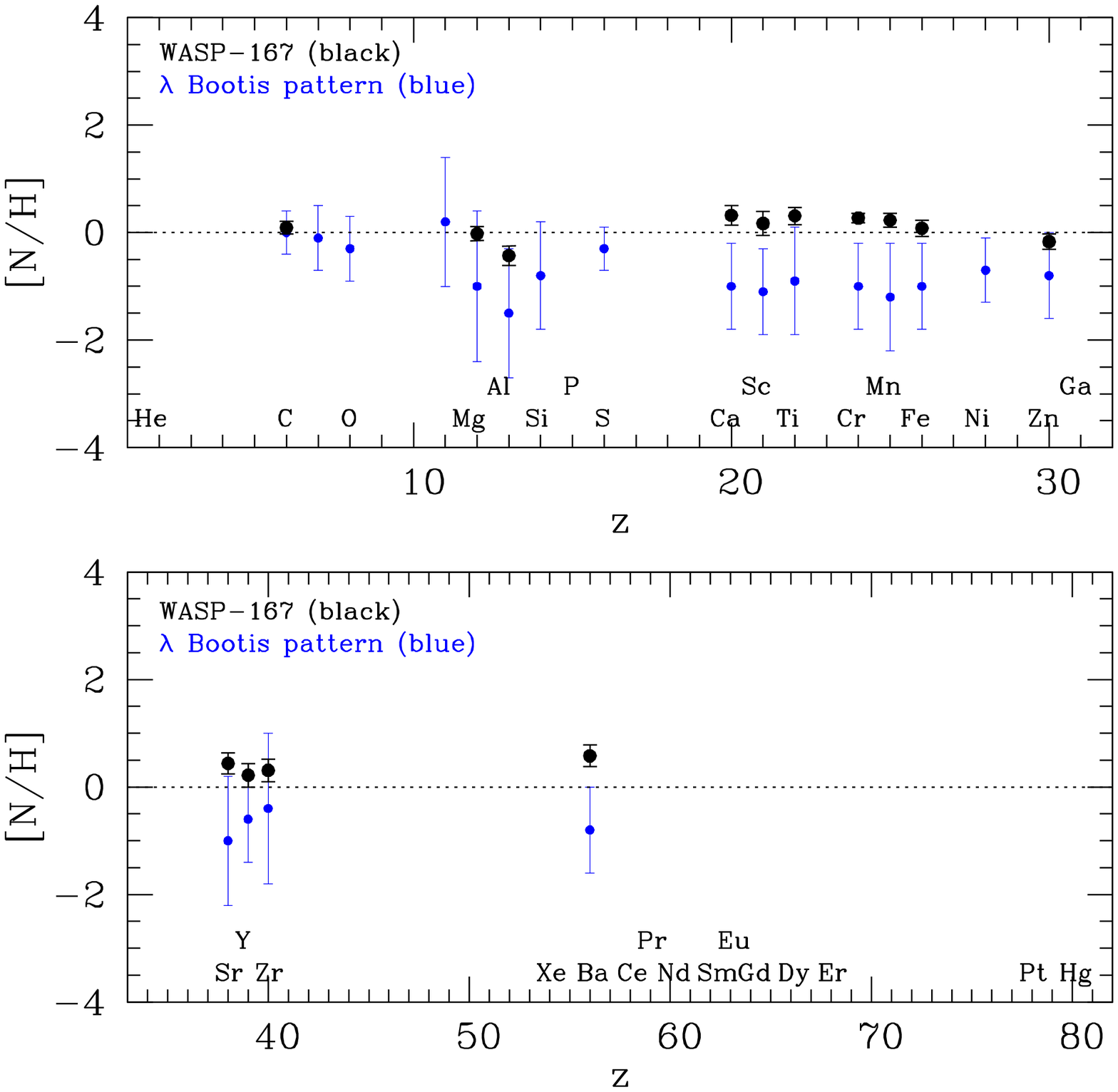}
\includegraphics[width=8cm]{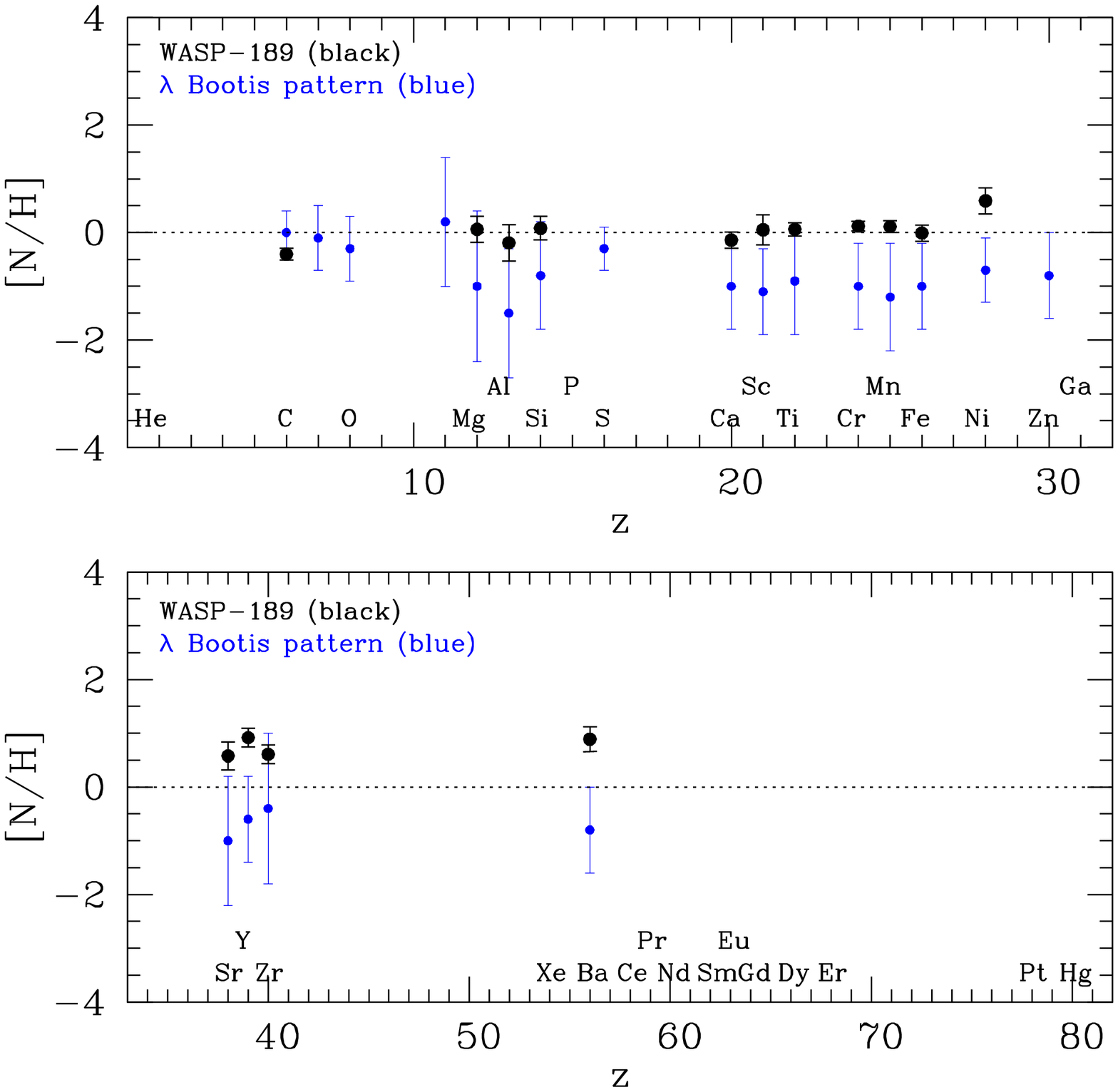}
\includegraphics[width=8cm]{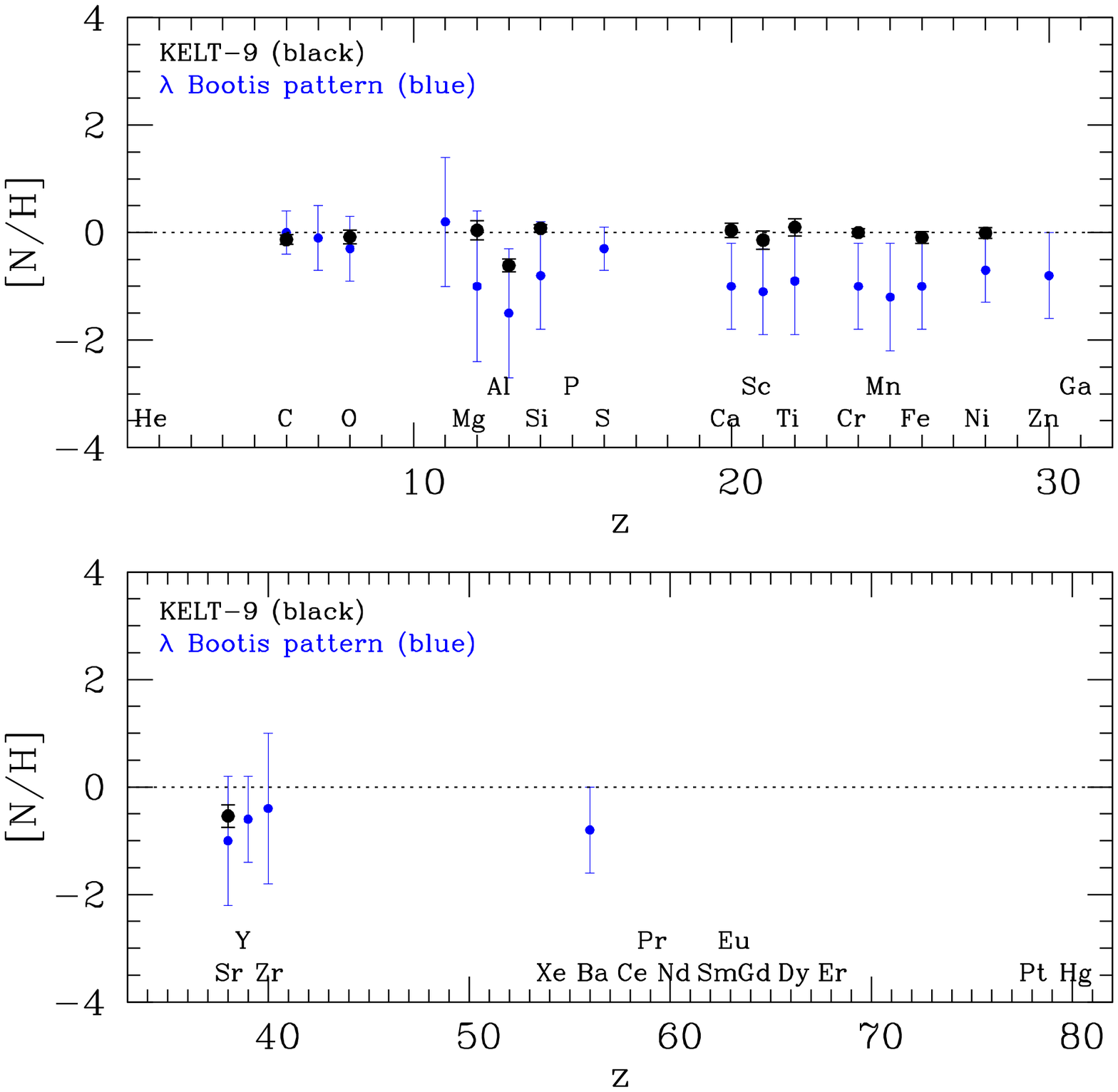}
\caption{Abundances of hot-Jupiter host stars (WASP-33, WASP-167, WASP-189 and KELT-9, in black)
compared to an average $\lambda$ Bo\"otis pattern (blue). 
We show two panels for each star, corresponding to elements with {z$<$32} and {z$>$32}.
See text for more details.}
\label{fig.hot-jup1}%
\end{figure*}

WASP-189: This object is listed in the \citet{renson-manfroid09}'s catalog of chemically peculiar stars with a "doubtful" A4m classification.
          Its suspected Am class was then ruled out \citep{anderson18}.
          This star presents Ca and Sc almost solar (different of Am stars), mostly solar abundances and some suprasolar s-process elements (Sr, Y, Zr and Ba),
          as we can see in the Fig. \ref{fig.hot-jup1}.

$\zeta$ Del: This star shows a clear $\lambda$ Bo\"otis pattern: subsolar metallic abundances (for example [Fe/H]=-0.68$\pm$0.12 dex),
             together with solar values of both [C/H] and [O/H] (-0.06$\pm$0.07 and 0.08$\pm$0.13 dex).

\section{Abundance pattern of hot-Jupiter stars}

We present in this section a number of Figures, comparing the abundances of early-type stars which host hot-Jupiter planets,
with the chemical pattern of $\lambda$ Bo\"otis stars.

\begin{figure*}
\centering
\includegraphics[width=8cm]{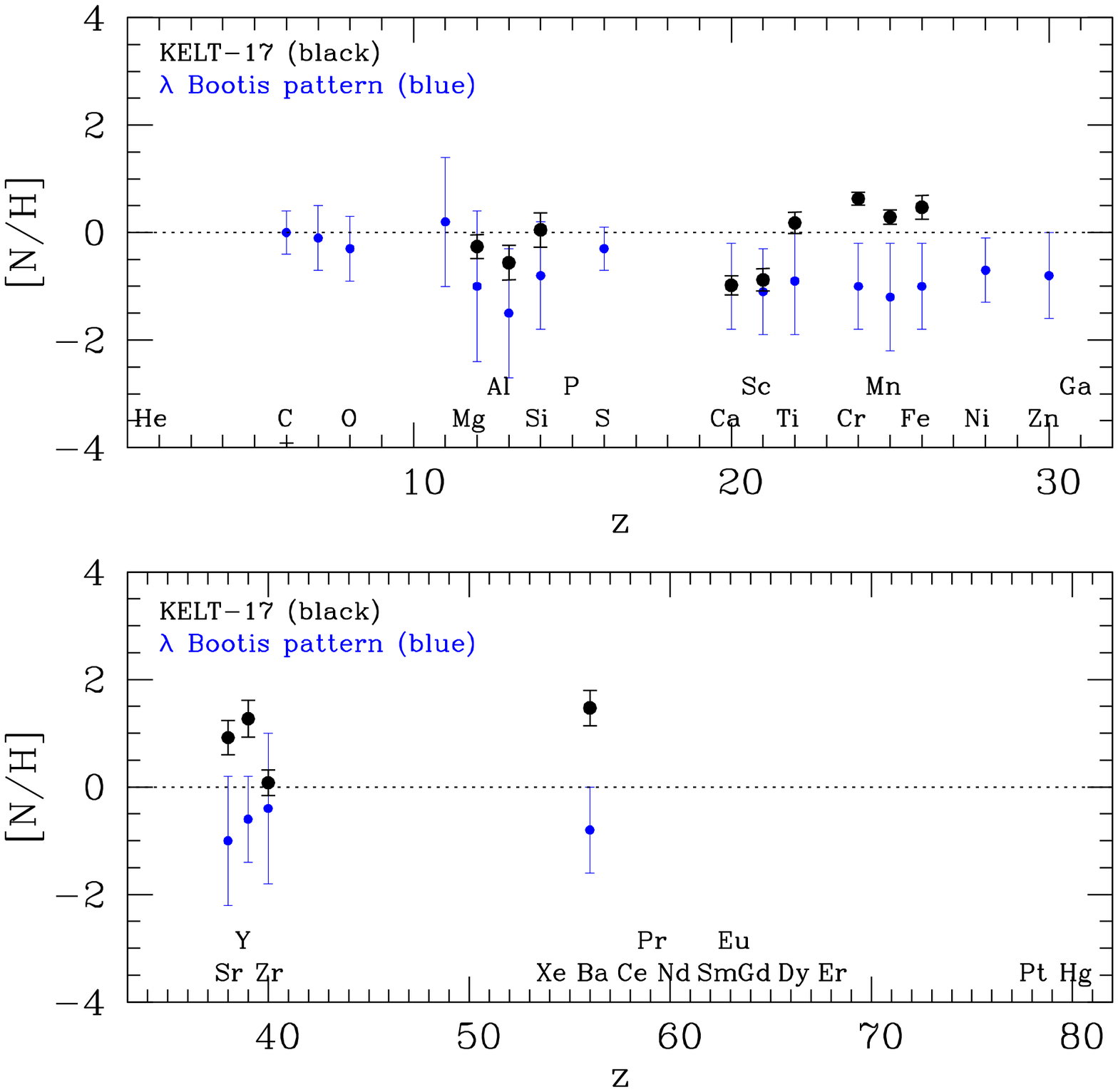}
\includegraphics[width=8cm]{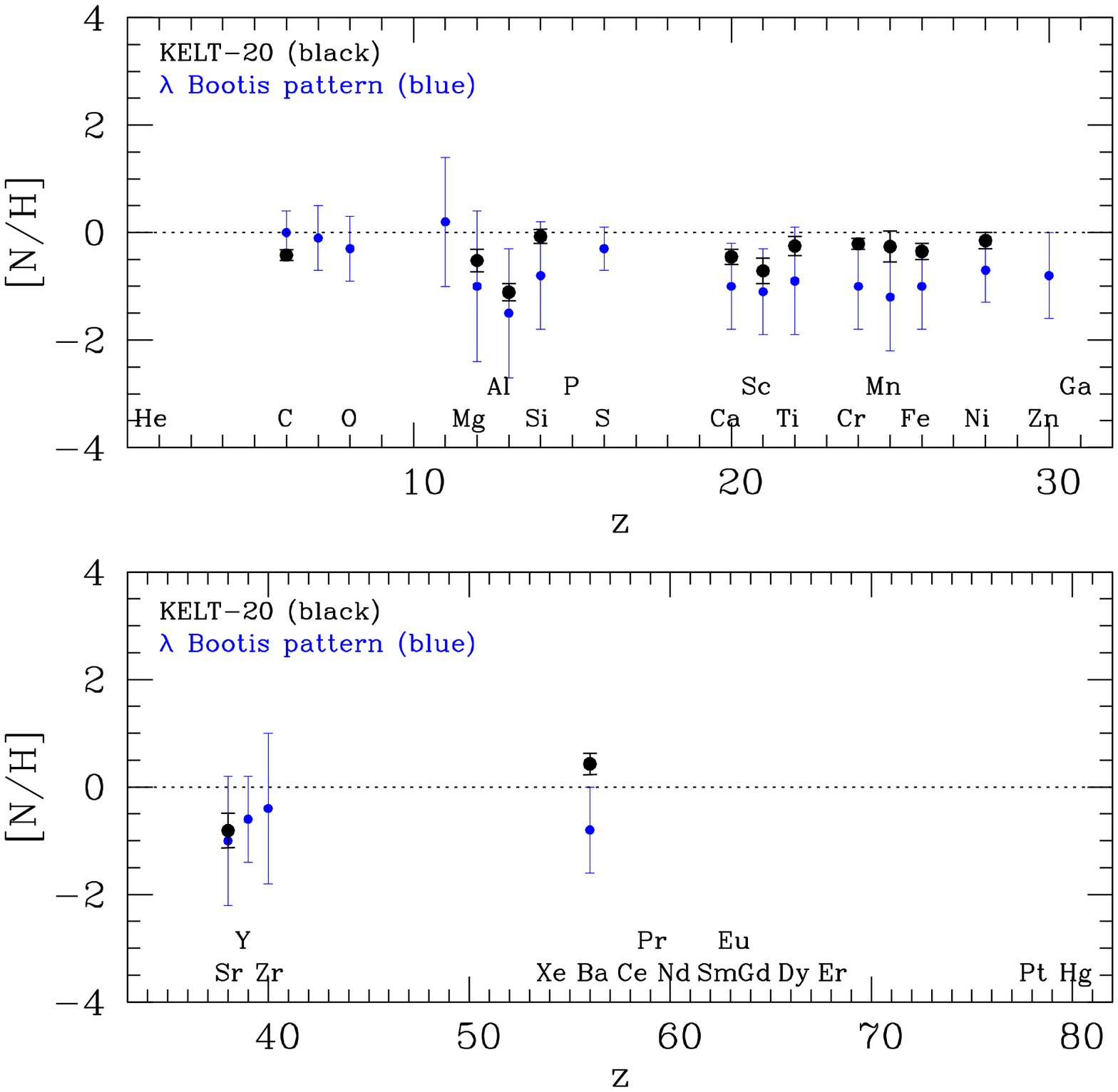}
\includegraphics[width=8cm]{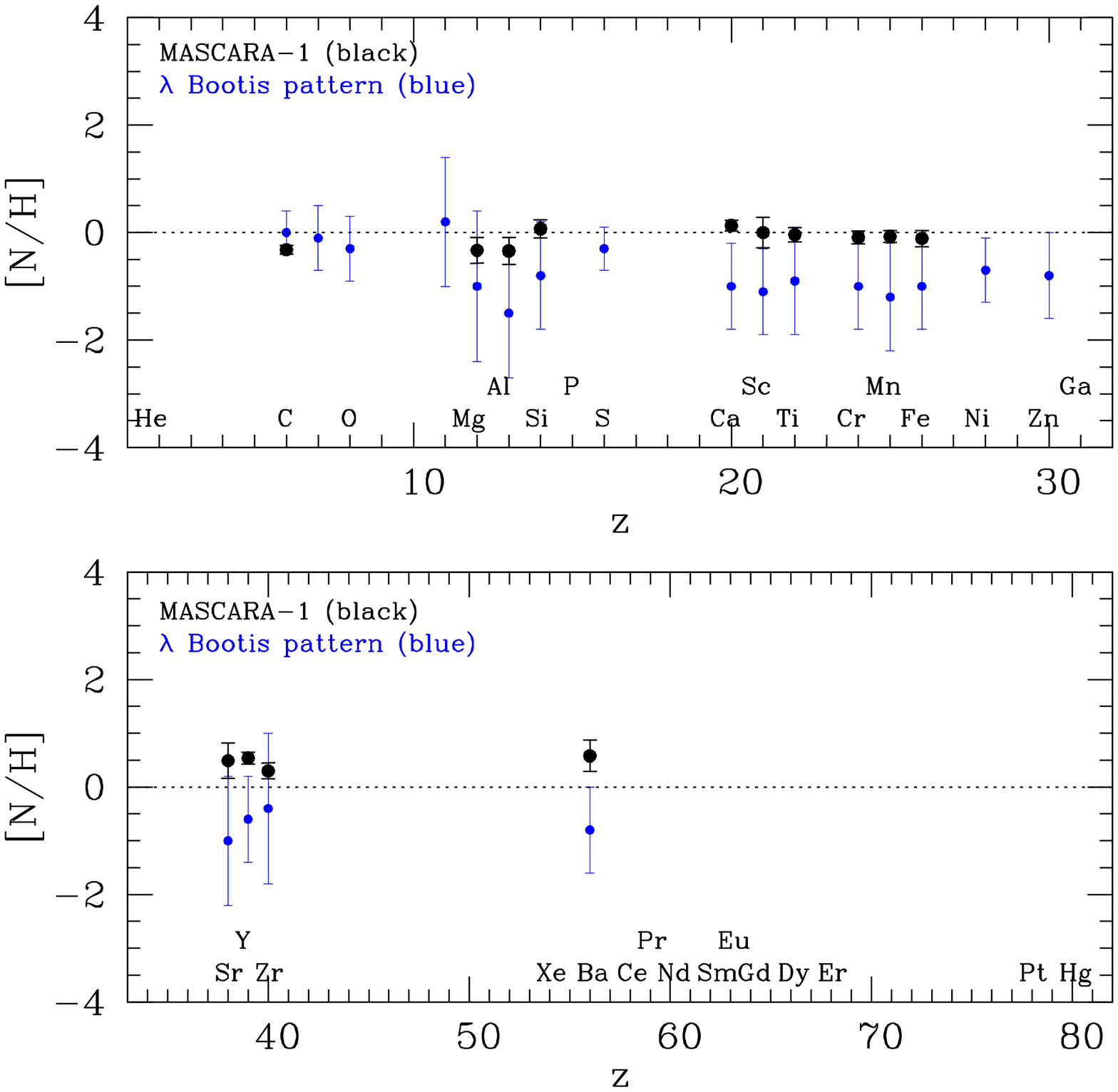}
\includegraphics[width=8cm]{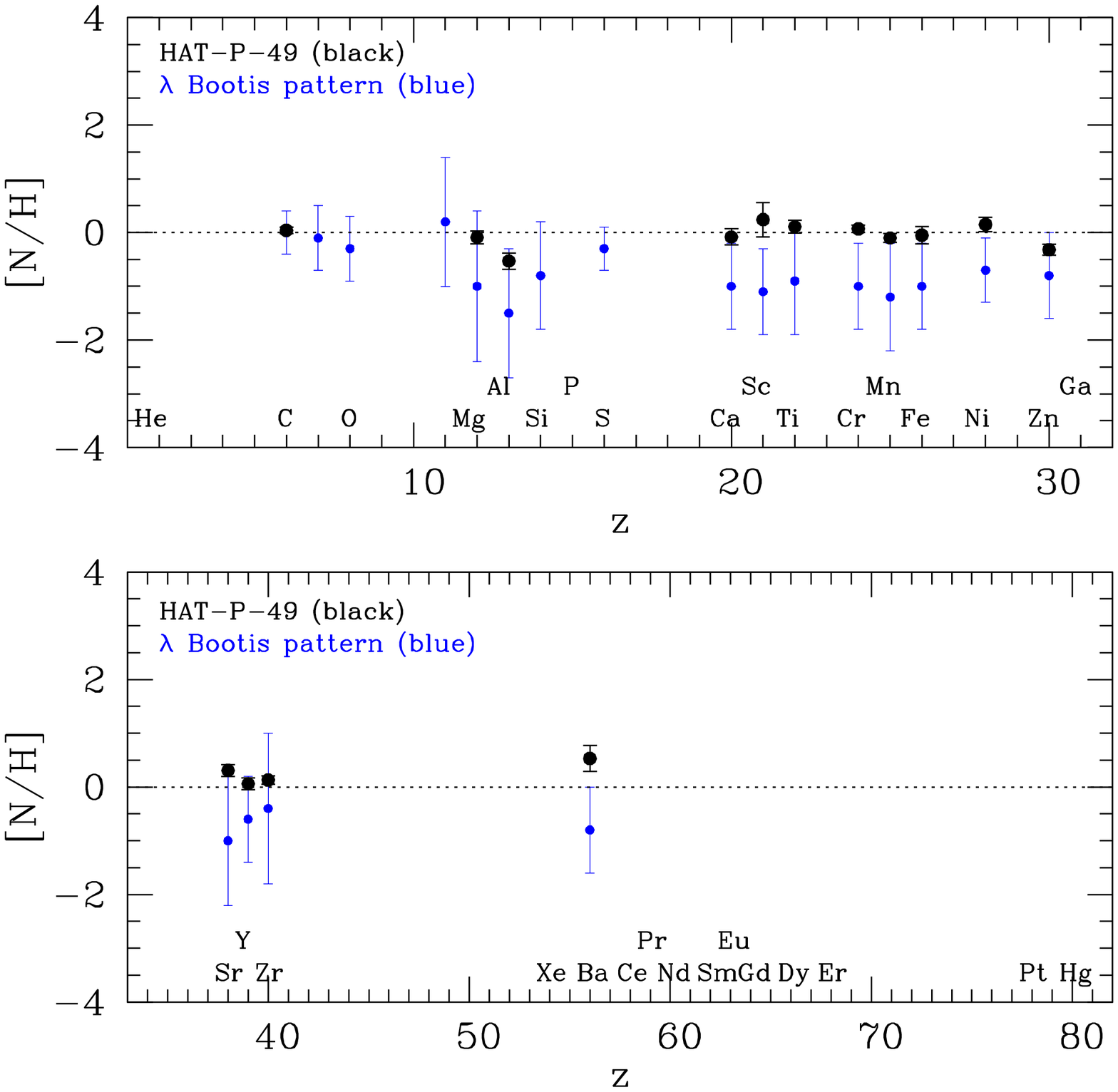}
\caption{Abundances of hot-Jupiter host stars (KELT-17, KELT-20, MASCARA-1 and HAT-P-49, in black)
compared to an average $\lambda$ Bo\"otis pattern (blue).
We show two panels for each star, corresponding to elements with {z$<$32} and {z$>$32}.
See text for more details.}
\label{fig.hot-jup2}%
\end{figure*}

\end{appendix}

\end{document}